\documentclass[twocolumn,aps,prb,showpacs,raggedbottom,nobalancelastpage,amssymb,superscriptaddress]{revtex4-1}

\usepackage{graphicx}
\usepackage{amsmath}
\usepackage{amssymb}
\usepackage{bm}
\usepackage[usenames]{color}
\usepackage{amsfonts}
\usepackage{appendix}
\usepackage{dsfont}
\usepackage{color}
\usepackage{braket}

\def \Fst{$1^\textrm{st}$ }
\def \Snd{$2^\textrm{nd}$ }
\def \Trd{$3^\textrm{rd}$ }

\usepackage{braket}
\usepackage{dcolumn}
\usepackage{wrapfig}
\usepackage{amstext}
\usepackage{gensymb}
\usepackage{cancel}
\usepackage{hyperref}
\usepackage[mathlines]{lineno}
\usepackage[english]{babel}

\newcommand\Ccancel[2][black]{\renewcommand\CancelColor{\color{#1}}\cancel{#2}}

\usepackage{xcolor}
\newsavebox\MBox
\newcommand\Cline[2][Red]{{\sbox\MBox{$#2$}%
		\rlap{\usebox\MBox}\color{#1}\rule[-1.2\dp\MBox]{\wd\MBox}{0.7pt}}}

\def\be{\begin{equation}}
\def\ee{\end{equation}}

\def\bs#1{\boldsymbol{#1}}

\begin{document}
	
	\title{Six-dimensional quantum Hall effect and three-dimensional topological pumps}
	
	\author{Ioannis Petrides}
	\affiliation{Institute for Theoretical Physics, ETH Zurich, 8093 Z{\"u}rich, Switzerland}
	\author{Hannah M. Price}
	\affiliation{\mbox{School of Physics and Astronomy, University of Birmingham, Edgbaston, Birmingham B15 2TT, United Kingdom}}
	\affiliation{INO-CNR BEC Center and Dipartimento di Fisica, Universit\`a di Trento, I-38123 Povo, Italy}
	\author{Oded Zilberberg}
	\affiliation{Institute for Theoretical Physics, ETH Zurich, 8093 Z{\"u}rich, Switzerland}

	\begin{abstract}
		Modern technological advances allow for the study of systems with additional synthetic dimensions. Using such approaches, higher-dimensional physics that was previously deemed to be of purely theoretical interest has now become an active field of research. In this work, we derive from first principles using a semiclassical equation-of-motion approach the bulk response of a six-dimensional Chern insulator. We find that in such a system a quantized bulk response appears with a quantization originating from a six-dimensional topological index: the \Trd Chern number. Alongside this unique six-dimensional response, we rigorously describe the lower even-dimensional Chern-type responses that can occur due to nonvanishing \Fst and \Snd Chern numbers in subspaces of the six-dimensional space. Last, we propose how to realize such a bulk response using three-dimensional topological charge pumps in cold atomic systems.
	\end{abstract} 
	
	
	\maketitle
	
	\section{Introduction}
	The introduction of topological concepts in physics has revolutionized our understanding of different phases of matter~\cite{RMP_TI, RMP_TI2, ozawareview}. Within this paradigm, systems are classified by global topological invariants that only take certain integer values and so cannot be smoothly varied. Instead, these invariants jump discontinuously across the topological phase transition connecting two distinct topological phases. A physical observable that depends on such a topological invariant is therefore ``topologically-protected" as it will be robust against perturbations that do not induce such a phase transition; this has important consequences, such as quantized bulk responses and robust edge states~\cite{schulz2000simultaneous,kellendonk2002edge}.
	
	A seminal example of a topological phase of matter is the two-dimensional (2D) quantum Hall (QH) system, in which the Hall conductance is precisely and robustly quantized in terms of fundamental constants~\cite{Klitzing:1980PRL}. In this system, the energy bands can be characterized by a topological invariant, called the first Chern number, which leads to the quantization of the Hall conductance~\cite{TKNN}. While this physics was first discovered for a 2D electronic material in a perpendicular magnetic field~\cite{Klitzing:1980PRL}, it is now increasingly relevant across a wide range of different platforms, thanks to the engineering of nonzero \Fst Chern numbers in, for example, 2D ultracold gases~\cite{Goldman:2016NatPhys,Goldman:2016NatPhys, cooper2018} and photonics~\cite{lu2016topological, khanikaev2017two, ozawareview}. 
	
	Remarkably, the 2D quantum Hall effect is just the first in a family of related quantized responses that can be accessed by changing the dimensionality of the system, i.e. the number of spatial dimensions along which a particle can move~\cite{Chiu:2016RMP}. The next new quantum Hall effect emerges as a nonlinear quantized response in a four-dimensional (4D) system, where the quantization is related to a 4D topological invariant called the \Snd Chern number~\cite{Avron1988,frohlich2000,IQHE4D,QiZhang,Price2015,prodan2016bulk}. It has been proposed to (i) explain the generation of magnetic fields in the early universe~\cite{frohlich2000}, (ii) reveal the topology of 2D quasicrystals~\cite{Kraus2013,kraus2016quasiperiodicity}, (iii) serve as a parent model for three-dimensional
	(3D) topological insulators\cite{QiZhang,li2013high,li2013topological,prodan2015virtual}, (iv) exhibit exotic quasiparticle excitations~\cite{IQHE4D}, and (v) be engineered directly in the laboratory by adding ``synthetic" dimensions to a system of atoms or photons~\cite{Price2015, Ozawa2016,Price2016}. In the latter, additional spatial dimensions are simulated by coupling sets of internal states such that particles move between different states just as they would hop between lattice sites along an extra direction~\cite{Boada2012,Celi:2012PRL, boada2015quantum, Mancini:2015Science, Stuhl:2015Science, luo2015quantum, Livi:2016PRL, yuan2016photonic, Ozawa2016, ozawa2017synthetic, Price:2017PRA,An:2017SciAdv}.  
	
	Higher-dimensional topological physics can also be probed experimentally by exploiting a powerful approach called topological charge pumping~\cite{Thouless83, Kunz1986, Kraus:2012a, Kraus:2012b, Kraus2013, Verbin:2015,prodan2015virtual, Lohse2016,Nakajima2016, Lohse2018, Zilberberg2018}. In a topological charge pump, a system is slowly and periodically ``pumped" over time, such that after each pump cycle, there is a quantized transport of particles across the system. The robust quantization of this transport can be related, through a mathematical mapping, to the quantization of the current response in a quantum Hall system with more spatial dimensions. Drawing on this correspondence, 1D topological pumps have been used to measure the \Fst Chern number~\cite{Lohse2016,Nakajima2016} and its corresponding boundary states~\cite{Kraus:2012a,Verbin:2015} usually associated with a 2D quantum Hall system. Recently, 2D topological pumps have been used to reconstruct the \Snd Chern number~\cite{Lohse2018} and the plethora of associated boundary phenomena~\cite{Zilberberg2018} of a 4D quantum Hall system for the first time. 
	
	In this paper, we develop these ideas further by showing how a 3D topological pump could be used to probe the six-dimensional (6D) quantum Hall effect. The latter has a quantized bulk response that emerges in a system with six or more spatial dimensions, and which is related to a 6D topological invariant: the \Trd Chern number~\cite{Nakahara, Chiu:2016RMP, Kunz1986, prodan2016bulk}. Unlike its 2D and 4D cousins~\cite{Price2015}, the 6D quantum Hall response only arises at third order in the perturbing electromagnetic fields, and can be understood from the interplay of an electric field with magnetic fields through two different planes. To illustrate the physics of this effect, we derive the 6D quantum Hall effect from a third-order semiclassical analysis, and demonstrate that all nontopological contributions to the current response vanish for a filled energy band, i.e.,  we obtain generalized 2D- and 4D-like \Fst and \Snd Chern number responses~\cite{Price2015} alongside a 6D topological \Trd Chern number response. Our results agree with algebraic K-theory derivations of \Trd Chern number bulk responses~\cite{prodan2016bulk}.
	
	We, further, demonstrate how the 6D quantum Hall effect can be experimentally accessible by introducing a minimal 6D model that can be mapped onto a 3D topological pump. Such a 3D pump could be realised, for example, by extending recent atomic experiments on 1D~\cite{Lohse2016,Nakajima2016} and 2D topological pumps~\cite{Lohse2018} to include 3D optical superlattices. 
	
	\subsection*{Outline}
	
	The structure of this paper is as follows: We begin in Sec.~\ref{sec:semiclassical} by introducing the relevant geometrical and topological properties of energy bands, before developing a third-order semiclassical approach to calculate the quantum Hall response in a system with six spatial dimensions. Then, in Sec.~\ref{sec:pumps}, we show how the 6D quantum Hall effect could be probed using a 3D topological pump. We illustrate this for an explicit model that has energy bands characterized by a nontrivial six-dimensional topological invariant, namely, the \Trd Chern number. 
	
	
	
	\begin{figure*}[ht]
		\includegraphics[width=\linewidth]{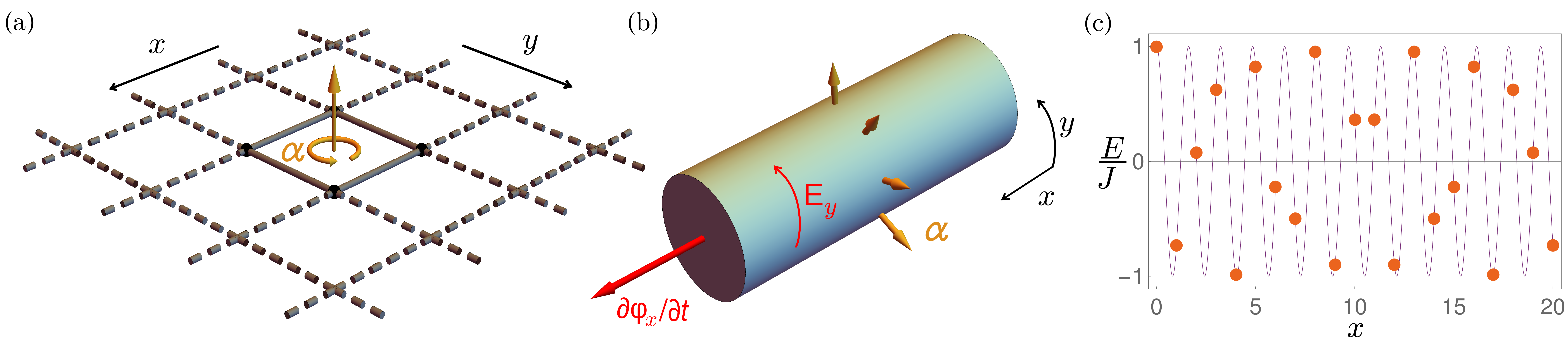}
		\caption{\label{fig1} Illustration of the relationship between a 2D quantum Hall system and a 1D topological charge pump. (a) The 2D quantum Hall effect on a square lattice, also referred to as the Harper-Azbel-Hofstadter (HAH) model~\cite{Harper:1955PPSA, Azbel:1964JETP, Hofstadter76} [cf.~Eq.~\eqref{HH}]. (b) Written in the Landau-gauge [as chosen in Eq.~\eqref{HH}], the HAH model can be written with periodic boundary conditions along the $y$ axis, i.e., you can wrap the 2D system onto a cylinder [see Eq.~\eqref{Eq:H2D}]. Using Faraday's induction law, an electric-field perturbation $E_y=\partial \phi_x/\partial t$ is generated by threading a magnetic flux through the cylinder. (c) Using dimensional reduction~\cite{Thouless83, Niu:1984, QiZhang, Kraus:2012a, Kraus:2012b}, the 2D HAH model is mapped onto a 1D topological charge pump model, where particles hop on a 1D periodic lattice in the presence of an on-site cosine potential [see Eq.~\eqref{Eq:H1D}]. As a function of a periodic modulation of the onsite potential, a quantized number of charges is pumped across the 1D system [cf.~Eqs.~\eqref{eq:pumpvelocity} and \eqref{eq:ChernNumber}]. Plotted is the onsite cosine potential with $\bar{\alpha} = 13/21$ and $\phi_x=0$. Disks mark the discrete sampling of the potential at lattice sites.}	
	\end{figure*}
	
	\section{A Semiclassical Approach to the 6D Quantum Hall Effect} \label{sec:semiclassical}
	
	In this section, we develop a semiclassical approach to derive the 6D quantum Hall effect. We first introduce the relevant geometrical and topological properties of six-dimensional energy bands in Sec.~\ref{sec:chern}, and then we discuss the semiclassical equations of motion for a wave packet moving with respect to such energy bands in Sec.~\ref{sec:semiclass}. We then apply this semiclassical approach in Sec.~\ref{sec:currenttot} to derive the total 6D quantum Hall current response of a system with a filled energy band. This extends and generalizes the semiclassical approaches previously developed for the 2D~\cite{Xiao2010} and the 4D quantum Hall effects~\cite{Price2015}.
	
	\subsection{Geometrical properties and topological invariants of a 6D quantum Hall system} \label{sec:chern}
	
	As we shall see in the following sections, the response of a 6D quantum Hall system stems from the geometrical properties and topological invariants of its energy bands, namely, from the Berry connection and Berry curvature~\cite{Berry1985,Xiao2010}, and the \Fst, \Snd and \Trd Chern numbers~\cite{TKNN, Nakahara, Chiu:2016RMP}. 
	
	To first define the relevant geometrical quantities, we begin from a single particle moving in a periodic potential, for which the eigenstates can be expressed using Bloch's theorem as $\ket{\chi_{n,\bs{k}}} \!=\! e^{i\bs{k}\cdot \bs{r}} \ket{n(\bs{k},\bs{r})}$, where  $\ket{n(\bs{k},\bs{r})}$ are the periodic Bloch functions and $\bs{k}$ is the crystal quasimomentum. The Bloch functions $\ket{n(\bs{k},\bs{r})}$ form energy bands within the Brillouin zone (BZ), with energies $\mathcal{E}_n({\bs k})$, labeled by the band index $n$. The key geometrical properties of the energy bands are encoded in their respective Berry connections and Berry curvatures~\cite{Berry1985}. For a single energetically isolated and nondegenerate energy band $n$, the latter can be expressed as an antisymmetric tensor with components: 
	\begin{eqnarray}
	\Omega^{\mu \nu} (\bs{k})&=&  \partial_{k_\mu} \mathcal{A}_{k_\nu }  -  \partial_{k_\nu} \mathcal{A}_{k_\mu}\,, \label{eq:berry2form}
	\end{eqnarray}
	where $\mathcal{A}_{k_\mu} = \bra{n (\bs{k},\bs{r})}{\partial_{k_\mu}}\ket{n (\bs{k},\bs{r})}$ is the Berry connection, and where the indices $\mu, \nu$ run over all six spatial coordinates. The Berry curvature \eqref{eq:berry2form} is analogous in structure to magnetic fields but in momentum space, i.e., the Berry connection acts like a magnetic-vector potential and the Berry curvature plays the role of the magnetic field~\cite{Berry1985, Nagaosa, bliokh2005spin, Price:2014PRL}. Similar to the magnetic quantities in this analogy, the Berry connection is gauge dependent, while the Berry curvature is gauge invariant and so can be extracted from physical observables~\cite{PricePRA2012,Wimmer:2017NatPhys,flaschner2016experimental, li2016bloch,PhysRevLett.111.220407,ozawa2014anomalous}.  
	More formally, the Berry curvature can be expressed as a differential two-form
	\begin{eqnarray}
	\Omega &=&\frac{1}{2} \Omega^{\mu \nu}(\bs{k}) \mathrm{d}\mathrm{k}_{\mu}\! \wedge \! \mathrm{d}\mathrm{k}_{\nu}\,,\label{eq:berry2form2}
	\end{eqnarray}
	where $\wedge$ is the antisymmetric wedge product. Note that expressions~\eqref{eq:berry2form} and \eqref{eq:berry2form2} can be generalized to include degeneracies between energy bands, in which case each component of the Berry curvature becomes a matrix~\cite{Xiao2010}. However, here we restrict ourselves to considering a single, isolated, nondegenerate energy band as stated above.  
	
	Crucially, the Berry curvature provides a basis for defining important topological invariants of an energy band, depending on the symmetries and dimensionality of the system~\cite{Chiu:2016RMP}. Here, we focus on noninteracting systems without additional local symmetries, in which case the key topological invariants are Chern numbers~\cite{TKNN, Nakahara, Chiu:2016RMP}. As topological invariants, these Chern numbers are global properties of the energy band that are constrained to take only integer values. They are then topologically robust to small perturbations and only change when the band gap to neighboring bands is closed. Consequently, they can lead to remarkably robust physical phenomena  such as the quantum Hall effects which we discuss here.  
	
	In a system with two spatial dimensions, only the \Fst Chern number is relevant. It can be defined as
	\begin{eqnarray}
	\nu_1=\!\frac{1}{2 \pi} \int_{\mathbb{T}^2} \mathrm{d}^2 \mathrm{k} \Omega^{xy} =\!\frac{1}{2 \pi} \int_{\mathbb{T}^2}  \Omega \,\in \mathbb{Z}
	\,,\label{eq:first_chern} 
	\end{eqnarray}
	where we chose the 2D system to lie in the $(xy)$ plane. The integral is taken over the entire 2D BZ, denoted here by $\mathbb{T}^2$ to emphasize its equivalence with a two-torus due to the periodicity of the crystal quasimomenta. Physically, the \Fst Chern number is the integer topological invariant that underlies the robust quantization of conductance found in the 2D quantum Hall effect~\cite{Klitzing:1980PRL,TKNN}. Experimentally, it has also been measured from the center-of-mass drift of an atomic cloud~\cite{Aidelsburger:2015NatPhys}, from dynamical vortex trajectories in a quenched cold-atom gas~\cite{tarnowski2017characterizing}, or from the heating rate of shaken systems~\cite{Tran2017,tran2018quantized,asteria2018measuring} and, as will be reviewed in more detail in Sec.~\ref{sec:pumps}, from 1D topological pumping~\cite{Lohse2016, Nakajima2016}. In addition, it has also been proposed to extract the \Fst Chern number from the steady state of driven-dissipative systems~\cite{ozawa2014anomalous,Salerno:2016PRB,Ozawa2016}. 
	
	Going up to four spatial dimensions, the \Snd Chern number emerges as a new topological invariant, defined as~\cite{Avron1988,IQHE4D,QiZhang,prodan2016bulk}
	\begin{align}
	\nu_2&=\frac{1}{32 \pi^2} \int_{\mathbb{T}^4} \mathrm{d}^4 \mathrm{k}  \epsilon_{\alpha \beta \gamma \delta }  \Omega^{\alpha \beta } \Omega^{\gamma \delta} \nonumber
	\\ &=\frac{1}{8 \pi^2} \int_{\mathbb{T}^4}  \Omega \wedge \Omega \,\in \mathbb{Z}\,, \label{eq:second_chern}
	\end{align}
	where $\mathbb{T}^4$ denotes the 4D BZ and where $ \epsilon_{\alpha \beta \gamma \delta }$ is the 4D Levi-Civita symbol, ensuring that this topological invariant vanishes in lower dimensions. The \Snd Chern number is responsible for the nonlinear 4D quantum Hall response of a system with four spatial dimensions~\cite{frohlich2000,IQHE4D,QiZhang, Price2015,prodan2016bulk}. It has recently been measured experimentally using two-dimensional topological pumps, which realize a dynamical version of the 4D quantum Hall effect~\cite{Lohse2018}, as well as in an effective parameter space associated with different internal states of a Bose-Einstein condensate~\cite{Sugawa:2016arXiv} (see also Refs.~[\onlinecite{Bardyn:2014, Mochol2018}] for related proposals). The associated topological boundary phenomena of two-dimensional pumps were studied using photonic waveguide arrays~\cite{Zilberberg2018}.
	
	In six spatial dimensions, the key topological invariant is the \Trd Chern number~\cite{Nakahara, Chiu:2016RMP,prodan2016bulk} 
	\begin{align}
	\nu_3 &=   \frac{1}{\left(2\pi\right)^3}\int_{\substack{\mathbb{T}^6}}\mathrm{d}^6 \mathrm{k}\frac{1}{2^3\cdot 3!}\epsilon_{\mu\nu\delta\epsilon\iota\rho}\Omega^{\mu\nu} \Omega^{\delta\epsilon} \Omega^{\iota\rho}\nonumber\\  
	&=\frac{1}{\left(2\pi\right)^3}\int_{\substack{\mathbb{T}^6}}\frac{1}{3!}\Omega\wedge \Omega\wedge \Omega \,\in \mathbb{Z}\,, \label{eq:3rd}
	\end{align}
	where the 6D BZ is denoted by $\mathbb{T}^6$ and where we have introduced the 6D Levi-Civita symbol, $\epsilon_{\mu\nu\delta\epsilon\iota\rho}$. From the 6D Levi-Civita symbol it can be seen that the \Trd Chern number is inherently a 6D topological invariant as it vanishes for systems with fewer than six spatial dimensions. As we shall show semiclassically in the following sections, the \Trd Chern number then underlies the 6D quantum Hall effect. Continuing further up in dimensionality, a new quantum Hall effect and a new Chern number emerge every time the number of dimensions is increased by two, where each successive Chern number can be defined as a higher wedge product of the Berry curvature differential form~\cite{Nakahara, Chiu:2016RMP,prodan2016bulk}. 
	
	Restricting ourselves to six dimensions, it is important to remember that the lower-dimensional topological invariants, namely, the \Fst and \Snd Chern numbers, can still be defined, but now with respect to the various two-dimensional planes and four-dimensional subvolumes of the system~\cite{Nakahara}. In total, each energy band in the 6D system is characterized by (i) a set of \Fst Chern numbers, associated with each possible 2D plane;  (ii) a set of \Snd Chern numbers, associated with each possible 4D subvolume; and (iii) a single \Trd Chern number, associated with the full 6D system.
	
	In the following, it will be convenient to introduce additional notation for contributions related to these \Fst and \Snd Chern numbers in a 6D system. In particular, we will use $\nu_1 ^{\mu\nu} $ to denote the \Fst Chern-number-like contribution coming from the $(\mu\nu)$ plane
	\begin{eqnarray}
	\nu_1 ^{\mu\nu} = \displaystyle \frac{1}{2\pi}\int_{\mathbb{T}^6 } \frac{\mathrm{d}^6 \mathrm{k}}{\left(2\pi\right)^4} \Omega^{\mu\nu}, \label{eq:1}
	\end{eqnarray}
	and $\nu_2 ^{\mu\nu\sigma\rho}$ to denote the \Snd Chern-number-like contribution coming from the $(\mu\nu\sigma\rho)$ subvolume
	\begin{align}
	\nu_2 ^{\mu\nu\sigma\rho} \!=\! \frac{1}{\left(2\pi\right)^2}\!\int_{\mathbb{T}^6} \frac{\text{d}^6 \mathrm{k}}{\left(2\pi\right)^2}\left(\Omega^{\mu\nu}\Omega^{\sigma\rho}+ \Omega^{\mu \sigma}\Omega^{\rho\nu}+ \Omega^{\mu \rho}\Omega^{\nu\sigma} \right)\,. \label{eq:2}
	\end{align}
	As can be seen, these expressions correspond to generalizing Eqs.~\eqref{eq:first_chern} and~\eqref{eq:second_chern}, respectively, to a 6D BZ. However, note that these are not integer-valued quantities as the integrals run over the full 6D BZ, instead of only a 2D or 4D closed manifold, respectively; consequently, these quantities depend both on the relevant lower-dimensional invariant as well as on the size of the perpendicular Brillouin zone to the selected 2D plane or 4D sub-volume~\cite{Price2015,Price2016}. 
	
	\subsection{Semiclassical equations of motion} \label{sec:semiclass}
	
	We now review how the geometrical properties of energy bands affect the semiclassical motion of a wave packet under perturbing electromagnetic fields~\cite{Niu1995, Sundaram1999, Xiao2010, Xiao2009, Gao2014, Price2015, Price2016}. 
	As these semiclassical equations of motion apply to systems with dimensions $d\geq 2$, the discussion will be general and applies also for a 6D system. We will, however, need to consider effects up to third order in the perturbing electromagnetic fields as it is only at this order that the 6D quantum Hall effect appears; this is in contrast to the previously studied 2D and 4D quantum Hall effects, which appear at first and second order in the external fields, respectively~\cite{TKNN,Price2015,prodan2016bulk}. 
	
	The semiclassical equations of motion describe a wave packet of charge $-e$ moving in the presence of weak electromagnetic perturbations: namely a weak electric field $\bs E= E_{\mu} \bs{e}^{\mu}$ and a weak magnetic field of strength $B_{\mu \nu}\!=\!\partial_{\mu}A_{\nu}\!-\!\partial_{\nu}A_{\mu}$, where $\bs A= A_{\mu} \bs{e}^{\mu}$ is the electromagnetic vector potential. These external fields are taken to be both spatially uniform and time independent. Note that any strong electromagnetic fields present are included intrinsically in the energy band structure, and so are captured by the band dispersion and the Berry curvature~\cite{Xiao2010, Price2016}. Hereafter, we take $\hbar\!=\!e\!=\!1$. 
	
	In the semiclassical description, the wave packet has a well-defined center-of-mass position $\bs{ r}_c\!=\! r_c^{\mu} \bs{e}_{\mu}$, and momentum $\bs{k}^c\!=\! k^c_{\mu} \bs{e}^{\mu}$. The wave packet is also assumed to move adiabatically, such that it can always be constructed out of the same subset of eigenstates throughout its motion. The choice of basis for the construction of the wave packet, therefore, determines the strength of external fields that can be considered. To illustrate this, we first review the usual semiclassical approach that is valid up to first order in the external perturbations~\cite{Niu1995,Chang1995}, before generalizing our discussion to higher orders.
	
	In a first-order approach, the full Hamiltonian is expanded as $H \approx H_c + H' + H''$, where $H_c$ is the full Hamiltonian evaluated at the center-of-mass position and where $H'$ ($H''$) are the first- (second-) order corrections due to the external electromagnetic fields~\cite{Sundaram1999,Gao2014}. Then, as the external fields are sufficiently weak, the wave packet can be built directly from the eigenstates $\ket{n(\bs{k},\bs{r})}$ of an isolated energy band of $H_c$ as~\cite{Niu1995,Chang1995}
	\begin{eqnarray}
	\ket{W_0} :=\displaystyle\int\limits_{\mathbb{T}^d} \mathrm{d}^d \mathrm{k} w(\bs{k},t) \ket{n (\bs{k},\bs{r})}\,, 
	\end{eqnarray}  
	where $\mathrm{d}^d \mathrm{k}$ is the volume element of the $d$-dimensional Brillouin zone, denoted by $\mathbb{T}^d$, and where $ w(\bs{k},t) $ is the momentum-space distribution function of the wave packet. The distribution function is chosen such that the center-of-mass position $\bs{r}_c$ and momentum $\bs{k}_c$ of the wave packet are defined as
	\begin{eqnarray}
	\bs{k}_c := \displaystyle\int_{\mathbb{T}^d} \mathrm{d}^d \mathrm{k} |w(\bs{k},t)|^2 \bs{k}\hspace{10pt}\&\hspace{10pt}\bs{r}_c := \bra{W_0}\hat{\bs{r}}\ket{W_0} \,, 
	\label{eq:CoM conditions}
	\end{eqnarray}  
	where hereafter the subscript $c$ is omitted. The semiclassical motion of this wave packet is then described by the first-order equations of motion~\cite{Niu1995,Chang1995}
	\begin{align}
	\dot{r}^\mu &= \frac{\partial  \mathcal{E}(\bs{k})}{ \partial {k_\mu} }  -\dot{k}_\nu\Omega^{\mu\nu}\,, \nonumber \\
	\dot{k}_{\mu} &= -\dot{r}^{\nu}  B_{\mu\nu} - E_{\mu}\,,
	\label{eq:EoM}
	\end{align}	
	where Einstein summation convention is assumed. As can be seen, the Berry curvature appears as an ``anomalous velocity" term in addition to the usual group velocity contribution~\cite{Karplus:1954PR} (appearing as the gradient of the energy band dispersion). This anomalous velocity can be understood as a momentum-space analog of the magnetic Lorentz force, in which the Berry curvature acts like a magnetic field in momentum space~\cite{Nagaosa, bliokh2005spin, Price:2014PRL}. This term has important physical consequences for semiclassical motion, and can be used to map out the distribution of the Berry curvature over an energy band~\cite{PricePRA2012,Cominotti2013,Wimmer:2017NatPhys}. 
	
	In order to consider higher orders in the perturbing fields, the wave packet must be instead constructed out of perturbed eigenstates. At second order, the appropriate basis is given by: $| \tilde{n}_0 \rangle = | {n} \rangle + | {n}' \rangle$, where $ | {n}' \rangle$ are the first-order eigenstate corrections. However, the equations of motion remarkably have the same form as in Eq.~\eqref{eq:EoM}, but with modified band dispersion and Berry curvature~\cite{Gao2014,Gao2015}. Going to third order, we construct our wave packet from the basis $| \tilde{n} \rangle = | {n} \rangle + | {n}' \rangle + | {n}'' \rangle$, where $ | {n}'' \rangle$ are the second-order eigenstate correction~\cite{supmat}. In this basis, the equations of motion have the same form as in first order [Eq.~\eqref{eq:EoM}] and second order~\cite{Gao2014}, except with a further modified band dispersion $\mathcal{\tilde{E}}(\bs{k})$ and Berry curvature $\tilde{\Omega}^{\mu \nu} $. The modifications consist of additional gauge-invariant contributions, which will vanish when considering the quantum Hall response of a filled energy band~\cite{supmat}. As the focus of this work is on quantized topological responses of filled bands, we will omit these corrections in what follows and use Eq.~\eqref{eq:EoM} directly. 
	
	With this simplification, we can find the wave-packet velocity to third order in the applied fields, by recursively solving the equations of motion as		
	\begin{widetext}
		\begin{eqnarray}
		\dot{r}^{\mu}(\bs{k}) \simeq \frac{\partial {\mathcal{E}}}{\partial k_{\mu}} 
		+ E_{\nu} {\Omega}^{\mu \nu} 
		+ \left[ \frac{\partial {\mathcal{E}}}{\partial k_\rho}
		+ E_{\sigma} {\Omega}^{\rho \sigma} + \left[ \frac{\partial {\mathcal{E}}}{\partial k_\delta}+ E_{\xi} {\Omega}^{\delta \xi} 
		+ \left[\frac{\partial {\mathcal{E}}}{\partial k_\epsilon} + ...\right] B_{\omega \epsilon} {\Omega}^{\delta \omega}\right] B_{\lambda \delta} {\Omega}^{\rho \lambda}\right] B_{\nu \rho} {\Omega}^{\mu \nu}\,,
		\label{eq:velocity}
		\end{eqnarray}
	\end{widetext}
	where all indices run over all $d$ spatial dimensions.

	\subsection{Quantum Hall response of a filled band} \label{sec:currenttot}
	
	To find the 6D quantum Hall response, we now need to consider the current density associated with a filled band of a system with six spatial dimensions. From our semiclassical equations, this can be calculated by taking the mean velocity in Eq.~\eqref{eq:velocity} and summing over states within a band, according to
	\be
	j^{\mu} =\frac{1}{V} \sum_{\bs k} \rho(\bs k) \, \dot{r}^\mu (\bs k)\,,\label{totalcurrent}
	\ee
	where $V$ is the real-space volume of the 6D system and $\rho(\bs k)$ is the distribution function for the band occupation. Hereafter, we will consider a filled band of spinless fermions for which $\rho(\bs k)\!=\!1$, although we note that all results can straightforwardly be applied to a uniformly filled band of bosons [$\rho(\bs k)\!=\! \rho$] by using~\cite{Price2016} $j^\mu (\rho)\!\rightarrow\! \rho j^\mu (\rho\!=\!1)$. 
	
	In the semiclassical approximation, the sum over occupied states is converted into an integral of quasimomenta ${\bs k}$ over the 6D Brillouin zone according to 
	\begin{equation}
	\frac{1}{V}\sum\limits_{k} \rho(\bs{k}) \dot{r}^\mu(\bs k)\longrightarrow  \displaystyle\int\limits_{\mathbb{T}^6} \frac{\mathrm{d}^6 \mathrm{k}}{\left(2\pi\right)^6} D_{\text{6D}}(\bs{r},\bs{k}) \dot{r}^\mu	(\bs k)\,,
	\label{eq:current response unexpanded}
	\end{equation}
	where we have introduced the \textit{modified} density of states $D_{\text{6D}}(\bs{r},\bs{k})$ for a six-dimensional system, which we now discuss in more detail. 
	
	\subsubsection{6D Modified Density of States}
	
	The modified density of states must be introduced in a semiclassical description to take into account the change in the number of available states in each energy band when both the Berry curvature and the external magnetic field are present~\cite{Xiao2005,Duval2005,Bliokh2006,Gosselin2006, Price2015, Price2016}. In the absence of either one of these corrections, the 6D phase-space density of states will simply be a constant given by: $D_{\text{6D}}({\bs r}, {\bs k})=\!1$. This can be understood classically from Liouville's theorem, which states that, if the dynamics are Hamiltonian, the phase-space volume element is conserved. 
	
	However, Liouville's theorem applies to \textit{canonical} rather than \textit{physical} co-ordinates, and these are not trivially related to one another when both a nonvanishing Berry curvature and external magnetic fields are present~\cite{Xiao2005,Duval2005,Bliokh2006,Gosselin2006, Price2015, Price2016}. To see this, we first consider a system subjected only to a magnetic field perturbation. In this case, the physical momentum ${\bs k}$ is related to the canonical momentum $ {\bs K}$ by minimal coupling ${\bs k} = {\bs K} - {\bs A} ({\bs r})$, where ${\bs A} ({\bs r})$ is the magnetic-vector potential. Alternatively, if we consider a system having a nonvanishing Berry curvature, the physical position ${\bs r}$ is related to the canonical position $ {\bs R}$ by ${\bs r} = {\bs R} + { \mathcal{A}} ({\bs k})$, where ${ \mathcal{A}} ({\bs k})$ is the Berry connection [cf.~Eq.~\eqref{eq:berry2form}]. In the presence of both non-vanishing Berry curvature and external magnetic field, both physical coordinates differ from the canonical coordinates, and generalized Peierls substitutions are required~\cite{Xiao2005,Duval2005,Bliokh2006,Gosselin2006}. These differences are then captured by the modified density of states, which can be understood in terms of the Jacobian of the transformation between physical and canonical coordinates, up to a multiplicative constant~\cite{Duval2005}. 
	
	Importantly, the modified density of states depends on the dimensionality of the system~\cite{Xiao2005, Price2015, Price2016}. To calculate it in 6D, we treat the equations of motion [Eq.~\ref{eq:EoM}] as classical equations and recast them in the form~\cite{Duval2005}
	\begin{eqnarray}
	\begin{array}{ccc}
	\omega_{ij} \dot{\xi}^j = \frac{\partial \mathit{h}}{\partial \xi^i} \,,
	\end{array}
	\label{eq:HAmiltonian formalism}
	\end{eqnarray}
	where $h$ is the classical Hamiltonian, $\xi^i$ are the collective phase-space physical coordinates and $\omega_{ij} $ is a symplectic matrix given by
	\begin{align}
	\omega = \left(\begin{matrix}
	-\underline{B} & -\mathds{I}_{6}\\
	\mathds{I}_{6}  & \underline{\Omega}
	\end{matrix}\right)\,, 
	\end{align}
	where $\mathds{I}_{6}$ is a size six identity matrix and $\underline{B}$, $\underline{\Omega}$ are $6\times 6$ antisymmetric matrices with components $B_{ij}$, $\Omega_{ij}$. The modified density of states can then be calculated as
	\begin{eqnarray}
	D_{\text{6D}}(\bs{r},\bs{k}) = \sqrt{\text{det}(\omega)}\,,
	\end{eqnarray} 
	to find
	\begin{align}
	\label{eq:dos}
	D_{6D}(\bs{r},\bs{k})&=1 +\frac{1}{2} B_{\mu\nu}{\Omega}^{\mu\nu} \\
	&+ \frac{1}{8^2\cdot 2}
	\left(\epsilon^{\mu\nu\rho\sigma\lambda\xi}B_{\rho\sigma}B_{\lambda\xi}\right)
	\left(\epsilon_{\mu\nu\gamma\omega\delta\iota}{\Omega}^{\gamma\omega}{\Omega}^{\delta\iota}\right) \nonumber \\
	& + \frac{1}{48^2 } \left(\epsilon^{\rho\sigma\xi\omega\delta\iota}B_{\rho\sigma}B_{\xi\omega}B_{\delta\iota}\right)
	\left(\epsilon_{\mu\nu\gamma\eta\kappa\alpha}{\Omega}^{\mu\nu}{\Omega}^{\gamma\eta}{\Omega}^{\kappa\alpha}\right)\,.\nonumber
	\end{align}
	Note that the last term will vanish in less than six dimensions due to the Levi-Civita symbols, whereas the second-to-last term will survive down to four dimensions as two of the indices in the Levi-Civita tensors are summed over. 
	
	\begin{figure*}[ht]
		\includegraphics[width=\linewidth]{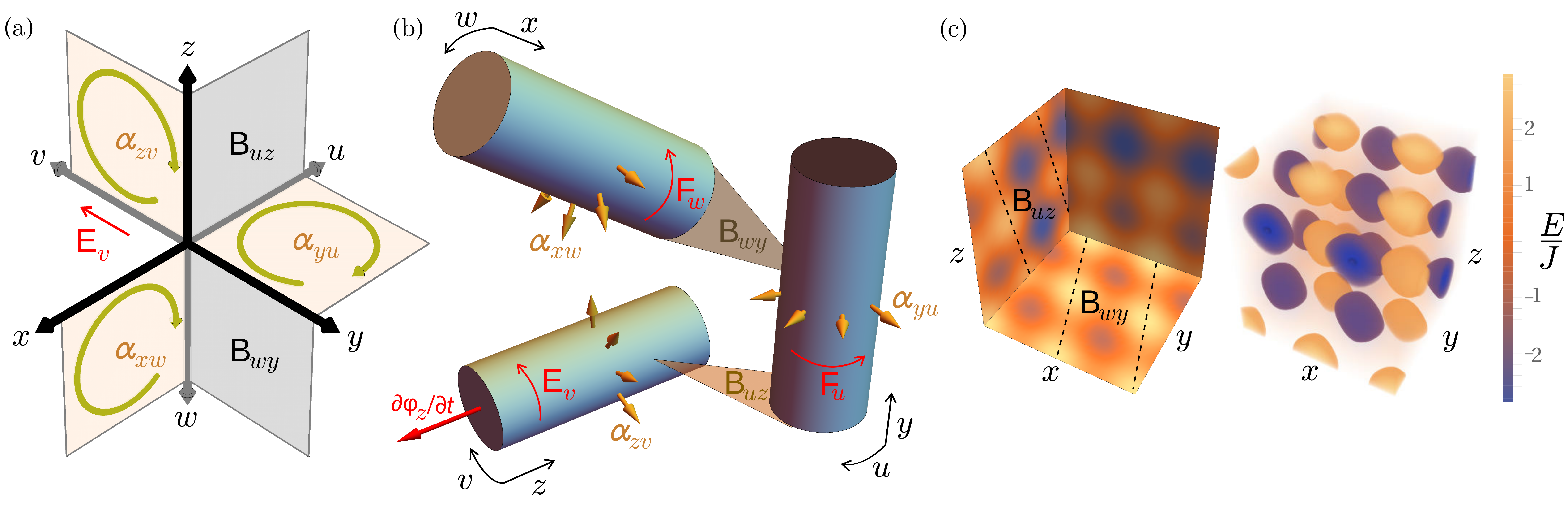}
		\caption{\label{fig2} Illustration of the relationship between a 6D quantum Hall system and a 3D topological charge pump. (a) The 6D quantum Hall effect on a 6D hypercube lattice with $\alpha_{xw}$, $\alpha_{yu}$, and $\alpha_{zv}$ threading the orthogonal $xw, yu$, and $zv$ planes, respectively [cf.~Eq.~\eqref{HH6}]. Note, we attempt to draw a 6D illustration using a 3D axes system: some imagination is required. An example of a \Trd Chern quantized Hall response involves the generation of $E_v, B_{uz}$, and $B_{wy}$ [see Eq.~\eqref{mainRes}]. (b) Written in the Landau-gauge [as chosen in Eq.~\eqref{HH6}], the model can be written with periodic boundary conditions in the $w,u$ and $v$ axes, i.e., you can wrap the 6D system onto three coupled cylinders [see Eq.~\eqref{HH6k}]. Using Faraday's induction law, an electric-field perturbation $E_v=\partial \phi_z/\partial t$ is generated by threading a magnetic flux through the $zv$ cylinder. The perturbing magnetic fields $B_{uz}$, and $B_{wy}$ generate Lorentz forces $F_u$ and $F_w$ in the $u$ and $w$ axes, respectively. (c) Using dimensional reduction, the 6D model \eqref{HH6} is mapped onto a 3D topological charge pump model where particles hop on a 3D periodic lattice in the presence of an ``egg-carton'' onsite potential composed of a sum of three cosine potentials in the orthogonal physical axes $x, y$, $z$ [see Eq.~\eqref{eq:Pump}]. For clarity, the surface potential is plotted (left) with guiding dashed lines showing the skewness due to the magnetic fields. The bulk ``egg-carton'' potential (right) is also shown for completeness. As in the 1D pump case, as a function of a periodic modulation of the onsite potential in the $z$ direction, a quantized number of charges is pumped along that axis. Spatial deformations of the potential couple the motion in the $z$ direction onto motion in the $y$ axis, which then induces motion in the $x$ axis, as expected from this Lorentz-type 3D pumping response [cf.~Eq.~\eqref{displacementl}].  }	
	\end{figure*}
	
	\subsubsection{Total current response}
	
	We are now ready to calculate the total current response \eqref{totalcurrent} of a 6D quantum Hall system by combining the mean velocity \eqref{eq:velocity} with the 6D modified density of states \eqref{eq:dos} and keeping terms up to third order in the perturbing electromagnetic fields. As this expression initially contains many terms, we will consider subsets of terms sequentially, according to their increasing order in the magnetic field strength. 
	
	{\it Order $\mathcal{O}(B^0)$}. At zeroth order in the magnetic field, there are only two terms appearing
	\begin{align}
	\int\limits_{\mathbb{T}^6} \frac{\mathrm{d}^6 \mathrm{k}}{\left(2\pi\right)^6} \left(  \frac{\partial \mathcal{E}}{\partial k_{\mu}} +  E_{\nu} \Omega^{\mu \nu}\right)= E_{\nu} \frac{\nu^{\mu \nu} _1}{2\pi} \,,
	\label{eq:1stCN}
	\end{align}
	where the first term vanishes after integration over the Brillouin zone due to the periodicity of the dispersion energy $ \mathcal{E}(k)$. The second term is related to the 2D quantum Hall effect as it depends on the \Fst Chern-number-like contribution introduced above [cf.~Eq.~{(\ref{eq:1})}]. An analogous effect can emerge in a 4D quantum Hall system, where the Berry curvature of a particular plane is integrated over the 4D BZ instead of the 6D BZ~\cite{Price2015,Price2016,prodan2016bulk}. 
	
	{\it Order $\mathcal{O}(B^1)$}. At first order in the magnetic field, there are two types of terms; the first of these depend on the group velocity and is
	\begin{align}
	&\int\limits_{\mathbb{T}^6} \frac{\mathrm{d}^6 \mathrm{k}}{\left(2\pi\right)^6}\left(\frac{1}{2} B_{\rho\sigma} \Omega^{\rho\sigma} \frac{\partial \mathcal{E}}{\partial k_{\mu}} + B_{\sigma\rho} \Omega^{\mu\sigma} \frac{\partial \mathcal{E}}{\partial k_{\rho}}\right)  \nonumber \\
	&=\int\limits_{\mathbb{T}^6} \frac{\mathrm{d}^6 \mathrm{k}}{\left(2\pi\right)^6}\frac{1}{2} B_{\sigma\rho}  \mathcal{E}\left(  \frac{\partial \Omega^{\rho\sigma}}{\partial k_{\mu}} -   \frac{\partial \Omega^{\mu\sigma}}{\partial k_{\rho}}+   \frac{\partial \Omega^{\mu\rho}}{\partial k_{\sigma}}\right)\,,
	\end{align}
	where we have used the antisymmetry of the magnetic-field components and where the equality is obtained through integration by parts. The terms in the parenthesis vanish due to the Bianchi identity
	\begin{eqnarray}
	\frac{\partial \Omega^{\rho\sigma}}{\partial k_{\mu}} -   \frac{\partial \Omega^{\mu\sigma}}{\partial k_{\rho}}+   \frac{\partial \Omega^{\mu\rho}}{\partial k_{\sigma}}=0\,. 
	\end{eqnarray}
	The remaining two terms at order $\mathcal{O}(B^{1})$ are
	\begin{multline}
	\int\limits_{\mathbb{T}^6} \frac{\mathrm{d}^6 \mathrm{k}}{\left(2\pi\right)^6}\left(\Omega^{\rho\nu}B_{\sigma \rho} \Omega^{\mu \sigma}{E}_\nu+ \frac{1}{2} B_{\sigma\rho}\Omega^{\sigma\rho}\Omega^{\mu\nu}{E}_\nu\right)  \\
	=\frac{1}{2} \frac{\nu_2 ^{\mu\nu\sigma\rho}}{(2\pi)^2}   B_{\sigma\rho}{E}_\nu
	\label{eq:2ndCNT}
	\end{multline} 
	where the prefactor $\frac{1}{2}$ takes care of overcounting from the Einstein summation. Equation~\eqref{eq:2ndCNT} is related to the 4D quantum Hall effect, i.e., it depends on the \Snd Chern-number-like contribution introduced above [cf.~Eq.~\eqref{eq:2}] and is second order in the applied electromagnetic fields (cf.~Ref.~[\onlinecite{Price2015}]). Note that there can be up to 10 independent terms coming from Eq.~\eqref{eq:2ndCNT}, corresponding to the number of unique four-dimensional subvolumes that can generate such a response in a given direction $\mu$.
	
	{\it Order $\mathcal{O}(B^2)$}. To simplify the expressions at second order in the magnetic field, we introduce the more compact notation of
	\begin{align}
	\mathbf{Q}^{\rho\sigma\lambda\xi}:=\left(\Omega^{\rho\sigma}\Omega^{\lambda\xi}+\Omega^{\lambda\rho}\Omega^{\sigma\xi}+\Omega^{\rho\xi}\Omega^{\sigma\lambda}\right)\,. 
	\end{align}
	At this order, we can split the contributions into two parts. The first set of second-order terms depends on the group velocity and using our compact notation is written as	

	\begin{align}
	\int\limits_{\mathbb{T}^6} \frac{\mathrm{d}^6 \mathrm{k}}{\left(2\pi\right)^6}\left( \frac{1}{8}\mathbf{Q}^{\rho\sigma\lambda\xi}B_{\rho\sigma}B_{\lambda\xi}\frac{\partial \mathcal{E}}{\partial k_{\mu}}+\frac{1}{2}\mathbf{Q}^{\mu\nu\sigma\lambda}B_{\nu\sigma}B_{\lambda\rho}\frac{\partial \mathcal{E}}{\partial k_{\rho}} \right)\,. 
	\label{eq:7T}
	\end{align}
	Using the antisymmetric properties of the tensors, we can rewrite Eq.~\eqref{eq:7T} as
	\begin{widetext}
		\begin{multline}
		\int\limits_{\mathbb{T}^6} \frac{\mathrm{d}^6 \mathrm{k}}{\left(2\pi\right)^6}\left( \displaystyle\frac{1}{8}B_{\nu\sigma}B_{\lambda\rho}\left[  \frac{\partial \mathcal{E}}{\partial k_{\rho}}\mathbf{Q}^{\mu\nu\sigma\lambda} + \frac{\partial \mathcal{E}}{\partial k_{\lambda}} \mathbf{Q}^{\mu\nu\rho\sigma} 
		+ \frac{\partial \mathcal{E}}{\partial k_{\sigma}}\mathbf{Q}^{\mu\lambda\rho\nu} + \frac{\partial \mathcal{E}}{\partial k_{\nu}}\mathbf{Q}^{\mu\rho\lambda\sigma} +\frac{\partial \mathcal{E}}{\partial k_{\mu}}\mathbf{Q}^{\nu\sigma\lambda\rho}  \right] \right)  \\
		=-\int\limits_{\mathbb{T}^6} \frac{\mathrm{d}^6 \mathrm{k}}{\left(2\pi\right)^6}\left( \displaystyle\frac{1}{8}B_{\nu\sigma}B_{\lambda\rho}\mathcal{E}\left[  
		\frac{\partial  \mathbf{Q}^{\mu\nu\sigma\lambda}}{\partial k_\rho}
		+ \frac{\partial    \mathbf{Q}^{\mu\nu\rho\sigma} }{\partial k_\lambda}
		+ \frac{\partial \mathbf{Q}^{\mu\lambda\rho\nu}}{\partial k_\sigma}
		+ \frac{\partial\mathbf{Q}^{\mu\rho\lambda\sigma}}{\partial k_\nu}
		+\frac{\partial \mathbf{Q}^{\nu\sigma\lambda\rho} }{\partial k_\mu} \right] \right) \,,\label{eq:term}
		\end{multline}
	\end{widetext}	
	where the second equality is obtained using integration by parts. Crucially, these terms vanish due to a generalized Bianchi identity for~\cite{supmat} $\mathbf{Q}^{\rho\sigma\lambda\xi}$
	\begin{eqnarray}
	\frac{\partial  \mathbf{Q}^{\mu\nu\sigma\lambda}}{\partial k_\rho} + \text{cycl}(\rho\mu\nu\sigma\lambda) = 0\,, 
	\label{eq:2nd BI}
	\end{eqnarray}
	where $\text{cycl}(\rho\mu\nu\sigma\lambda)$ denotes all cyclic permutations of indices for the quantity ${\partial  \mathbf{Q}^{\mu\nu\sigma\lambda}}/{\partial k_\rho}$. Similar terms in systems of four, or more, dimensions will vanish due to the above identity.
	
	The remaining terms at second order in the magnetic field are
	\begin{align}
	\int\limits_{\mathbb{T}^6}  &\frac{\mathrm{d}^6\mathrm{k} }{\left( 2\pi\right)^6}\Bigg( \frac{1}{8}B_{\rho\sigma}B_{\lambda\xi}{E}_\nu\left(\Omega^{\rho\sigma}\Omega^{\lambda\xi}-\Omega^{\rho\lambda}\Omega^{\sigma\xi}+\Omega^{\rho\xi}\Omega^{\sigma\lambda}\right)\Omega^{\mu\nu} \nonumber \\
	&		+B_{\delta\epsilon}B_{\iota\rho}{E}_\nu\Omega^{\epsilon\nu}\Omega^{\rho\delta}\Omega^{\mu\iota}
	+\frac{1}{2}B_{\epsilon\iota}B_{\rho\sigma}{E}_\nu\Omega^{\nu\epsilon}\Omega^{\rho\sigma}\Omega^{\mu\iota}\Bigg) \nonumber \\
	&=\frac{1}{8}\frac{\nu_3}{(2\pi)^3}\epsilon^{\mu\nu\delta\epsilon\iota\rho} B_{\delta\epsilon}B_{\iota\rho}{E}_\nu\,, \label{eq:3rdCN}
	\end{align}
	where the $\frac{1}{8}$ factor takes care of overcounting from the Einstein summation, and where $\nu_3$ is the \Trd Chern number introduced above [cf.~Eq.~\eqref{eq:3rd}]. This is the quantum Hall response which emerges in systems with six dimensions. It depends on the six-dimensional topological invariant and quantizes the third-order response in the perturbing electromagnetic fields. This response appears only in systems with six or more dimensions while in 4D it vanishes due to the antisymmetry of the Levi-Civita tensor (cf.~Ref.~[\onlinecite{Price2015}]).
	
	{\it Order $\mathcal{O}(B^3)$}. For consistency reasons, we consider also terms that are third order in the magnetic field. The relevant terms are~\cite{supmat}
	\begin{align}
	\int\limits_{\mathbb{T}^6} \! &\frac{\mathrm{d}^6 \mathrm{k}}{\left(  2\pi\right)^6}\!  \left(\! \frac{1}{8}\frac{\partial \mathcal{E}}{\partial k_{\rho}} B_{\nu\rho}\Omega^{\mu\nu}B_{\xi\omega}B_{\lambda\iota} \mathbf{Q}^{\xi\omega\lambda\iota} \right. \nonumber \\
	& \!+\!\frac{\partial \mathcal{E}}{\partial k_{\epsilon}} B_{\omega\epsilon}\Omega^{\delta\omega}B_{\lambda\delta}\Omega^{\rho\lambda}B_{\nu\rho}\Omega^{\mu\nu} \nonumber \\
	& \!+\! \frac{1}{48}\frac{\partial \mathcal{E}}{\partial k_{\mu}} B_{\rho\sigma}B_{\xi\omega}B_{\delta\iota}\mathbf{Q}^{\rho\sigma\xi\omega\delta\iota}   \nonumber \\
	&\left. +\frac{1}{2}\frac{\partial \mathcal{E}}{\partial k_{\sigma}} B_{\epsilon\eta}\Omega^{\epsilon\eta}B_{\nu\rho}\Omega^{\mu\nu}B_{\lambda\sigma}\Omega^{\rho\lambda}  \right)\,,
	\end{align}
	where we have introduced another shorthand notation $
	\mathbf{Q}^{\mu\nu\rho\xi\omega\lambda} := \Omega^{\mu\nu}\mathbf{Q}^{\rho\xi\omega\lambda}
	+\Omega^{\mu\rho}\mathbf{Q}^{\xi\nu\omega\lambda}
	+\Omega^{\mu\lambda}\mathbf{Q}^{\xi\omega\nu\rho}
	+\Omega^{\mu\xi}\mathbf{Q}^{\nu\rho\omega\lambda}
	+\Omega^{\mu\omega}\mathbf{Q}^{\xi\nu\lambda\rho}
	$. 
	
	Further manipulation of these antisymmetric tensors and once more using integration by parts~\cite{supmat}, it can be shown that these terms vanish under a generalized Bianchi identity for $	\mathbf{Q}^{\mu\nu\rho\xi\omega\lambda}$
	\begin{eqnarray}
	\frac{\partial  \mathbf{Q}^{\mu\nu\xi\omega\lambda\iota}}{\partial k_\rho}+ \text{cycl}(\rho\mu\nu\xi\omega\lambda\iota) = 0 \,,  
	\label{eq:3rd BI}
	\end{eqnarray} 
	where $\text{cycl}(\rho\mu\nu\xi\omega\lambda\iota)$ denotes all cyclic permutations of indices for the quantity ${\partial \mathbf{Q}^{\mu\nu\xi\omega\lambda\iota}}/{\partial k_\rho}$.
	
	{\it Final result}. Collecting together all the nonvanishing terms [cf.~Eqs.~\eqref{eq:1stCN},~\eqref{eq:2ndCNT}, and \eqref{eq:3rdCN}] for the current response of a fully occupied band, we obtain  
	\begin{multline} \label{mainRes}
	j^\mu =\frac{\nu^{\mu \nu} _1 }{2\pi}E_{\nu}
	+ 
	\frac{1}{2}  \frac{\nu_2 ^{\mu\nu\sigma\rho}}{(2\pi)^2}  B_{\sigma\rho}{E}_\nu
	\\+
	\frac{1}{8}\frac{\nu_3}{(2\pi)^3}\epsilon^{\mu\nu\delta\epsilon\iota\rho} B_{\delta\epsilon}B_{\iota\rho}{E}_\nu
	\end{multline}
	up to third order in perturbing external fields. The induced current has three topological contributions: (i) a first-order correction related to the \Fst Chern number. Such a term arises in systems with two or more dimensions and in 2D it corresponds to the well-known quantum Hall effect~\cite{Klitzing:1980PRL}. (ii) A second-order correction related to the \Snd Chern number, manifesting only in systems with dimensionality greater or equal to four (cf.~Ref.~[\onlinecite{Price2015}]). (iii) A third-order correction which is proportional to the \Trd Chern number, present only in systems with six or more dimensions (cf.~Ref.~[\onlinecite{prodan2016bulk}]). We note that higher-order corrections to the current response will vanish due to the fact that Chern numbers of order higher than the physical dimensions are zero because of antisymmetry.
	
	As can be seen from Eq.~(\ref{mainRes}), there are many possible choices for the orientations of the magnetic- and electric-field perturbations, which will all lead to a \Trd Chern-type current density response of the same magnitude. However, from the semiclassical analysis, it can be seen that these various responses can have different microscopic origins~\cite{Price2015, Price2016}, depending on whether the relevant terms come from the particle density [via the modified density of states in Eq.~\eqref{eq:dos}] or from a Lorentz-type response [via the mean velocity in Eq.~\eqref{eq:velocity}] or from a combination of the two. For this reason, a particular \Trd Chern number response can be classified as a density-type, Lorentz-type, or mixed density-Lorentz-type response; these can be distinguished by looking at center-of-mass observables, which are related to the particle density of the filled band, as well as to the current density (cf.~Ref.~[\onlinecite{Price2016}]).

	\section{Topological pumps} \label{sec:pumps}
	
	We have derived the general bulk response of a 6D QH system in Eq.~\eqref{mainRes}, and now turn to discuss how such a response could be probed experimentally. One avenue towards exploring such effects is to engineer ``synthetic dimensions", in which sets of internal states are parametrically coupled so as to simulate motion along extra spatial dimensions~\cite{Price2015, Price2016, Boada2012,Celi:2012PRL, boada2015quantum, Mancini:2015Science,Stuhl:2015Science,luo2015quantum, Livi:2016PRL,yuan2016photonic, Ozawa2016, ozawa2017synthetic, Price:2017PRA,An:2017SciAdv}. The 6D quantum Hall response could then be observed directly in the current density or, depending on the type of response, in center-of-mass observables, such as the center-of-mass drift of an atomic cloud~\cite{Aidelsburger:2015NatPhys, Price2015, Price2016} or the driven-dissipative steady state~\cite{ozawa2014anomalous, Ozawa2016}. However, building a system with effectively six spatial dimensions is technologically challenging, as it would require adding and controlling at least three synthetic dimensions in addition to the real spatial dimensions of the physical system. 
	
	An alternative and powerful avenue towards realizing such higher-dimensional topological responses involves adiabatic and periodic scanning over auxiliary dimensions, so-called ``topological pumping''~\cite{Thouless83, Kunz1986, Kraus:2012a, Kraus:2012b, Kraus2013, Verbin:2015, Lohse2016,Nakajima2016, Lohse2018, Zilberberg2018,prodan2015virtual}. We shall now briefly review in Sec.~\ref{sec:1dpump} the route to generating a 1D  topological charge pump starting from the 2D quantum Hall effect, before discussing the generalization of topological pumping to higher dimensions in Sec.~\ref{sec:3D}. In particular, we shall show how this concept can be extended to realize 3D topological pumps with a quantized \Trd Chern number response.
	
	\subsection{1D topological pumps} \label{sec:1dpump}
	Let us start by considering a 2D QH system [see Fig.~\ref{fig1}(a)]. 
	As an example, we shall consider the Harper-Azbel-Hofstadter (HAH) model~\cite{Harper:1955PPSA, Azbel:1964JETP, Hofstadter76}, where particles move on a 2D square lattice in the presence of a perpendicular magnetic field
	\begin{multline}
	\label{HH}
	H_{\rm 2D}	= \sum_{{\bf r}} H_{xy} \\
	=-J \sum_{{\bf r}}
	[
	\hat{c}^\dagger_{{\bf r}+ a\mathbf{e}_{x}}\hat{c}_{{\bf r}}
	+
	e^{i2\pi \bar{\alpha} x}\hat{c}^\dagger_{{\bf r} +a \mathbf{e}_{y}}\hat{c}_{{\bf r}}
	+
	\mathrm{h.c.}]\,, 
	\end{multline}
	where $\hat{c}_{{\bf r}}$ is the annihilation operator of a particle at position ${\bf r} = (x, y)$, $J$ is the nearest-neighbor hopping amplitude with $\mathbf{e}_{x}, \mathbf{e}_{y}$ denoting unit vectors in the $x, y$ directions, and where $a$ is the lattice spacing. A magnetic flux $\alpha=2\pi\bar{\alpha}$ in units of the magnetic flux quantum $\Phi_0$ threads each plaquette of the model and is written in the Landau gauge using the Peierls substitution~\cite{Peierls:1933ZPhys}. We choose this model as it leads to an experimentally-relevant 1D pumping model, variants of which have been realized in cold atoms~\cite{Lohse2016,Nakajima2016} and photonic waveguides arrays~\cite{Kraus:2012a,Verbin:2015}. It also emphasizes the generality of our analysis, as in its semiclassical limit $\alpha\ll 1$, the HAH model describes the continuum QH limit with a Landau-level spectrum, while in other regimes it can describe a plethora of physics ranging from graphene-like two-band effects~\cite{Zilberberg2014} to quasiperiodic phenomena~\cite{Kraus:2012a, Kraus:2012b, Kraus2013, kraus2016quasiperiodicity}. 
	
	As this is a 2D QH system, homogeneously filling a band of the HAH model leads to a quantized Hall conductance, proportional to the \Fst Chern number~\cite{TKNN}. To see how this is related to 1D topological pumping, we place the HAH model on a cylinder with periodic boundary conditions in the $y$ direction. Thanks to our chosen Landau gauge, we can then proceed by Fourier transforming the model only in the $y$ direction to obtain [see Fig.~\ref{fig1}(b)]
	\begin{align} \label{Eq:H2D}
	H_{\rm 2D} &= \sum_{x,k_y} H_{x k_y}& \notag\\
	&= - J \sum_{x,k_y} \left( \hat{c}_{x,k_y}^\dag \hat{c}_{x +a  \mathbf{e}_{x},k_{y}} + \text{h.c.} 
	\right. \notag \\
	&  \hspace{50pt} \left.+ 2 \cos\left(2\pi \bar{\alpha} x - k_{y} a \right) \hat{c}_{x,k_y}^\dag \hat{c}_{x,k_y} \right )\,.
	\end{align}
	By applying the procedure of ``dimensional reduction'', this 2D QH system can be directly related to a 1D pump. More specifically, dimensional reduction corresponds to taking the cylinder's circumference to zero and reinterpreting the momentum $k_{y}$ in terms of an external parameter $\varphi$, i.e., reducing the dimensionality of the Hamiltonian by one dimension, makes the creation/annihilation operators $k_{y}$ independent and removes the sum in Eq.~\eqref{Eq:H2D}~\cite{Thouless83, Niu:1984, QiZhang, Kraus:2012a, Kraus:2012b}. The resulting 1D model
	\begin{align} \label{Eq:H1D}
	H_{x}\! =\! - J \sum_{x} &\left( \hat{c}_{x}^\dag \hat{c}_{x + a \mathbf{e}_{x}} + \text{h.c.} 
	+ 2 \cos\left(2\pi \bar{\alpha} x - \varphi \right) \hat{c}_{x}^\dag \hat{c}_{x} \right )\!
	\end{align}
	describes a particle hopping on a 1D lattice in the presence of an onsite spatially varying potential, which is controlled externally through the parameter $\varphi$ [see Fig.~\ref{fig1}(c)].
	
	The 1D model \eqref{Eq:H1D} can be adiabatically pumped by slowly changing the external parameter $\varphi(t)$ over time, i.e., by temporally modulating the onsite energy in a periodic fashion. At each time $t$, we can find the 1D bands of the system in terms of Bloch functions $\ket{n ({k}_{x},\varphi(t), x)}$ and eigenenergies $\mathcal{E}_{n} (k_{x},\varphi(t))$. Similar to the semiclassical approach of Sec.~\ref{sec:semiclassical}, the adiabatic motion of a wave packet with respect to a given nondegenerate, instantaneous energy band can be captured by the semiclassical equations of motion~\cite{Xiao2010}  
	\begin{align} \label{eq:pumpvelocity}
	\dot{x} = \frac{\partial \mathcal{E} (k_{x},\varphi)}{\partial k_{x}}+\Omega^{x \varphi}  \dot{\varphi}\,, 
	\end{align}
	where ${x}$ and $k_{x}$ now denote the center-of-mass position and momentum of the wave packet, respectively. The Berry curvature of the instantaneous band is now given by $\Omega^{x \varphi}= i \left( \langle \partial_{\varphi} n | \partial_{k_{x}} n \rangle  - \langle \partial_{k_{x}} n | \partial_{\varphi}  n \rangle \right)$, where we have dropped the band index $n$. As in Eq.~(\ref{eq:EoM}), the first term on the right-hand side describes the usual group velocity, while the second term is the anomalous velocity~\cite{Karplus:1954PR, Xiao2010} that is now controlled by the pumping rate $\dot{\varphi}$. Note that one can understand the connection between this 1D anomalous velocity and the standard 2D response through Faraday's induction law, in which an electric field perturbation is generated by threading a magnetic flux through the aforementioned cylinder $\dot{\varphi}=E_y$ [see Fig.~\ref{fig1}(b)]. 
	In other words,  an electric field perturbation corresponds, in the pumping limit, to a time-dependent modulation of the potential.
	
	As in a 2D QH system, the topological response is associated with a filled or homogeneously populated bulk band of the 1D pump. To proceed, a similar approach to Sec.~\ref{sec:currenttot} can be applied to Eq.~(\ref{eq:pumpvelocity}), except now the summation is over a 1D BZ as the model is one dimensional. The periodicity of the eigenenergies $\mathcal{E}_{n} (k_{x},\varphi(t))$ over this 1D BZ guarantees that the group velocity contribution for a filled band sums to zero. However, in contrast to a 2D quantum Hall system, the current response of the 1D pump is not itself topological as the Berry curvature is only integrated over a single momentum, and so is not related to the \Fst Chern number. 
	
	Instead, the topological behavior emerges when we consider particle transport of a filled band over a full pump cycle, i.e., integrating the current response over time such that $\varphi$ varies from $0$ to $2\pi$. Then the center-of-mass drift of a filled band is given by $\delta x_{\text{COM}} =  a\nu_1 / \alpha$, where $\nu_1$ is the topological \Fst Chern number associated with the pumping process~\cite{Thouless83, Niu:1984, Kraus:2012a, Kraus:2012b, Lohse2016}
	\begin{equation}
	\label{eq:ChernNumber}
	\nu_1 = \frac{1}{2 \pi} \int_{\rm \mathbb{T}^1} \int_0^{2 \pi} \Omega^{{x} \varphi} \ \mathrm{d} \varphi \mathrm{d}\mathrm{k}_{x}\,
	\end{equation}
	and $a/\alpha$ is the length of the 1D unit cell. This topological displacement has been directly measured in a 1D topological pump of cold atoms~\cite{Lohse2016,Nakajima2016}, while the corresponding boundary phenomena have been experimentally probed in photonic waveguide arrays~\cite{Kraus:2012a,Verbin:2015}. 
	
	\subsection{Higher-dimensional topological pumps} \label{sec:3D}
	
	Having seen the connection between 2D QH physics and 1D topological pumps, we are in a position to discuss higher-dimensional topological pumps and how these relate to higher-dimensional quantum Hall effects. A key additional ingredient for these higher-dimensional QH responses is the inclusion of magnetic field perturbations [cf.~e.g.,~Eq.\eqref{mainRes}], on top of the electric field responsible for the 2D QH response. In a topological pump, as discussed above, the analog of the latter is a time-dependent modulation of the onsite potential. As we shall show below, the analog of a magnetic field perturbation is then a static deformation of the onsite potential~\cite{Kraus2013}, as was experimentally realized in Ref.~\onlinecite{Lohse2018} for a 2D topological pump that probed the \Snd Chern-type pumping response. In this section, we shall first introduce a minimal model for the 6D quantum Hall effect and then explain how it can be implemented in a 3D topological pump so as to realize a \Trd Chern-type response. 
	
	We start by considering a minimal 6D QH model composed of three copies of the HAH model in orthogonal planes [cf.~Eq.~\eqref{HH} and see Fig.~\ref{fig2}(a)]
	\begin{align}
	H_{\rm 6D} =&\sum_{{\bf r}}\left( H_{xw}+H_{yu}+H_{zv}\right)\,, 
	\label{HH6}
	\end{align}
	where now the spinless electrons move on a 6D hypercubic lattice with positions ${\bf r} =(x,y,z,w,u,v)$, nearest-neighbor hopping amplitudes $J$ and where each plaquette in the $xw$, $yu$ and $zv$ plane is threaded by a magnetic flux, $\alpha_{xw}$, $\alpha_{yu}$, $\alpha_{zv}$, respectively. As in Eq.~\eqref{HH}, each copy of the HAH model is in the Landau gauge of a particular plane, taking now $\mathbf{e}_{\mu}$ as a 6D unit vector along the $\mu$ direction. 
	
	The energy spectrum of $H_{\rm 6D}$ is given by a Minkowski sum over the energy bands of the three constituent Hamiltonians
	\begin{multline}
	\mathcal{E}  = \Big\{\mathcal{E}_{xw} + \mathcal{E}_{yu} + \mathcal{E}_{zv} \,\,| \\ \mathcal{E}_{xw} \in \sigma(H_{xw}), \mathcal{E}_{yu} \in \sigma(H_{yu}), \mathcal{E}_{zv} \in \sigma(H_{zv})\Big\}
	\end{multline} 
	where $\sigma(H)$ denotes the spectrum of the Hamiltonian $H$. Correspondingly, the eigenstates of $H_{\rm 6D}$ are product states of eigenstates from the three constituent Hamiltonians. Therefore, there are nonvanishing Berry curvatures only within the $xw$, $yu$ and $zv$ planes, i.e., the curvature in a plane $\mu\nu$ associated with a given energy band can be written as $\Omega^{\mu \nu}=\Omega^{xw}\delta_{\mu x}\delta_{\nu w} + \Omega^{yu}\delta_{\mu y}\delta_{\nu u} + \Omega^{zv}\delta_{\mu z}\delta_{\nu v}$. Consequently, in this model the \Snd and \Trd Chern numbers can be expressed as products of corresponding \Fst Chern numbers, e.g.,~$\nu_3 = \nu_{xw}\nu_{yu}\nu_{zv}$. 
	
	To reduce the 6D QH model to a 3D topological pump, we apply the procedure of dimensional reduction as introduced above [see Fig.~\ref{fig2}(b)]. In this chosen gauge, we apply periodic boundary conditions in three directions $w$, $u$, and $v$, and Fourier transform to find
	\begin{align}
	H_{\rm 6D} =&\sum_{\Gamma}\left( H_{x k_w}+H_{y k_u}+H_{z k_v}\right)\,
	\label{HH6k}
	\end{align}
	[c.f. Eq.~\eqref{Eq:H1D}], while remembering that, in 6D, all operators and sums now run over the full set of $\Gamma=(x,y,z,k_w,k_u,k_v)$. From this, we can read off the corresponding  3D pump model as
	\begin{align} \label{Eq:H3D}
	H_{\rm 3D} \!=\!  - J \sum_{{\bf r}'} &  \left(    c^\dagger _{{\bf r}'+a \mathbf{e}_x } c_{{\bf r}'} 
	+  c^\dagger _{{\bf r}'+a\mathbf{e}_y } c_{{\bf r}'} 	+  c^\dagger _{{\bf r}'+a\mathbf{e}_z} c_{{\bf r}'} +\text{h.c.}  \right. \notag \\
	& + \left[ 2 \cos(2\pi \bar{\alpha}_{xw} x -\varphi_x) + 2 \cos(2\pi \bar{\alpha}_{yu} y - \varphi_y) \right. \notag \\
	& \left. \left.+ 2\cos(2\pi \bar{\alpha}_{zv} z- \varphi_z)\right] c^\dagger _{{\bf r}'} c_{{\bf r}'}\right),
	\end{align}
	which is now describing a particle hopping on a 3D lattice, at positions ${\bf r}' =(x,y,z)$, in the presence of an onsite spatially varying potential that is controlled by three external parameters, $\varphi_x, \varphi_y$, and $\varphi_z$ [see Fig.~\ref{fig2}(c)]. Such a model could be realized in a 3D optical superlattice for ultracold atoms, where the period of the superlattice potential in the different directions reflects the number of magnetic flux quanta $\alpha_{xw}$, $\alpha_{yu}$, and $\alpha_{zv}$, of the original 6D model, similar to the recent 1D and 2D topological pump experiments of Refs.~[\onlinecite{Lohse2016,Nakajima2016, Lohse2018}]. 
	
	In order to observe a topological response, we now need to consider how to include appropriate perturbations in the 3D topological pump model [Eq.~\eqref{Eq:H3D}]. From Eq.~(\ref{mainRes}), we observe that we can incorporate a plethora of magnetic field perturbations through different planes in the 6D QH model, in order to study the various \Trd Chern number responses. However, some of these magnetic field perturbations are irrelevant to the 3D pump because we cannot observe currents in the reduced dimensions $w, u$, and $v$, i.e., we do not need to consider magnetic perturbations involving $B_{\mu \nu}$ with both $\mu, \nu \in\{w,u,v\}$. The remaining range of possible magnetic-field perturbations involve at least one index that is a real dimension in the 3D pump, and can be written in a gauge that allows us to proceed with the dimensional reduction procedure. Then, we obtain a general model that incorporates all possible magnetic field perturbations in 6D as spatial deformations in the dimensionally-reduced 3D-pump model
	\begin{align} \label{eq:Pump}
	&H_{\rm 3D} \!=\!  - J \sum_{{\bf r}'}   \left(    c^\dagger _{{\bf r}'+a\mathbf{e}_x } c_{{\bf r}'} 
	+  c^\dagger _{{\bf r}'+a\mathbf{e}_y } c_{{\bf r}'} 	+  c^\dagger _{{\bf r}'+a\mathbf{e}_z} c_{{\bf r}'} +\text{h.c.}  \right. \notag \\
	& + \left[ 2 \cos(2\pi ( ( \bar{\alpha}_{xw}+ \bar{B}_{wx} )x +\bar{B}_{wy} y + \bar{B}_{wz} z)  -\varphi_x)  \right. \notag \\ 
	&+ 2 \cos(2\pi ( ( \bar{\alpha}_{yu} + \bar{B}_{uy}) y + \bar{B}_{uz} z + \bar{B}_{ux} x ) - \varphi_y)   \\
	& \left. \left.+ 2\cos(2\pi (( \bar{\alpha}_{zv}+ \bar{B}_{vz}) z + \bar{B}_{vy} y + \bar{B}_{vx} x) - \varphi_z)\right] c^\dagger _{{\bf r}'} c_{{\bf r}'}\right),\notag
	\end{align}
	where $ \bar{B}_{\mu \nu} = {B}_{\mu \nu} a^2 / \Phi_0$. In terms of the original 6D model, these possible perturbations can be divided into two types: (i) those such as $\bar{B}_{wx}$ which are in the same plane as an intrinsic strong magnetic flux in Eq.~\eqref{HH6} and which therefore affect the particle density of a filled band, and (ii) those such as $\bar{B}_{wy}$, which are in a plane in which there are no strong magnetic fluxes and therefore leads to a Lorentz force on moving particles~\cite{Price2016, Lohse2018}. In terms of the 3D pump, the analog of these perturbations is to (i) modify the period of the potential along a particular direction and (ii) couple different directions within an on-site potential term. A perturbation of the latter type was recently experimentally realized in a 2D topological pump for ultracold atoms by introducing a small tilt angle in the 2D optical superlattice~\cite{Lohse2018}. 
	
	Through an appropriate combination of these perturbing fields [cf.~Eq.~(\ref{mainRes})], a quantized \Trd Chern-type bulk displacement may be observed. Note that depending on which perturbations are involved, this can be identified as a density-type, Lorentz-type, or mixed density-Lorentz-type response, as introduced above. While these different responses lead to the same average current density, they can be distinguished from center-of-mass observables, such as the center-of-mass displacement of a filled band after a pump cycle~\cite{Price2016}.  
	
	As an example of how a \Trd Chern-type response can be probed in a 3D topological pump, we consider the 3D pump model (\ref{eq:Pump}) with only two nonzero spatial perturbations $\bar{B}_{zu}$ and $\bar{B}_{yw}$, and where $\varphi_z$ is pumped adiabatically and periodically in time. Such a model could be realized in a 3D optical superlattice of cold atoms, where the superlattice is tilted by small angles in the $xy$ and $zw$ planes. 
	In terms of the original 6D model, the analogous electromagnetic perturbations would lead to a quantized current response (\ref{mainRes}) of 
	\begin{align}
	\label{eq:current response of minimal model}
	j^x &= \frac{\nu_3}{\left(2\pi \right)^3} B_{uz} B_{wy} E_v\,,\\
	j^y &= \frac{\nu_2 ^{yvuz}}{\left(2\pi \right)^2} B_{uz} E_v\,, \nonumber \\
	j^z &= \frac{\nu_1 ^{zv}}{2\pi } E_v\,, \nonumber\\
	j^w &= j^u = j^v = 0 \,, \nonumber	
	\end{align} 
	where the pumping of $\varphi_z$ is analogous to the electric field $E_v$ and where the \Trd Chern number only enters the current density along $j^x$. This corresponds to a Lorentz-type response, as both magnetic perturbations enter the current density through the mean velocity (\ref{eq:velocity}). Consequently, this topological response can be clearly measured both from the current density or from center-of-mass observables~\cite{Price2016}. 
	
	In the 3D topological pump, the corresponding center-of-mass displacement of a filled band after a pump cycle in $\varphi_z$ is
	\begin{align}
	& \delta x_{\text{COM}}  = \frac{\nu_3 \bar{B}_{uz} \bar{B}_{wy}}{\alpha_{xw}\alpha_{yu}\alpha_{zv}}a, \nonumber \\
	& \delta y_{\text{COM}}  = \frac{\nu_2^{yz} \bar{B}_{uz}}{\alpha_{zv}\alpha_{yu}}a , \nonumber \\
	& \delta z_{\text{COM}} = \frac{\nu_1^{z} }{\alpha_{zv}}a ,
	\label{displacementl}
	\end{align} 
	where the \Trd Chern number can be extracted from the center-of-mass displacement, $\delta x_{\text{COM}}$, and the topological invariants of the pump cycle are defined as
	\begin{eqnarray}
	\nu_3 &=& \frac{1}{8\pi^3} \int_{\mathbb{T}^3} \int_0^{2 \pi}  \Omega^{{x} \varphi_x}  \Omega^{{y} \varphi_y} \Omega^{{z} \varphi_z}  \mathrm{d} \varphi_x \mathrm{d} \varphi_y\mathrm{d} \varphi_z   \mathrm{d} \mathrm{k}_x \mathrm{d} \mathrm{k}_y \mathrm{d} \mathrm{k}_z, \nonumber \\
	\nu^{yz}_2 &=& \frac{1}{4 \pi^2} \int_{\mathbb{T}^2} \int_0^{2 \pi}\Omega^{{y} \varphi_y} \Omega^{{z} \varphi_z}   \mathrm{d} \varphi_y\mathrm{d} \varphi_z   \mathrm{d} \mathrm{k}_y \mathrm{d} \mathrm{k}_z,  \nonumber \\
	\nu^{z}_1 &=& \frac{1}{2 \pi} \int_{\mathbb{T}^1} \int_0^{2 \pi} \Omega^{{z} \varphi_z} \ \mathrm{d} \varphi_z \mathrm{d}\mathrm{k}_{z}\,. 
	\end{eqnarray}
	%
	
	
	\section{Conclusions}
	We have derived the bulk responses induced in a six-dimensional topological Chern insulator under electromagnetic perturbations up to third order and shown that these are related to the topological indices of the occupied bands. In so doing, we show that there is a nonlinear quantized topological response, which is absent in lower dimensions and which is proportional to the \Trd Chern number: a 6D topological invariant. While the existence of this 6D topological invariant has been derived mathematically~\cite{Nakahara,prodan2016bulk} and postulated to exist by symmetry arguments~\cite{RMP_TI,RMP_TI2}, our semiclassical analysis provides the microscopic interpretation of how it manifests in a 6D quantum Hall effect. Thanks to recent technological advances, this higher-dimensional topological response could be probed by using synthetic dimensions to effectively engineer a system with six spatial dimensions or by using topological pumping to scan over extra dimensions with time modulation. 
	
	As a concrete experimental proposal, we have constructed a minimal 6D model that will exhibit a \Trd Chern number response, and also shown, using dimensional reduction, how this model can be mapped onto a 3D topological pump. Such a mapping can assist cold-atomic experiments to directly probe the six-dimensional topology in the laboratory, building on recent experiments which realized one-dimensional and two-dimensional topological pumps using optical superlattices~\cite{Lohse2016,Nakajima2016, Lohse2018}. Going further, it will be of great interest to study topological edge states and the effect of interparticle interactions in both the 6D quantum Hall effect and the 3D topological pump. So far, the role of many-body interactions is largely unexplored in such systems, and these may yet hold promise for finding new exotic quasiparticle excitations~\cite{IQHE4D}. 
	
	\textit{Note added.} Recently, we became aware of a similar
	work on higher-dimensional QH effects~\cite{Lee2018} that, following the online appearance of our paper, also uses the same generalized Bianchi identities [cf.~Eqs.~\eqref{eq:2nd BI} and \eqref{eq:3rd BI}], which were initially derived here.
	
	\acknowledgments
	
	H.M.P. received funding from the Royal Society and
	from the European Union Horizon 2020 research and innovation
	program under the Marie Sklodowska-Curie
	Grant Agreement No 656093: ÒSynOpticÓ. I.P. and O.Z.
	acknowledge financial support from the Swiss National Foundation. %

	%

\newpage
\cleardoublepage
\setcounter{figure}{0}
\renewcommand{\figurename}{Supplementary Material Figure}

\onecolumngrid
\begin{center}
\textbf{\normalsize Supplemental Material for}\\
\vspace{3mm}
\textbf{\large The 6D quantum Hall effect and 3D topological pumps}
\vspace{4mm}

{ Ioannis\ Petrides,$^{1}$ Hannah M. Price,$^{2,3}$, and Oded Zilberberg$^{1}$}\\
\vspace{1mm}
\textit{\small $^{1}$Institute for Theoretical Physics, ETH Z\"urich, 8093 Z\"urich, Switzerland\\
$^{2}$School of Physics and Astronomy, University of Birmingham, Edgbaston, Birmingham B15 2TT, United Kingdom\\
$^{3}$INO-CNR BEC Center and Dipartimento di Fisica, Universit\`a di Trento, I-38123 Povo, Italy
}

\vspace{5mm}
\end{center}
\setcounter{equation}{0}
\setcounter{section}{0}
\setcounter{figure}{0}
\setcounter{table}{0}
\setcounter{page}{1}
\makeatletter
\renewcommand{\bibnumfmt}[1]{[#1]}
\renewcommand{\citenumfont}[1]{#1}

\setcounter{enumi}{1}
\renewcommand{\theequation}{\Roman{enumi}.\arabic{equation}}

\section{Semiclassical Equations of Motion up to Third Order in Perturbing Electromagnetic Fields}\label{ap:EoM}
In this Section, we provide additional details on the derivation of the semiclassical equations of motion up to third-order in applied electromagnetic fields.

\subsection{Semiclassical equations of motion of a wave packet}\label{apsemic}

As discussed in the main text, the semiclassical equations of motion can be derived for a wave packet with a well-defined center-of-mass position and momentum. In particular, this wave packet should be built from an appropriate set of energetically-isolated eigenstates, such that the wave packet motion is adiabatic with respect to this manifold of states~\cite{Niu1995, Sundaram1999, Xiao2010, Xiao2009, Gao2014}. To find the appropriate basis at third-order in the electromagnetic perturbations, we begin from a negatively charged particle moving in the presence of a magnetic vector potential $ A_{\mu}(\hat{\bs{r}}) = \hat{\bs{r}}^\nu B_{\mu\nu}$ and scalar potential $\phi( \hat{\bs{r}})=\mathbf{E}\cdot \hat{\bs{r}}$. Then the full quantum Hamiltonian can be written as
		\begin{eqnarray}
		\hat{H} = \left[\hat{\mathbf{p}} - \mathbf{A}(\hat{\bs{r}})\right]^2 + \phi( \hat{\bs{r}}) + V(\hat{\bs{r}})\,,
		\end{eqnarray}
where $V(\hat{\bs{r}})$ is a spatially-periodic potential. 

The Hamiltonian can be expanded a small distance $\delta \hat{\bs{r}} = \hat{\bs{r}}-\bs{r}_c$ around the center of mass position $\bs{r}_c$ as:
		\begin{eqnarray}
		\begin{array}{rl}
		\hat{H}&= \big[\hat{\mathbf{p}} - \mathbf{A}(\bs{r}_c + \delta \hat{\bs{r}})\big]^2 +\phi \left(\bs{r}_c + \delta \hat{\bs{r}}\right)+ V(\hat{\bs{r}}) = \hat{H}_c +\hat{H}' + \hat{H}''\,,
		\end{array}
		\end{eqnarray}
		where
		\begin{eqnarray}
		\begin{array}{c}
		\hat{H}_c =  \big[\hat{\mathbf{p}} - \mathbf{A}(\bs{r}_c)\big]^2 +\phi (\bs{r}_c) +  V(\hat{\bs{r}}), 
		 \qquad
		 \hat{H}' = -\big\{\delta \hat{\bs{r}}^{\rho} B_{\mu\rho},\hat{p}^\mu\big\} + E_{\mu} \delta \hat{\bs{r}}^{\mu}, 
		 \qquad
		\hat{H}'' = \delta \hat{\bs{r}}^{\nu} B^{\mu}\hspace{1pt}_{\nu}\delta \hat{\bs{r}}^{\rho} B_{\mu\rho}
		\end{array}
		\end{eqnarray}
consist of $\hat{H}_c$, the local Hamiltonian, as well as $\hat{H}' $ and $\hat{H}''$, which are respectively the first-order and second-order corrections to $\hat{H}_c$. As derived using perturbation theory in Section~\ref{ap:pert in estates}, the eigenstates of $\hat{H}$ can be expanded as		
\begin{eqnarray}
		\ket{\tilde{n}} \approx \ket{n} +\ket{n'} + \ket{n''} + ... \,,
		\end{eqnarray}
where $\ket{n}$ are the eigenstates of $\hat{H}_c $ and 	$\ket{n'}$ ($\ket{n''}$) are the first-order (second-order) corrections. In order to find the semiclassical equations of motion up to third order in the external fields, the wave-packet must be built from these perturbed eigenstates as:
\begin{eqnarray}
	\ket{\tilde{W}_0} :=\displaystyle\int\limits_{\mathbb{T}^d} \mathrm{d}^d \mathrm{k} {w}(\bs{k},t) \ket{\tilde{n} (\bs{k},\bs{r})}, 
	\end{eqnarray}    
where $ {w}(\bs{k},t) =  \mathtt{w}_{\bs{k},t} e^{i\theta(\bs{k} , t)} $ is now the distribution function of the wave packet with respect to the perturbed eigenstates, $\mathtt{w}_{\bs{k},t}$ denotes the weight of state $\ket{\tilde{n} (\bs{k},\bs{r})}$ at time t and $\theta(\bs{k} , t) = \int_{0}^{t}\mathcal{\tilde{E}}(\bs{k})\mathrm{d}t' + \gamma(t)$ is the sum of the dynamical phase given by the temporal integral over the perturbed energy dispersion $ \mathcal{\tilde{E}}$, and the geometrical phase $\gamma = -i\int_{0}^{t} \tilde{\mathcal{A}}_{t}\mathrm{d}t'$ with $\tilde{\mathcal{A}}_{t} := i \bra{\tilde{n} (\bs{k})}\frac{\mathrm{d}}{\mathrm{d}t} \ket{\tilde{n} (\bs{k})}$. The center-of-mass position of this wave-packet in two or more spatial dimensions is then~\cite{Sundaram1999}
\begin{eqnarray}\begin{array}{cl}
		\bs{r}_c &= \displaystyle\int \mathrm{d}^d \mathrm{k}  \left[|w(\bs{k},t)|^2 \nabla \theta + |w(\bs{k},t)|^2\tilde{\mathcal{A}}+ \mathtt{w}^* _{\bs{k},t} \nabla \mathtt{w}_{\bs{k},t} \right]\\\\
		&= \nabla \theta(\bs{k}_c , t) + \mathcal{\tilde{A}} |_{\bs{k}_c}
		\end{array}\,,
		\label{eq:position}
		\end{eqnarray}
		where $\tilde{\mathcal{A}}$ is the Berry connection of the perturbed states, which will be derived in terms of unperturbed states in Sec.~\ref{ap:pert in estates}, and we have assumed a narrow distribution of the wave-packet in momentum space.  The term proportional to $ \mathtt{w}^* _{\bs{k},t} \nabla \mathtt{w}^{} _{\bs{k},t}$ vanishes due to the fact that the wave-packet has zero gradient at $\bs{k}_c$, i.e., the center-of-mass momentum corresponds to the turning point of the distribution (its maximum). 
		
Taking the time derivative of Eq.~(\ref{eq:position}), the evolution of the wave-packet's center-of-mass position is
		\begin{align}
		\label{eq:dot{r}_c}
\dot{{r}}_c ^{\mu} &=\Big( {\partial_{k_\mu} \tilde{\mathcal{E}}(\bs{k}) } - \partial_{k_\mu} \mathcal{\tilde{A}}_{t}+ \frac{\mathrm{d}}{\mathrm{d}t} \mathcal{\tilde{A}}_\mu\Big)\Big|_{\bs{k}=\bs{k}_c}\notag\\ 
&= {\partial_{k_\mu} \tilde{\mathcal{E}}(\bs{k}_c) }  -\dot{{k}}_\nu \tilde{\Omega}^{\mu\nu}|_{\bs{k}=\bs{k}_c}\,,
		\end{align}
where $\tilde{\Omega}^{\mu \nu} (\bs k)=  \partial_{k_\mu}\tilde{ \mathcal{A}}_{\nu }  -  \partial_{k_\nu} \tilde{\mathcal{A}}_{\mu}$, and where the second line is obtained from the triple scalar product. 
		
		The equation of motion for the center-of-mass momentum $\bs{k}_c$ follows a similar derivation. We first make a gauge transformation such that the electromagnetic vector potential $ \mathbf{A}(\bs{r}_c  +\delta\bs{r},t)$ appears as a time-dependent phase accumulated by the wave-packet~\cite{Niu1995}
		\begin{multline}
		\bra{\tilde{W}_0}\left[\left[\mathbf{\hat{p}} + \bs{k} - \mathbf{A}(\bs{r}_c  +\delta\bs{r},t)\right]^2 + V(\bs{r}_c  +\delta\bs{r}) \right] \ket{\tilde{W}_0} \\
		=\bra{\tilde{W}_0}e^{i\int\limits_0 ^{\bs{r}} \mathbf{A}(\bs{r}_c  +\delta\bs{r}',t)\mathrm{d}(\delta\bs{r}')}\left[\left[\mathbf{\hat{p}} + \bs{k} \right]^2 + V(\bs{r}_c  +\delta\bs{r}) \right] e^{-i\int\limits_0 ^{\bs{r}} \mathbf{A}(\bs{r}_c  +\delta\bs{r}',t)\mathrm{d}(\delta\bs{r}')}\ket{\tilde{W}_0} \,.
		\end{multline}
Since we can always choose a gauge
where the vector potential vanishes at the center-of-mass position $\bs{r}_c$, the value of the moving wave-packet near $\bs{r}_c$ can be approximated as
 \begin{eqnarray}
 e^{-i\int\limits_0 ^{\bs{r}}  \mathbf{A}(\bs{r}_c  +\delta\bs{r}',t)\mathrm{d}(\delta\bs{r}')}\ket{\tilde{W}_0}\sim  e^{-i \mathbf{A}(\bs{r}_c ,t)\bs{r}} \ket{\tilde{W}_0}\,.
 \end{eqnarray}

The time derivative of the expectation value of momentum is given by 
		\begin{eqnarray}
		\dot{\bs{k}}_c = \frac{\mathrm{d}}{\mathrm{d}t} \bra{\tilde{W}}-i\nabla_{\bs{r}} \ket{\tilde{W}}\,,
		\end{eqnarray}
where $\ket{\tilde{W}}:=  e^{-i \mathbf{A}(\bs{r}_c,t) \bs{r}}\ket{\tilde{W}_0}$. Following standard differentiation rules we therefore obtain
		\begin{eqnarray}\begin{array}{c}
		\begin{array}{cl}	\dot{\bs{k}}_c &= -\dot{ \mathbf{A}}(\bs{r}_c,t) - \frac{\mathrm{d}}{\mathrm{d}t} \mathcal{\tilde{A}} _{\bs{r}} |_{\bs{r}=\bs{r}_c}\\\\
		&   = -\dot{ \mathbf{A}}(\bs{r}_c,t) \,,
		\end{array}
		\end{array}
		\end{eqnarray}
		where the second line follows due to the fact that there is no Berry connection $\tilde{\mathcal{A}} _{\bs{r}}= i \bra{\tilde{n} (\bs{k})}\nabla_{\bs{r}} \ket{\tilde{n} (\bs{k})}$ in real-space in the systems we consider.
		
		Finally, by including both the electric field $\mathbf{E}$ and magnetic field $\mathbf{B}$ in the vector potential $ A_\mu(\bs{r},t) = E_\mu t +  r^\nu B_{\mu\nu}$, we obtain the two equations of motion for the wave-packet's centre-of-mass position and momentum (dropping the subscript $c$ for simplicity)
		\begin{align}
		\dot{r}^\mu &= \partial_{k_\mu} \tilde{\mathcal{E}}(\bs{k})  -\dot{k}_\nu\tilde{\Omega}^{\mu\nu} \,,\\
		\dot{k}_{\mu} &= -\dot{r}^{\nu}  B_{\mu\nu} - E_{\mu}\,. 
		\end{align}
These equations of motion have the same form as the usual semiclassical equations of motion at first order~\cite{Niu1995} or second order~\cite{Gao2014} in the electromagnetic perturbations, except with a further modified Berry curvature and energy band dispersion. To understand whether these corrections affect the quantum Hall physics, we now consider how the perturbed Berry curvature, $\tilde{\Omega}$, is related to the original Berry curvature, $\Omega$, defined with respect to $H_c$.  
 		
\subsection{Perturbative corrections to the eigenstates and the Berry curvature}\label{ap:pert in estates}

In order to derive the corrections to the Berry curvature, we first consider how the eigenstates themselves are affected by the higher-order terms. We use standard perturbation theory for a Hamiltonian expanded up to second order 
	\begin{eqnarray}
	\tilde{H} \approx H + \lambda H' + \lambda ^2 H'' \,,
	\end{eqnarray}
and seek solutions of the form
	\begin{eqnarray}
	\ket{\tilde{n}} \approx \ket{n} +\lambda\ket{n'} + \lambda^2\ket{n''}\,.
	\end{eqnarray}
	Here $\ket{n}$ is the eigenstate of the unperturbed Hamiltonian, i.e., $H\ket{n} = \mathcal{E}_n \ket{n}$, where now we keep the band index on the energy dispersion to quantify band-mixing effects. The desired eigenstates for constructing the wave-packet can be found by setting $\lambda=1$. Writing the eigenvalue equation 
	\begin{eqnarray}
	\Big( H + \lambda H' + \lambda ^2 H'' \Big)\Big(\ket{n} +\lambda\ket{n'} + \lambda^2\ket{n''}\Big)  \overset{!}{=} \Big(\mathcal{E}_n +\lambda \mathcal{E}' _n + \lambda^2 \mathcal{E}'' _n\Big)\Big(\ket{n} +\lambda\ket{n'} + \lambda^2\ket{n''}\Big) \,,
	\end{eqnarray}
	and grouping terms with the same order in $\lambda$ we obtain 
	\begin{eqnarray}
	\begin{array}{lc}
	\lambda^0:& H\ket{n} = \mathcal{E}_n \ket{n}\,,\\\\
	\lambda^1: & H'\ket{n} + H \ket{n'} = \mathcal{E}_n \ket{n'} + \mathcal{E}' _n \ket{n}\,,\\\\
	\lambda^2: & H'' \ket{n} + H'\ket{n'} + H \ket{n''} = \mathcal{E}_n \ket{n''} + \mathcal{E}' _n \ket{n'} + \mathcal{E}'' _n \ket{n}\,,\\\\
	\lambda^3 & H''\ket{n'} + H'\ket{n''} + H\ket{n'''} =  \mathcal{E}_n \ket{n'''} +  \mathcal{E}'_n \ket{n''} +\mathcal{E}''_n \ket{n'} +  \mathcal{E}'''_n \ket{n}\,.
	\end{array}
	\label{eq:perturbation series}
	\end{eqnarray} 
	The above equations have to be satisfied individually while the perturbed state remains normalised to unity
	\begin{eqnarray}
	\left(\bra{n} +\lambda\bra{n'} + \lambda^2\bra{n''}\right)\left(\ket{n} +\lambda\ket{n'} + \lambda^2\ket{n''}\right) \overset{!}{=} 1 \,.
	\end{eqnarray} 
	Expanding the above equation and grouping terms with the same order in $\lambda$ we obtain
	\begin{eqnarray}
	\begin{array}{lc}
	\lambda^0:& \braket{n|n} =1\,,\\\\
	\lambda^1: &\braket{n|n'} + \braket{n'|n} = 0\,,\\\\
	\lambda^2: &\braket{n|n''} + \braket{n'|n'} + \braket{n''|n} = 0\,,\\\\
	\lambda^3: &\braket{n|n'''} + \braket{n'|n''} + \braket{n''|n'} + \braket{n'''|n} = 0\,.\\\\
	\end{array}
	\label{eq:normalisation series}
	\end{eqnarray}
	
	We present the solutions order by order in $\lambda$:
	
	\paragraph{\textbf{$\mathcal{O}(\lambda^0)$ solutions:}}
	
	Are defined by the unperturbed Hamiltonian
	\begin{eqnarray}
	H\ket{n} = \mathcal{E}_n \ket{n}\,.
	\end{eqnarray}
	
	\paragraph{\textbf{$\mathcal{O}(\lambda^1)$ solutions:}}
	
	We re-write the first-order corrections to the eigenstates as a sum of all unperturbed states
	\begin{eqnarray}
	\ket{n'} = \sum\limits_{m} c_{mn} \ket{m}\,,
	\end{eqnarray} 
	 with weights $c_{mn}$. Substituting this in the $\mathcal{O}(\lambda^1)$ condition in \eqref{eq:perturbation series}, we obtain
	\begin{eqnarray}
	H'\ket{n} + \sum\limits_{m} H c_{mn}\ket{m} = \sum\limits_{m} c_{mn} \mathcal{E}_n \ket{m} + \mathcal{E}' _n \ket{n}\,.
	\end{eqnarray}
	We multiply from the left with an arbitrary state $\bra{m}$ (which is not equal to our unperturbed state $\bra{n}$) and solve for $c_{nm}$
	\begin{eqnarray}
	c_{mn} = \frac{\bra{m}H'\ket{n}}{\mathcal{E}_n - \mathcal{E}_m}\equiv \frac{H' _{mn}}{\mathcal{E}_n - \mathcal{E}_m}\hspace{20pt}\text{ for }m\ne n\,.
	\end{eqnarray}
	The remaining $c_{nn}$ coefficient can be easily derived from the $\mathcal{O}(\lambda^1)$ in \eqref{eq:normalisation series} as $c_{nn} = 0$. To conclude, the 1st order corrections are given by
	\begin{eqnarray}\begin{array}{c}
	\displaystyle\boxed{\ket{n'} = \sum\limits_{m\ne n } c_{mn} \ket{m}\hspace{15pt}	\text{with      }\hspace{15pt}c_{mn} = \frac{H' _{mn}}{\mathcal{E}_n - \mathcal{E}_m}\hspace{15pt}	\text{and      }\hspace{15pt}c_{nn} = 0}\,.
	\end{array}
	\label{eq:1st order corr}
	\end{eqnarray}
	It is worth noting that the first order corrections induce a shift on the energy of the wavepacket through the inter-band mixing given by the $c_{mn}$ coefficients.
	
	\paragraph{\textbf{$\mathcal{O}(\lambda^2)$ solutions:}}
	
	We re-write the second-order corrections to the eigenstates as a sum of all unperturbed states
	\begin{eqnarray}
	\ket{n''} = \sum\limits_{m} c' _{mn} \ket{m}\,,
	\end{eqnarray} 
	 with weights $c' _{mn}$. Substituting this in the $\mathcal{O}(\lambda^2)$ condition in \eqref{eq:perturbation series}, we obtain
	\begin{eqnarray}
	H''\ket{n} + \sum\limits_{m\ne n} H' c_{mn}\ket{m} + \sum\limits_{m} H c' _{mn}\ket{m} = \sum\limits_{m} c' _{mn} \mathcal{E}_n \ket{m} + \sum\limits_{m\ne n} c _{mn} \mathcal{E}' _n \ket{m} + \mathcal{E}'' _n \ket{n}\,.
	\end{eqnarray}
	We multiply from the left with an arbitrary state $\bra{m}$ (which is not equal to our unperturbed state $\bra{n}$) and solve for $c' _{nm}$
	\begin{eqnarray}
	c' _{mn} = \frac{H'' _{mn}}{\mathcal{E}_n - \mathcal{E}_m}  - \frac{H' _{nn}H' _{mn}}{(\mathcal{E}_n - \mathcal{E}_m)^2}+ \sum\limits_{m'\ne n }\frac{H' _{mm'}H' _{m'n}}{(\mathcal{E}_n - \mathcal{E}_m)(\mathcal{E}_n - \mathcal{E}_m')} \hspace{20pt}\text{ for }m\ne n \,.
	\end{eqnarray}
	The remaining $c'_{nn}$ coefficient can be found from the $\mathcal{O}(\lambda^2)$ condition in \eqref{eq:normalisation series} as 
	\begin{eqnarray}
	c' _{nn} = -\frac{1}{2}\sum\limits_{m\ne n } c^* _{mn} c_{mn}\,.
	\end{eqnarray}
	To conclude, the 2nd order corrections are given by
	\begin{eqnarray}\boxed{
		\begin{array}{ccl}
		&\displaystyle\ket{n''} = \sum\limits_{m } c'_{mn} \ket{m}&\\\\
		\text{with      }&\hspace{15pt}c' _{mn} =  \frac{H'' _{mn}}{\mathcal{E}_n - \mathcal{E}_m}  - \frac{H' _{nn}H' _{mn}}{(\mathcal{E}_n - \mathcal{E}_m)^2}+ \sum\limits_{m'\ne n }\frac{H' _{mm'}H' _{m'n}}{(\mathcal{E}_n - \mathcal{E}_m)(\mathcal{E}_n - \mathcal{E}_m')} &\hspace{20pt}\text{ for }m\ne n\\\\
		&c' _{nn} = -\frac{1}{2}\sum\limits_{m\ne n } c^* _{mn} c_{mn}&\hspace{20pt}\text{ for }m = n
		\end{array}}
	\label{eq:2nd order corr}
	\end{eqnarray}
	It is important to note that the second order corrections renormalise the ground state through the intra-band mixing coefficients $c' _{nn}$, in addition to an energy shift from the inter-band mixing coefficients.
	
	\paragraph{\textbf{$\mathcal{O}(\lambda^3)$ solutions:}}
	
	We re-write the third order corrections to the eigenstates as a sum of all unperturbed states
	\begin{eqnarray}
	\ket{n'''} = \sum\limits_{m} c'' _{mn} \ket{m}\,,
	\end{eqnarray} 
	 with weights $c'' _{mn}$. Substituting this in the $\mathcal{O}(\lambda^3)$ condition in \eqref{eq:perturbation series} we obtain
	\begin{eqnarray}
	\begin{array}{rl}
	&\sum\limits_{m\ne n} H'' c_{mn}\ket{m} + \sum\limits_{m} H' c'_{mn}\ket{m} + \sum\limits_{m} H c'' _{mn}\ket{m} \\\\
	=& \sum\limits_{m} c'' _{mn} \mathcal{E}_n \ket{m} + \sum\limits_{m} c' _{mn} \mathcal{E}' _n \ket{m}+ \sum\limits_{m\ne n} c _{mn} \mathcal{E}'' _n \ket{m} + \mathcal{ E}''' _n \ket{n}\,.
	\end{array}
	\end{eqnarray}
	We multiply from the left with an arbitrary state $\bra{m}$ (which is not equal to our unperturbed state $\bra{n}$) and solve for $c'' _{nm}$
	\begin{eqnarray}
	c'' _{mn} = -\frac{H'' _{nn}c_{mn}}{\mathcal{E}_n - \mathcal{E}_m}  -\frac{H' _{nn}c' _{mn}}{\mathcal{E}_n - \mathcal{E}_m} + \sum\limits_{m'\ne n }\frac{H'' _{mm'}c _{m'n}}{(\mathcal{E}_n - \mathcal{E}_m)} + \sum\limits_{m' }\frac{H' _{mm'}c' _{m'n}}{(\mathcal{E}_n - \mathcal{E}_m)} \hspace{20pt}\text{ for }m\ne n\,.
	\end{eqnarray}
	The remaining $c''_{nn}$ coefficient can be found from the $\mathcal{O}(\lambda^3)$ condition in \eqref{eq:normalisation series}	
	\begin{eqnarray}
	c'' _{nn} = -\sum\limits_{m\ne n } c^* _{mn} c' _{mn}\,.
	\end{eqnarray}
	To conclude, the 3nd order corrections are given by
	\begin{eqnarray}\boxed{
		\begin{array}{ccl}
		&\displaystyle\ket{n'''} = \sum\limits_{m } c''_{mn} \ket{m}&\\\\
		\text{with      }&\hspace{15pt}c'' _{mn} = -\frac{H'' _{nn}c_{mn}}{\mathcal{E}_n - \mathcal{E}_m}  -\frac{H' _{nn}c' _{mn}}{\mathcal{E}_n - \mathcal{E}_m} + \sum\limits_{m'\ne n }\frac{H'' _{mm'}c _{m'n}}{(\mathcal{E}_n - \mathcal{E}_m)} + \sum\limits_{m' }\frac{H' _{mm'}c' _{m'n}}{(\mathcal{E}_n - \mathcal{E}_m)} &\hspace{20pt}\text{ for }m\ne n\\\\
		&	c'' _{nn} = -\sum\limits_{m\ne n } c^* _{mn} c' _{mn}&\hspace{20pt}\text{ for }m = n
		\end{array}}
	\label{eq:3rd order corr}
	\end{eqnarray}
	It is important to note that again the ground state is renormalised by the third order corrections through the $c'' _{nn}$ term, in addition to an energy shift from the inter-band mixing coefficients.
	
	\subsection{Perturbative corrections to the Berry connection }\label{ap:Perturbative corrections to the Berry connection}

	The Berry connection of the $n$-th isolated band is defined as
	\begin{eqnarray}\begin{array}{rl}	
	\tilde{\mathcal{A}} &= \bra{\tilde{n}}i\nabla \ket{\tilde{n}}\approx \mathcal{A} + \lambda a' + \lambda^2 a''  + \lambda^3 a'''  +...\,,
	\end{array}
	\end{eqnarray}
	where the corrections are given by
	\begin{eqnarray}
	\begin{array}{lc}
	\lambda^0:& \mathcal{A} = \bra{{n}}i\nabla \ket{{n}} \,,\\\\
	\lambda^1: & a' = \bra{{n}}i\nabla \ket{{n'}} + \bra{{n'}}i\nabla \ket{{n}}\,,\\\\
	\lambda^2: &a''  = \bra{{n}}i\nabla \ket{{n''}}+ \bra{{n'}}i\nabla \ket{{n'}} + \bra{{n''}}i\nabla \ket{{n}}\,,\\\\
	\lambda^3: &a'''  = \bra{{n}}i\nabla \ket{{n'''}}+ \bra{{n'}}i\nabla \ket{{n''}}+ \bra{{n''}}i\nabla \ket{{n'}} + \bra{{n'''}}i\nabla \ket{{n}}\,.\\\\
	\end{array}
	\label{eq:Berry series}
	\end{eqnarray}
Correspondingly, the Berry curvature is modified to
	\begin{eqnarray}
	 \tilde{\Omega}^{\mu\nu} \approx \Omega^{\mu\nu} + \lambda(\partial_{\mu}a'_{\nu} -\partial_{\nu}a'_{\mu})+ \lambda^2(\partial_{\mu}a''_{\nu} -\partial_{\nu}a''_{\mu})+ \lambda^3(\partial_{\mu}a'''_{\nu} -\partial_{\nu}a'''_{\mu}) +...
	\end{eqnarray}
	Below we show that the perturbative corrections to the Berry connection are gauge invariant quantities. This is a necessary property to show that their contribution to the current response vanishes when considering a fully occupied band.\\
	
	\paragraph{\textbf{$\mathcal{O}(\lambda^1)$ corrections:}}
	Using the above results we can write the first-order correction as
	\begin{eqnarray}
	a' = \sum\limits_{m\ne n }\bra{n} i\nabla c_{mn}\ket{m} +h.c. 
	\end{eqnarray}
	Under a gauge transformation $\ket{n} \longrightarrow e^{i\chi_n}\ket{n}$, the corrections to the Berry connection transform as follows (see Sec.~\ref{ap:Gauge Trans} for the transformation of $c_{mn}$)
	\begin{eqnarray}
	\begin{array}{rl}
	\displaystyle a'  \longrightarrow &\sum\limits_{m\ne n }\bra{n}e^{-i\chi_{n}} i\nabla e^{i\Delta\chi_{mn}}c_{mn}e^{i\chi_{m}}\ket{m} +h.c=\sum\limits_{m\ne n }\bra{n} i\nabla c_{mn}\ket{m}  - \nabla \chi_{n} c_{nn} +h.c= a' \,,
	\end{array}
	\end{eqnarray}
	where we have used the fact that $c_{nn}=0$ [see Eq.~(\ref{eq:1st order corr})]. Hence, $a'$ is a gauge invariant term.\\
	
	\paragraph{\textbf{$\mathcal{O}(\lambda^2)$ corrections:}}
	The second-order correction to the Berry connection can be written as
	\begin{eqnarray}
	a''  = \sum\limits_{m } \bra{n}i\nabla c' _{mn}\ket{m} + \sum\limits_{{m\ne n\atop l\ne n} }\bra{m}c^*_{mn}i\nabla c_{ln} \ket{l}+ \sum\limits_{m } \bra{m}c'^* _{mn} i\nabla\ket{n}\,.
	\end{eqnarray}
	Under a gauge transformation $\ket{n} \longrightarrow e^{i\chi_n}\ket{n}$, it transform as follows (see also Sec.~\ref{ap:Gauge Trans} for the individual transformations)
	\begin{eqnarray}
	\begin{array}{rl}
	\displaystyle a''   \longrightarrow & \sum\limits_{m } \bra{n}e^{-i\chi_{n}}\nabla e^{i\Delta\chi_{mn}}c' _{mn}e^{i\chi_{m}}\ket{m} 
	+ \sum\limits_{{m\ne n\atop l\ne n} }\bra{m}e^{-i\chi_{m}}c^*_{mn}e^{-i\Delta\chi_{mn}}\nabla e^{i\Delta\chi_{ln}}c_{ln} e^{i\chi_{l}}\ket{l}\\\\
	&\hspace{20pt}+ \sum\limits_{m } \bra{m}e^{-i\chi_{m}}c'^* _{mn}e^{-i\Delta\chi_{mn}} \nabla e^{i\chi_{n}}\ket{n}\\\\
	
	&= a''   - \nabla\chi_{n} c' _{nn} - \sum\limits_{m\ne n }\nabla\chi_{n}c^* _{mn} c_{mn}- \nabla\chi_{n} c'^* _{nn}\,.\\\\

	\end{array}
	\end{eqnarray}
	Using the definition of $c'_{nn}$ [see Eq.~(\ref{eq:2nd order corr})], the gauge-dependent terms can be shown to cancel 
	\begin{eqnarray}
	\begin{array}{rl}
	&\displaystyle- \nabla\chi_{n} c' _{nn} - \sum\limits_{m\ne n }\nabla\chi_{n}c^* _{mn} c_{mn}- \nabla\chi_{n} c'^* _{nn}=\displaystyle+ \frac{1}{2}\nabla\chi_{n} \sum\limits_{m\ne n } c^* _{mn} c_{mn} - \sum\limits_{m\ne n }\nabla\chi_{n}c^* _{mn} c_{mn}+ \frac{1}{2}\nabla\chi_{n} \sum\limits_{m\ne n } c^* _{mn} c_{mn} =0\,, \nonumber
	\end{array}
	\end{eqnarray}
making $a'' $ a gauge invariant quantity.\\
	
	\paragraph{\textbf{$\mathcal{O}(\lambda^3)$ corrections:}}
	The third order correction can be written as
	\begin{eqnarray}
	\displaystyle a'''  = \sum\limits_{m }\bra{n}i\nabla c'' _{mn}\ket{m} 
	+ \sum\limits_{{m\ne n\atop l} }\bra{m}c^*_{mn}i\nabla c'_{ln} \ket{l}
	+ \sum\limits_{{m\ne n\atop l} }\bra{l}c'^*_{ln}i\nabla c_{mn} \ket{m}
	+ \sum\limits_{m }\bra{m} c''^* _{mn}i \nabla \ket{n}\,.
	\end{eqnarray}
	Under a gauge transformation $\ket{n} \longrightarrow e^{i\chi_n}\ket{n}$, it transform as follows (see also Sec.~\ref{ap:Gauge Trans})
	\begin{eqnarray}
	\begin{array}{rl}
	\displaystyle \sum\limits_{m }\bra{n}i\nabla c'' _{mn}\ket{m}  \longrightarrow &\displaystyle\sum\limits_{m }\bra{n}e^{-i \chi_{n}}\nabla e^{i\Delta \chi_{mn}}c'' _{mn}e^{i \chi_{m}}\ket{m} =\sum\limits_{m }\bra{n}i\nabla c'' _{mn}\ket{m}  -   \nabla\chi_{n}   c'' _{nn}\\\\
	=&\displaystyle\sum\limits_{m }\bra{n}i\nabla c'' _{mn}\ket{m} +  \nabla\chi_{n} \sum\limits_{m\ne n }c^* _{mn}  c' _{mn}\,,\\\\

	\pagebreak
	\displaystyle \sum\limits_{{m\ne n\atop l} }\bra{m}\left(c^*_{mn}i\nabla c'_{ln} \right)\ket{l} \longrightarrow & \displaystyle\sum\limits_{{m\ne n\atop l} } \bra{m}e^{-i\chi_{m}}e^{-i\Delta\chi_{mn}} c^*_{mn}i\nabla \left(e^{i\Delta\chi_{ln}} c'_{ln} e^{i\chi_{l}}\ket{l}\right)=\displaystyle\sum\limits_{{m\ne n\atop l} } \bra{m}e^{-i\chi_{n}} c^*_{mn}i\nabla  \left(c'_{ln} e^{i\chi_{n}}\ket{l}\right)\\\\
	=&\displaystyle\sum\limits_{{m\ne n\atop l} } \bra{m} c^*_{mn}i\nabla c_{ln}\ket{l} - 
	\sum\limits_{{m\ne n} } \nabla\chi_n  c^*_{mn} c'_{mn} \,,\\\\
	
	\displaystyle \sum\limits_{m }\bra{m} c''^* _{mn} i\nabla \ket{n}  \longrightarrow &\displaystyle\sum\limits_{m }\bra{m}e^{-i\chi_{m}} c''^* _{mn}e^{-i\Delta\chi_{mn}} \nabla e^{i\chi_{n}}\ket{n} =\displaystyle\sum\limits_{m }\bra{m} c''^* _{mn}i \nabla \ket{n} -   \nabla\chi_{n}   c''^* _{nn}\\\\
	=&\displaystyle\sum\limits_{m }\bra{m} c''^* _{mn} i\nabla \ket{n}+  \nabla\chi_{n} \sum\limits_{m\ne n }c'^* _{mn}  c _{mn}\,,\\\\
	
	\displaystyle \sum\limits_{{m\ne n\atop l} }\bra{l}\left(c'^*_{ln}i\nabla c_{mn} \right)\ket{m} \longrightarrow &\displaystyle \sum\limits_{{m\ne n\atop l} } \bra{l}e^{-i\chi_{l}}e^{-i\Delta\chi_{ln}} c'^*_{ln}i\nabla \left(e^{i\Delta\chi_{mn}} c_{mn} e^{i\chi_{m}}\ket{m}\right)=\displaystyle\sum\limits_{{m\ne n\atop l} } \bra{l}e^{-i\chi_{n}} c'^*_{ln}i\nabla  \left(c_{mn} e^{i\chi_{n}}\ket{m}\right)\\\\
	=&\displaystyle\sum\limits_{{m\ne n\atop l} } \bra{l} c'^*_{ln}i\nabla c_{mn}\ket{m} - 
	\sum\limits_{{m\ne n} } \nabla\chi_n  c'^*_{mn} c_{mn} \,,\\\\
	\end{array}
	\end{eqnarray}
	where in the second equalities we have used the definition of $c''_{nn}$ [Eq.~(\ref{eq:3rd order corr})]. Summing the above terms makes the gauge dependent terms to cancel, therefore $a''' $ is a gauge invariant quantity.

	\subsection{Effect on fully occupied bands}\label{ap:effect on fully occ bands}

	The contribution of the corrections to the current response come through quantities integrated over the Brillouin Zone (see main text). For simplicity, suppose that an isolated band has energy $\tilde{\mathcal{E}} = \mathcal{E}+\mathcal{\epsilon}$ and Berry connection $\tilde{\mathcal{A}}=\mathcal{A}+a$ where $\epsilon$ and $a$ are gauge invariant quantities. At third-order, the correction to the Berry connection will come from those corrections defined above.
	
	\paragraph{\Fst Chern number:}
	Given a 2-dimensional plane, denoted by $(\mu,\nu)$-coordinates, the contributions to the current response from the \Fst Chern number are proportional to the integral of the corresponding Berry curvature $\Omega^{\mu\nu} =\epsilon^{\mu\nu} \partial_{k_\mu}\tilde{ \mathcal{A}}_{\nu }  $ over the 2D plane. Denoting $\tilde{\Omega}^{\mu\nu}\mathrm{dk}_\mu \mathrm{dk}_\nu\equiv \nabla\times \tilde{\mathcal{A}}$, the contributions (up to pre-factors) are given by
	\begin{eqnarray}
	\begin{array}{rl}
	\displaystyle\int\limits_{\mathbb{T}^2} \nabla\times \tilde{\mathcal{A}} &= \displaystyle\int\limits_{\mathbb{T}^2} \nabla\times \mathcal{A} + \int\limits_{ \mathbb{T}^2} \nabla\times a\\\\
	&= \displaystyle\int\limits_{\mathbb{T}^2} \nabla\times \mathcal{A} + \int\limits_{\partial \mathbb{T}^2} a\,,\\\\
	\end{array}
	\end{eqnarray}
	where the second line is obtained because $a$ is a gauge-invariant quantity and therefore the integral over the boundary of the chosen 2D Brillouin Zone is well defined. However, since $a$ is periodic, its integral over the boundary vanishes and therefore does not contribute to the \Fst Chern number
	\begin{eqnarray}
	\begin{array}{rl}
	\displaystyle\int\limits_{\mathbb{T}^2} \nabla\times \tilde{\mathcal{A}} &= \displaystyle\int\limits_{\mathbb{T}^2} \nabla\times \mathcal{A}\,. \\\\
	\end{array}
	\end{eqnarray}
	
	\paragraph{\Snd Chern number:}
	Given a 4-dimensional sub-volume, denoted by the $(\mu,\nu,\rho,\sigma)$-coordinates, the contributions to the current response from the \Snd Chern number are proportional to
	\begin{eqnarray}
	\begin{array}{rl}
	\displaystyle\int\limits_{\mathbb{T}^4} \left(\nabla\times \tilde{\mathcal{A}} \right)\wedge\left(\nabla\times \tilde{\mathcal{A}} \right)&= \displaystyle\int\limits_{\mathbb{T}^4} \left(\nabla\times {\mathcal{A}} \right)\wedge\left(\nabla\times {\mathcal{A}} \right) + \int\limits_{\mathbb{T}^4} \left(\nabla\times {\mathcal{A}} \right)\wedge\left(\nabla\times {a} \right)+ \int\limits_{\mathbb{T}^4} \left(\nabla\times {a} \right)\wedge\left(\nabla\times {\mathcal{A}} \right) + ...\\\\
	\end{array}
	\end{eqnarray}
	where $\left(\nabla\times \tilde{\mathcal{A}} \right)\wedge\left(\nabla\times \tilde{\mathcal{A}} \right) \equiv \epsilon^{\nu\rho\mu\sigma} \partial_{k_\nu}\tilde{\mathcal{A}}_{\rho} \partial_{k_\mu}{\tilde{\mathcal{A}}}_\sigma \mathrm{dk}_\mu \mathrm{dk}_\nu \mathrm{dk}_\rho \mathrm{dk}_\sigma $.
	The last two terms can be written as
	\begin{eqnarray}
	\begin{array}{ll}
	
	\int\limits_{\mathbb{T}^4} \left(\nabla\times {\mathcal{A}} \right)\wedge\left(\nabla\times {a} \right) 
	&= \int\limits_{\mathbb{T}^4}\nabla\times \left(\left(\nabla\times {\mathcal{A}} \right)\wedge {a} \right) 
	- \int\limits_{\mathbb{T}^4} \left(\nabla\times\left(\nabla\times {\mathcal{A}} \right) \right)\wedge {a}\\\\
	&= \int\limits_{\mathbb{T}^4}\nabla\times \left(\left(\nabla\times {\mathcal{A}} \right)\wedge {a} \right) \\\\
	&=\int\limits_{\partial\mathbb{T}^4} \left(\nabla\times {\mathcal{A}} \right)\wedge {a}\\\\
	&=0\,,
	\end{array}
	\label{eq:2ndCN corr}
	\end{eqnarray}
	where the second line is obtained using Bianchi's identity to realize that 
	\begin{eqnarray}
	\begin{array}{rl}
	\displaystyle\left(\nabla\times\left(\nabla\times {\mathcal{A}} \right) \right)\wedge {a} &\displaystyle\equiv \epsilon^{\mu\nu\rho\sigma} (\partial_{k_\mu}\partial_{k_\nu}\mathcal{A}_{\rho} ){a}_\sigma \mathrm{dk}_\mu \mathrm{dk}_\nu \mathrm{dk}_\rho \mathrm{dk}_\sigma \\\\
	&\displaystyle=\frac{1}{2}\epsilon^{\mu\nu\rho\sigma} (\partial_{k_\mu}\Omega^{\nu\rho} ){a}_\sigma \mathrm{dk}_\mu \mathrm{dk}_\nu \mathrm{dk}_\rho \mathrm{dk}_\sigma\\\\
	&\displaystyle=\frac{1}{6}\epsilon^{\mu\nu\rho\sigma} (\partial_{k_\mu}\Omega^{\nu\rho}+\partial_{k_\rho}\Omega^{\mu\nu}+\partial_{k_\nu}\Omega^{\rho\mu} ){a}_\sigma \mathrm{dk}_\mu \mathrm{dk}_\nu \mathrm{dk}_\rho \mathrm{dk}_\sigma\\\\
	&=0\,.
	\end{array}
	\end{eqnarray}	
	The third line of Eq.~(\ref{eq:2ndCN corr}) is obtained because $\left(\nabla\times {\mathcal{A}} \right)\wedge {a}$ is a gauge-invariant quantity and, hence, its integral over the boundary of  the chosen 4D Brillouin zone is well defined. However, this term is periodic, which makes the integral over the boundary vanish. Thus, the perturbative corrections of a fully occupied band give zero contribution to the \Snd Chern number
	\begin{eqnarray}
	\begin{array}{rl}
	\displaystyle\int\limits_{\mathbb{T}^4} \left(\nabla\times \tilde{\mathcal{A}} \right)\wedge\left(\nabla\times \tilde{\mathcal{A}} \right)&= \displaystyle\int\limits_{\mathbb{T}^4} \left(\nabla\times {\mathcal{A}} \right)\wedge\left(\nabla\times {\mathcal{A}} \right) \,.
	\end{array}
	\end{eqnarray}	
	\paragraph{Dispersion:}
	The contributions from the dispersion energy are proportional to
	\begin{eqnarray}
	\begin{array}{rl}
	\displaystyle\int\limits_{\mathbb{T}^d} \nabla\tilde{\mathcal{E}} &=\displaystyle\int\limits_{\mathbb{T}^d} \nabla{\mathcal{E}} + \displaystyle\int\limits_{\mathbb{T}^d} \nabla{\mathcal{\epsilon}}  \\\\
	& =\displaystyle\int\limits_{\partial\mathbb{T}^d} {\mathcal{E}} + \displaystyle\int\limits_{\partial\mathbb{T}^d} {\mathcal{\epsilon}} = 0 \,, \\\\
	\end{array}
	\end{eqnarray}
	where the second line is obtained because both $\mathcal{E}$ and $\epsilon$ are gauge-invariant quantities, and hence the integral over the boundary of the Brillouin Zone is well defined, but as both $\mathcal{E}$ and $\epsilon$ are periodic, the integral over the boundary vanishes.

\subsection{Gauge transformations}\label{ap:Gauge Trans}

In this subsection, we list for completeness the gauge transformation identities used above.
	\begin{align}
	A_{nm} &\longrightarrow \bra{n}e^{-i \chi_{n}}i\nabla e^{i\chi_{m}}\ket{m}
	=A_{nm}e^{i\Delta \chi_{nm}} -  \nabla \chi_m \delta_{n,m}\,,\\
	\bs{r}_{c} &\longrightarrow \bra{n}e^{-i \chi_{n}}i\nabla e^{i\chi_{n}}\ket{n}= \bs{r}_c -  \nabla \chi_n \,,\\
	\hat{H}' _{mn} &\longrightarrow \bra{m}e^{-i \chi_{m}}\hat{H}' e^{i\chi_{n}}\ket{n}=\bra{m}e^{-i \chi_{m}} \Big(
	-\big\{(\delta \hat{\bs{r}}^{\rho}+ \nabla_{k_\rho} \chi_n) B_{\mu\rho},\hat{p}^\mu\big\} + E_{\mu} (\delta \hat{\bs{r}} ^{\mu}+ \nabla_{k_\mu} \chi_n)
	  \Big) e^{i\chi_{n}}\ket{n}= e^{i\Delta \chi_{mn}} \hat{H}' _{mn} \,,\\
	\hat{H}'' _{mn} &\longrightarrow \bra{m}e^{-i \chi_{m}}\hat{H}'' e^{i\chi_{n}}\ket{n}=\bra{m}e^{-i \chi_{m}}\Big[(\delta \hat{\bs{r}}^{\nu} + \nabla_{k_\nu} \chi_n) B^{\mu}\hspace{1pt}_{\nu}(\delta \hat{\bs{r}}^{\rho} + \nabla_{k_\rho} \chi_n) B_{\mu\rho}\Big]e^{i\chi_{n}}\ket{n}=  e^{i\Delta \chi_{mn}} \hat{H}'' _{mn} \,,\\
	 c_{mn} &\longrightarrow \frac{1}{\mathcal{E}_n - \mathcal{E}_m}\bra{m}e^{-i \chi_{m}}\hat{H}' e^{i\chi_{n}}\ket{n}\,\\
	&=\frac{1}{\mathcal{E}_n - \mathcal{E}_m}\bra{m}e^{-i \chi_{m}}\Big(
	-\big\{(\delta \hat{\bs{r}}^{\rho}+ \nabla_{k_\rho} \chi_n) B_{\mu\rho},\hat{p}^\mu\big\} + E_{\mu} (\delta \hat{\bs{r}} ^{\mu}+ \nabla_{k_\mu} \chi_n)
	\Big)\ket{n}=\displaystyle \frac{e^{i\Delta \chi_{mn}} \hat{H}' _{mn} }{\mathcal{E}_n - \mathcal{E}_m}=\displaystyle e^{i\Delta \chi_{mn}} c_{mn}\,,\nonumber\\
	c' _{mn} &\longrightarrow  \frac{e^{i\Delta \chi_{mn}} H'' _{mn}}{\mathcal{E}_n - \mathcal{E}_m}  - \frac{e^{i\Delta \chi_{nn}} H' _{nn}e^{i\Delta \chi_{mn}} H' _{mn}}{(\mathcal{E}_n - \mathcal{E}_m)^2}+ \sum\limits_{m'\ne n }\frac{e^{i\Delta \chi_{mm'}} H' _{mm'}e^{i\Delta \chi_{m'n}} H' _{m'n}}{(\mathcal{E}_n - \mathcal{E}_m)(\mathcal{E}_n - \mathcal{E}_m')}\\
	&=e^{i\Delta \chi_{mn}} \frac{ H'' _{mn}}{\mathcal{E}_n - \mathcal{E}_m}  - e^{i\Delta \chi_{mn}}\frac{ H' _{nn} H' _{mn}}{(\mathcal{E}_n - \mathcal{E}_m)^2}+ \sum\limits_{m'\ne n }e^{i\Delta \chi_{mn}}\frac{ H' _{mm'} H' _{m'n}}{(\mathcal{E}_n - \mathcal{E}_m)(\mathcal{E}_n - \mathcal{E}_m')}=e^{i\Delta \chi_{mn}}c' _{mn}\,,\nonumber\\
	c''_{mn} &\longrightarrow   -\frac{H'' _{nn}e^{i\Delta \chi_{mn}} c_{mn}}{\mathcal{E}_n - \mathcal{E}_m}  -\frac{H' _{nn}e^{i\Delta \chi_{mn}} c' _{mn}}{\mathcal{E}_n - \mathcal{E}_m} + \sum\limits_{m'\ne n }\frac{e^{i\Delta \chi_{mm'}} H'' _{mm'}e^{i\Delta \chi_{m'n}} c _{m'n}}{(\mathcal{E}_n - \mathcal{E}_m)} + \sum\limits_{m' }\frac{e^{i\Delta \chi_{mm'}} H' _{mm'}e^{i\Delta \chi_{m'n}} c' _{m'n}}{(\mathcal{E}_n - \mathcal{E}_m)}\\
	&= -\frac{H'' _{nn}e^{i\Delta \chi_{mn}} c_{mn}}{\mathcal{E}_n - \mathcal{E}_m}  -\frac{H' _{nn}e^{i\Delta \chi_{mn}} c' _{mn}}{\mathcal{E}_n - \mathcal{E}_m} + \sum\limits_{m'\ne n }\frac{e^{i\Delta \chi_{mn}} H'' _{mm'} c _{m'n}}{(\mathcal{E}_n - \mathcal{E}_m)} + \sum\limits_{m' }\frac{e^{i\Delta \chi_{mn}} H' _{mm'}c' _{m'n}}{(\mathcal{E}_n - \mathcal{E}_m)}=e^{i\Delta \chi_{mn}}	c''_{mn}\,.\nonumber
	\end{align}

	\section{Generalized Bianchi Identities }\label{ap:Generalised BI}
	\setcounter{enumi}{2} 
\setcounter{equation}{0}
In the main text, we use generalised Bianchi identities to show that non-topological contributions to the current vanish. In this section, we prove these identities. We begin from the Bianchi identity for the quantity $ \mathbf{Q}^{\nu\rho\sigma\alpha}$,
	\begin{eqnarray}
\frac{\partial  \mathbf{Q}^{\mu\nu\sigma\lambda}}{\partial k_\rho} + \text{cycl}(\rho\mu\nu\sigma\lambda) = 0\,. 
	 \label{eq:2nd BI sup}
	\end{eqnarray}
This identity can be shown by expanding out the terms as:
	\begin{eqnarray}
	\begin{array}{rl}
	&  \frac{\partial }{\partial k_{\mu}} \mathbf{Q}^{\nu\rho\sigma\alpha} + \frac{\partial }{\partial k_{\alpha}} \mathbf{Q}^{\mu\nu\rho\sigma} + \frac{\partial }{\partial k_{\sigma}} \mathbf{Q}^{\alpha\mu\nu\rho}+ \frac{\partial }{\partial k_{\rho}} \mathbf{Q}^{\sigma\alpha\mu\nu}+ \frac{\partial }{\partial k_{\nu}} \mathbf{Q}^{\rho\sigma\alpha\mu} \\\\
	
	=&\displaystyle\Cline[black]{\frac{\partial \Omega^{\rho\sigma} }{\partial k_{\mu}}\Omega^{\nu\alpha} } + \Cline[red]{\frac{\partial \Omega^{\nu\alpha} }{\partial k_{\mu}}\Omega^{\rho\sigma} } 
	+ \Cline[green]{\frac{\partial \Omega^{\rho\alpha} }{\partial k_{\mu}}\Omega^{\sigma\nu}} + \Cline[blue]{\frac{\partial \Omega^{\sigma\nu} }{\partial k_{\mu}}\Omega^{\rho\alpha} }
	+\Cline[cyan]{\frac{\partial \Omega^{\nu\rho} }{\partial k_{\mu}}\Omega^{\sigma\alpha}} +\Cline[magenta]{\frac{\partial \Omega^{\sigma\alpha} }{\partial k_{\mu}}\Omega^{\nu\rho}} \\\\
	
	&\displaystyle+\Cline[yellow]{\frac{\partial \Omega^{\rho\sigma} }{\partial k_{\alpha}}\Omega^{\mu\nu}} +\Cline[red]{\frac{\partial \Omega^{\mu\nu} }{\partial k_{\alpha}}\Omega^{\rho\sigma}} 
	+\Cline[gray]{\frac{\partial \Omega^{\sigma\nu} }{\partial k_{\alpha}}\Omega^{\mu\rho}} +\Cline[green]{\frac{\partial \Omega^{\mu\rho} }{\partial k_{\alpha}}\Omega^{\sigma\nu}} 
	+\Cline[magenta]{\frac{\partial \Omega^{\sigma\mu} }{\partial k_{\alpha}}\Omega^{\rho\nu}} +\Cline[brown]{\frac{\partial \Omega^{\rho\nu} }{\partial k_{\alpha}}\Omega^{\sigma\mu}} \\\\
	
	&\displaystyle+\Cline[magenta]{\frac{\partial \Omega^{\alpha\mu} }{\partial k_{\sigma}}\Omega^{\nu\rho}}+\Cline[lime]{\frac{\partial \Omega^{\nu\rho} }{\partial k_{\sigma}}\Omega^{\alpha\mu}}
	+\Cline[gray]{\frac{\partial \Omega^{\alpha\nu} }{\partial k_{\sigma}}\Omega^{\rho\mu}}+\Cline[black]{\frac{\partial \Omega^{\rho\mu} }{\partial k_{\sigma}}\Omega^{\alpha\nu}}
	+\Cline[yellow]{\frac{\partial \Omega^{\alpha\rho} }{\partial k_{\sigma}}\Omega^{\mu\nu}}+\Cline[blue]{\frac{\partial \Omega^{\mu\nu} }{\partial k_{\sigma}}\Omega^{\alpha\rho}}\\\\
	
	&\displaystyle+\Cline[yellow]{\frac{\partial \Omega^{\sigma\alpha} }{\partial k_{\rho}}\Omega^{\mu\nu}} +\Cline[cyan]{\frac{\partial \Omega^{\mu\nu} }{\partial k_{\rho}}\Omega^{\sigma\alpha} }
	+\Cline[lime]{\frac{\partial \Omega^{\sigma\nu} }{\partial k_{\rho}}\Omega^{\alpha\mu} }+\Cline[green]{\frac{\partial \Omega^{\alpha\mu} }{\partial k_{\rho}}\Omega^{\sigma\nu} }
	+\Cline[black]{\frac{\partial \Omega^{\sigma\mu} }{\partial k_{\rho}}\Omega^{\nu\alpha}} +\Cline[brown]{\frac{\partial \Omega^{\nu\alpha} }{\partial k_{\rho}}\Omega^{\sigma\mu}} \\\\
	
	&\displaystyle+\Cline[lime]{\frac{\partial \Omega^{\rho\sigma} }{\partial k_{\nu}}\Omega^{\alpha\mu}} +\Cline[red]{\frac{\partial \Omega^{\alpha\mu} }{\partial k_{\nu}}\Omega^{\rho\sigma} }
	+\Cline[brown]{\frac{\partial \Omega^{\rho\alpha} }{\partial k_{\nu}}\Omega^{\mu\sigma}} +\Cline[blue]{\frac{\partial \Omega^{\mu\sigma} }{\partial k_{\nu}}\Omega^{\rho\alpha} }
	+\Cline[cyan]{\frac{\partial \Omega^{\rho\mu} }{\partial k_{\nu}}\Omega^{\sigma\alpha}} +\Cline[gray]{\frac{\partial \Omega^{\sigma\alpha} }{\partial k_{\nu}}\Omega^{\rho\mu} }\\\\
	=&0\,,
	\end{array}
	\label{eq:BI2}
	\end{eqnarray}
	where the color coding represents terms which combine to give the original Bianchi identity and, hence, vanish. For example the black colors give
	\begin{eqnarray}
	\begin{array}{rl}
	&\frac{\partial \Omega^{\rho\sigma} }{\partial k_{\mu}}\Omega^{\nu\alpha} + \frac{\partial \Omega^{\rho\mu} }{\partial k_{\sigma}}\Omega^{\alpha\nu} + \frac{\partial \Omega^{\sigma\mu} }{\partial k_{\rho}}\Omega^{\nu\alpha}=\Omega^{\nu\alpha}\left(\frac{\partial \Omega^{\rho\sigma} }{\partial k_{\mu}} + \frac{\partial \Omega^{\mu\rho} }{\partial k_{\sigma}}+ \frac{\partial \Omega^{\sigma\mu} }{\partial k_{\rho}}\right)=0	\,.\end{array}
	\end{eqnarray}

In the main text, we also used a generalised Bianchi identity for the quantity $\mathbf{Q}^{\mu\nu\xi\omega\lambda\iota}$:
\begin{eqnarray}
\frac{\partial  \mathbf{Q}^{\mu\nu\xi\omega\lambda\iota}}{\partial k_\rho}+ \text{cycl}(\rho\mu\nu\xi\omega\lambda\iota) = 0 \,. 
	\label{eq:3rd BI sup}
	\end{eqnarray} 
 To show this, we need the definition of this quantity as well as its antisymmetry in order to analyse each term separately 
\begin{eqnarray}
\begin{array}{l}
\bullet \,\,\displaystyle \partial_\rho \mathbf{Q}^{\mu\nu\xi\omega\lambda\iota}=\partial_\rho \left[\Omega^{\mu\nu}\mathbf{Q}^{\xi\omega\lambda\iota}
+\Omega^{\mu\xi}\mathbf{Q}^{\omega\lambda\iota\nu}
+\Omega^{\mu\omega}\mathbf{Q}^{\lambda\iota\nu\xi}
+\Omega^{\mu\lambda}\mathbf{Q}^{\iota\nu\xi\omega}
+\Omega^{\mu\iota}\mathbf{Q}^{\nu\xi\omega\lambda}\right]\\\\

\bullet \,\,\displaystyle \partial_\mu \mathbf{Q}^{\nu\xi\omega\lambda\iota\rho}=\partial_\mu \left[\Omega^{\xi\rho}\mathbf{Q}^{\omega\lambda\iota\nu}
+\Omega^{\xi\omega}\mathbf{Q}^{\lambda\iota\nu\rho}
+\Omega^{\xi\lambda}\mathbf{Q}^{\iota\nu\rho\omega}
+\Omega^{\xi\iota}\mathbf{Q}^{\nu\rho\omega\lambda}
+\Omega^{\xi\nu}\mathbf{Q}^{\rho\omega\lambda\iota}\right]\\\\

\bullet \,\,\displaystyle \partial_\nu \mathbf{Q}^{\xi\omega\lambda\iota\rho\mu}=\partial_\nu \left[\Omega^{\mu\xi}\mathbf{Q}^{\lambda\omega\iota\rho}
+\Omega^{\mu\lambda}\mathbf{Q}^{\omega\iota\rho\xi}
+\Omega^{\mu\omega}\mathbf{Q}^{\iota\rho\xi\lambda}
+\Omega^{\mu\iota}\mathbf{Q}^{\rho\xi\lambda\omega}
+\Omega^{\mu\rho}\mathbf{Q}^{\xi\lambda\omega\iota}\right]\\\\

\bullet \,\,\displaystyle \partial_\xi \mathbf{Q}^{\omega\lambda\iota\rho\mu\nu}=\partial_\xi \left[\Omega^{\mu\omega}\mathbf{Q}^{\lambda\iota\rho\nu}
+\Omega^{\mu\lambda}\mathbf{Q}^{\iota\rho\mu\omega}
+\Omega^{\mu\iota}\mathbf{Q}^{\rho\nu\omega\lambda}
+\Omega^{\mu\rho}\mathbf{Q}^{\nu\omega\lambda\iota}
+\Omega^{\mu\nu}\mathbf{Q}^{\omega\lambda\iota\rho}\right]\\\\

\bullet \,\,\displaystyle \partial_\omega \mathbf{Q}^{\lambda\iota\rho\mu\nu\xi}=\partial_\omega \left[\Omega^{\mu\lambda}\mathbf{Q}^{\rho\iota\nu\xi}
+\Omega^{\mu\rho}\mathbf{Q}^{\iota\nu\xi\lambda}
+\Omega^{\mu\iota}\mathbf{Q}^{\nu\xi\lambda\rho}
+\Omega^{\mu\nu}\mathbf{Q}^{\xi\lambda\rho\iota}
+\Omega^{\mu\xi}\mathbf{Q}^{\lambda\rho\iota\nu}\right]\\\\

\bullet \,\,\displaystyle \partial_\lambda \mathbf{Q}^{\iota\rho\mu\nu\xi\omega}=\partial_\lambda \left[\Omega^{\mu\iota}\mathbf{Q}^{\rho\nu\xi\omega}
+\Omega^{\mu\rho}\mathbf{Q}^{\nu\xi\omega\iota}
+\Omega^{\mu\nu}\mathbf{Q}^{\xi\omega\iota\rho}
+\Omega^{\mu\xi}\mathbf{Q}^{\omega\iota\rho\nu}
+\Omega^{\mu\omega}\mathbf{Q}^{\iota\rho\nu\xi}\right]\\\\

\bullet \,\,\displaystyle \partial_\iota \mathbf{Q}^{\rho\mu\nu\xi\omega\lambda}=\partial_\iota \left[\Omega^{\mu\rho}\mathbf{Q}^{\xi\nu\omega\lambda}
+\Omega^{\mu\xi}\mathbf{Q}^{\nu\omega\lambda\rho}
+\Omega^{\mu\nu}\mathbf{Q}^{\omega\lambda\rho\xi}
+\Omega^{\mu\omega}\mathbf{Q}^{\lambda\rho\xi\nu}
+\Omega^{\mu\lambda}\mathbf{Q}^{\rho\xi\nu\omega}\right]\\\\
\end{array}
\end{eqnarray}
Following standard differentiation rules, each term is transformed into a derivative on $\Omega^{\xi\iota}$ times $\mathbf{Q}^{\lambda\omega\mu\rho}$ plus a derivative on $\mathbf{Q}^{\lambda\omega\mu\rho}$ times $\Omega^{\xi\iota}$, i.e.,
\begin{eqnarray}
\partial_\nu\left[\Omega^{\xi\iota}\mathbf{Q}^{\lambda\omega\mu\rho}\right] = \partial_\nu\left(\Omega^{\xi\iota}\right)\mathbf{Q}^{\lambda\omega\mu\rho} + \Omega^{\xi\iota}\partial_\nu\left(\mathbf{Q}^{\lambda\omega\mu\rho}\right)\,.
\end{eqnarray}
We first collect terms proportional to $\Omega$'s
\begin{eqnarray}
\begin{array}{ccl}
\displaystyle\bigcirc=&+&\Omega^{\mu\nu}\left[
\partial_\rho \left( \mathbf{Q}^{\xi\omega\lambda\iota}   \right)
+\partial_\xi \left( \mathbf{Q}^{\omega\lambda\iota\rho}   \right)
+\partial_\omega \left( \mathbf{Q}^{\xi\lambda\rho\iota}   \right)
+\partial_\lambda \left( \mathbf{Q}^{\xi\omega\iota\rho}   \right)
+\partial_\iota \left( \mathbf{Q}^{\omega\lambda\rho\xi}   \right)
\right]\\\\

&+&\Omega^{\mu\xi}\left[
\partial_\rho \left( \mathbf{Q}^{\omega\lambda\iota\nu}   \right)
+\partial_\nu \left( \mathbf{Q}^{\lambda\omega\iota\rho}   \right)
+\partial_\omega \left( \mathbf{Q}^{\lambda\rho\iota\nu}   \right)
+\partial_\lambda \left( \mathbf{Q}^{\omega\iota\rho\nu}   \right)
+\partial_\iota \left( \mathbf{Q}^{\nu\omega\lambda\rho}   \right)
\right]\\\\

&+&\Omega^{\mu\omega}\left[
\partial_\rho \left( \mathbf{Q}^{\lambda\iota\nu\xi}   \right)
+\partial_\nu \left( \mathbf{Q}^{\iota\rho\xi\lambda}   \right)
+\partial_\xi \left( \mathbf{Q}^{\lambda\iota\rho\nu}   \right)
+\partial_\lambda \left( \mathbf{Q}^{\iota\rho\nu\xi}   \right)
+\partial_\iota \left( \mathbf{Q}^{\lambda\rho\xi\nu}   \right)
\right]\\\\

&+&\Omega^{\mu\nu}\left[
\partial_\rho \left( \mathbf{Q}^{\iota\nu\xi\omega}   \right)
+\partial_\nu \left( \mathbf{Q}^{\omega\iota\rho\xi}   \right)
+\partial_\xi \left( \mathbf{Q}^{\iota\rho\nu\omega}   \right)
+\partial_\omega \left( \mathbf{Q}^{\rho\iota\nu\xi}   \right)
+\partial_\iota \left( \mathbf{Q}^{\rho\xi\nu\omega}   \right)
\right]\\\\

&+&\Omega^{\mu\iota}\left[
\partial_\rho \left( \mathbf{Q}^{\nu\xi\omega\lambda}   \right)
+\partial_\nu \left( \mathbf{Q}^{\rho\xi\lambda\omega}   \right)
+\partial_\xi \left( \mathbf{Q}^{\rho\nu\omega\lambda}   \right)
+\partial_\omega \left( \mathbf{Q}^{\nu\xi\lambda\rho}   \right)
+\partial_\lambda \left( \mathbf{Q}^{\rho\nu\xi\omega}   \right)
\right]\\\\

&+&\Omega^{\mu\rho}\left[
\partial_\nu \left( \mathbf{Q}^{\xi\lambda\omega\iota}   \right)
+\partial_\xi \left( \mathbf{Q}^{\nu\omega\lambda\iota}   \right)
+\partial_\omega \left( \mathbf{Q}^{\iota\nu\xi\lambda}   \right)
+\partial_\lambda \left( \mathbf{Q}^{\nu\xi\omega\iota}   \right)
+\partial_\iota \left( \mathbf{Q}^{\xi\nu\omega\lambda}   \right)
\right]\\\\

&+&\Omega^{\xi\rho}	\partial_\mu \left( \mathbf{Q}^{\omega\lambda\iota\nu}   \right)
+\Omega^{\xi\omega}\partial_\mu \left( \mathbf{Q}^{\lambda\iota\nu\rho}   \right)
+\Omega^{\xi\lambda}\partial_\mu \left( \mathbf{Q}^{\iota\nu\rho\omega}   \right)
+\Omega^{\xi\iota}\partial_\mu \left( \mathbf{Q}^{\nu\rho\omega\lambda}   \right)
+\Omega^{\xi\nu}\partial_\mu \left( \mathbf{Q}^{\rho\omega\lambda\iota}   \right)
\\\\
=&+&\Omega^{\xi\rho}	\partial_\mu \left( \mathbf{Q}^{\omega\lambda\iota\nu}   \right)
+\Omega^{\xi\omega}\partial_\mu \left( \mathbf{Q}^{\lambda\iota\nu\rho}   \right)
+\Omega^{\xi\lambda}\partial_\mu \left( \mathbf{Q}^{\iota\nu\rho\omega}   \right)
+\Omega^{\xi\iota}\partial_\mu \left( \mathbf{Q}^{\nu\rho\omega\lambda}   \right)
+\Omega^{\xi\nu}\partial_\mu \left( \mathbf{Q}^{\rho\omega\lambda\iota}   \right)
\\\\
=&+&\partial_\mu \left( \Omega^{\omega\lambda}  \right)\mathbf{Q}^{\xi\rho\iota\nu}
+\partial_\mu \left( \Omega^{\iota\nu}  \right)\mathbf{Q}^{\xi\lambda\rho\omega} 
+\partial_\mu \left( \Omega^{\iota\omega}  \right)\mathbf{Q}^{\xi\rho\lambda\nu} 
+\partial_\mu \left( \Omega^{\nu\lambda}  \right)\mathbf{Q}^{\xi\rho\omega\iota} 
+\partial_\mu \left( \Omega^{\omega\nu}  \right)\mathbf{Q}^{\lambda\iota\xi\rho} \\\\
&+&\partial_\mu \left( \Omega^{\lambda\iota}  \right)\mathbf{Q}^{\omega\nu\xi\rho} 
+\partial_\mu \left( \Omega^{\nu\rho}  \right)\mathbf{Q}^{\lambda\iota\xi\omega}
+\partial_\mu \left( \Omega^{\rho\iota}  \right)\mathbf{Q}^{\lambda\nu\xi\omega}
+\partial_\mu \left( \Omega^{\lambda\rho}  \right)\mathbf{Q}^{\iota\nu\xi\omega}
+\partial_\mu \left( \Omega^{\rho\omega}  \right)\mathbf{Q}^{\iota\nu\xi\lambda}
\end{array}			\,,
\label{eq:circle}
\end{eqnarray}
where the second equality is obtained using the identity $\partial_\mu\mathbf{Q}^{\iota\nu\xi\lambda} + cycl(\mu\iota\nu\xi\lambda) = 0$ [cf.~Eq.~\eqref{eq:BI2}], while the third equality is obtained simply by applying the derivative and regrouping the terms. 

We are left with terms proportional to $\mathbf{Q}$'s
\begin{eqnarray}
\begin{array}{ll}
\bigstar=&+\mathbf{Q}^{\xi\omega\lambda\iota} \left(   \partial_\rho \Omega^{\mu\nu} + \partial_\nu \Omega^{\rho\mu}  \right)
+\mathbf{Q}^{\omega\lambda\iota\nu}  \Ccancel[red]{\left(   \partial_\rho \Omega^{\mu\xi} + \partial_\xi \Omega^{\rho\mu} + \partial_\mu \Omega^{\xi\rho}  \right)}\\\\
&+\mathbf{Q}^{\lambda\iota\nu\xi} \left(   \partial_\rho \Omega^{\mu\omega} + \partial_\omega \Omega^{\rho\mu}   \right)
+\mathbf{Q}^{\iota\nu\xi\omega} \left(   \partial_\rho \Omega^{\mu\lambda} + \partial_\lambda \Omega^{\rho\mu}   \right)
+\mathbf{Q}^{\nu\xi\omega\lambda} \left(   \partial_\rho \Omega^{\mu\iota} + \partial_\iota \Omega^{\rho\mu}   \right)\\\\
&+\mathbf{Q}^{\lambda\omega\iota\rho} \Ccancel[red]{\left(   \partial_\nu \Omega^{\mu\xi} + \partial_\xi \Omega^{\nu\mu}  + \partial_\mu \Omega^{\xi\nu}   \right)}
+\mathbf{Q}^{\omega\iota\rho\xi} \left(   \partial_\nu \Omega^{\mu\lambda} + \partial_\lambda \Omega^{\nu\mu}   \right)\\\\
&+\mathbf{Q}^{\iota\rho\xi\lambda} \left(   \partial_\nu \Omega^{\mu\omega} + \partial_\omega \Omega^{\nu\mu}   \right)
+\mathbf{Q}^{\rho\xi\lambda\omega} \left(   \partial_\nu \Omega^{\mu\iota} + \partial_\iota \Omega^{\nu\mu}   \right)
+\mathbf{Q}^{\rho\iota\nu\xi} \left(   \partial_\omega \Omega^{\mu\lambda} + \partial_\lambda \Omega^{\omega\mu}   \right)\\\\
&
+\mathbf{Q}^{\lambda\iota\rho\nu}\Ccancel[red]{ \left(   \partial_\xi \Omega^{\mu\omega} + \partial_\omega \Omega^{\xi\mu}+ \partial_\mu \Omega^{\omega\xi}   \right)}
+\mathbf{Q}^{\iota\rho\nu\omega} \Ccancel[red]{\left(   \partial_\xi \Omega^{\mu\lambda} + \partial_\lambda \Omega^{\xi\mu}+ \partial_\mu \Omega^{\lambda\xi}   \right)}\\\\
&
+\mathbf{Q}^{\rho\nu\omega\lambda}\Ccancel[red]{ \left(   \partial_\xi \Omega^{\mu\iota} + \partial_\iota \Omega^{\xi\mu}+ \partial_\mu \Omega^{\iota\xi}   \right)}
+\mathbf{Q}^{\nu\xi\lambda\rho} \left(   \partial_\omega \Omega^{\mu\iota} + \partial_\iota \Omega^{\omega\mu}  \right)\\\\
&
+\mathbf{Q}^{\rho\nu\xi\omega} \left(   \partial_\lambda \Omega^{\mu\iota} + \partial_\iota \Omega^{\lambda\mu}  \right)\\\\

=&+\mathbf{Q}^{\xi\omega\lambda\iota} \left(   \partial_\rho \Omega^{\mu\nu} + \partial_\nu \Omega^{\rho\mu}  \right)
+\mathbf{Q}^{\lambda\iota\nu\xi} \left(   \partial_\rho \Omega^{\mu\omega} + \partial_\omega \Omega^{\rho\mu}   \right)
+\mathbf{Q}^{\iota\nu\xi\omega} \left(   \partial_\rho \Omega^{\mu\lambda} + \partial_\lambda \Omega^{\rho\mu}   \right)\\\\
&+\mathbf{Q}^{\nu\xi\omega\lambda} \left(   \partial_\rho \Omega^{\mu\iota} + \partial_\iota \Omega^{\rho\mu}   \right)
+\mathbf{Q}^{\omega\iota\rho\xi} \left(   \partial_\nu \Omega^{\mu\lambda} + \partial_\lambda \Omega^{\nu\mu}   \right)
+\mathbf{Q}^{\iota\rho\xi\lambda} \left(   \partial_\nu \Omega^{\mu\omega} + \partial_\omega \Omega^{\nu\mu}   \right)\\\\
&+\mathbf{Q}^{\rho\xi\lambda\omega} \left(   \partial_\nu \Omega^{\mu\iota} + \partial_\iota \Omega^{\nu\mu}   \right)
+\mathbf{Q}^{\rho\iota\nu\xi} \left(   \partial_\omega \Omega^{\mu\lambda} + \partial_\lambda \Omega^{\omega\mu}   \right)

+\mathbf{Q}^{\nu\xi\lambda\rho} \left(   \partial_\omega \Omega^{\mu\iota} + \partial_\iota \Omega^{\omega\mu}  \right)\\\\
&		+\mathbf{Q}^{\rho\nu\xi\omega} \left(   \partial_\lambda \Omega^{\mu\iota} + \partial_\iota \Omega^{\lambda\mu}  \right)\\\\
\end{array}\,,
\label{eq:star}
\end{eqnarray}
where the crossed terms vanish due to Bianchi's identity $\partial_\rho \Omega^{\mu\nu} + cycl(\rho\mu\nu) = 0$. Adding Eqs.~(\ref{eq:circle}) and (\ref{eq:star}), we obtain 
\begin{eqnarray}
\begin{array}{rl}
\bigcirc + \bigstar = & +\mathbf{Q}^{\xi\omega\lambda\iota} \left( \partial_\rho \Omega^{\mu\nu}  + \partial_\nu \Omega^{\rho\mu}+ {\color{red} \partial_\mu \Omega^{\nu\rho}} \right)
+\mathbf{Q}^{\lambda\iota\nu\xi} \left(   \partial_\rho \Omega^{\mu\omega} + \partial_\omega \Omega^{\rho\mu}+ {\color{red} \partial_\mu \Omega^{\omega\rho}}   \right)\\\\
&+\mathbf{Q}^{\iota\nu\xi\omega} \left(   \partial_\rho \Omega^{\mu\lambda} + \partial_\lambda \Omega^{\rho\mu}  + {\color{red} \partial_\mu \Omega^{\lambda\rho}} \right)
+\mathbf{Q}^{\nu\xi\omega\lambda} \left(   \partial_\rho \Omega^{\mu\iota} + \partial_\iota \Omega^{\rho\mu}  + {\color{red} \partial_\mu \Omega^{\iota\rho}} \right)\\\\
&+\mathbf{Q}^{\omega\iota\rho\xi} \left(   \partial_\nu \Omega^{\mu\lambda} + \partial_\lambda \Omega^{\nu\mu} + {\color{red} \partial_\mu \Omega^{\lambda\nu}}  \right)
+\mathbf{Q}^{\iota\rho\xi\lambda} \left(   \partial_\nu \Omega^{\mu\omega} + \partial_\omega \Omega^{\nu\mu}  + {\color{red} \partial_\mu \Omega^{\omega\nu}}  \right)\\\\
&+\mathbf{Q}^{\rho\xi\lambda\omega} \left(   \partial_\nu \Omega^{\mu\iota} + \partial_\iota \Omega^{\nu\mu} + {\color{red} \partial_\mu \Omega^{\iota\nu}}   \right)
+\mathbf{Q}^{\rho\iota\nu\xi} \left(   \partial_\omega \Omega^{\mu\lambda} + \partial_\lambda \Omega^{\omega\mu}+ {\color{red} \partial_\mu \Omega^{\lambda\omega}}    \right)\\\\
&
+\mathbf{Q}^{\nu\xi\lambda\rho} \left(   \partial_\omega \Omega^{\mu\iota} + \partial_\iota \Omega^{\omega\mu} + {\color{red} \partial_\mu \Omega^{\iota\omega}}  \right)
+\mathbf{Q}^{\rho\nu\xi\omega} \left(   \partial_\lambda \Omega^{\mu\iota} + \partial_\iota \Omega^{\lambda\mu}+ {\color{red} \partial_\mu \Omega^{\iota\lambda}}   \right)\\\\
=&0
\end{array}\,,
\end{eqnarray} 
where the red terms originate from Eq.~\eqref{eq:circle} and the black terms from Eq.~(\ref{eq:star}). The entire set of terms is zero because each group of terms in parentheses is a variant of Bianchi's identity, leading to the generalised identity stated above. 

	\section{Order $\mathcal{O}(B^3)$}\label{ap:3rd order corre}
	\setcounter{enumi}{3}
\setcounter{equation}{0}

In this Section, we explicitly show that the non-topological terms at third order in the magnetic field strength vanish. These terms are all proportional to the derivative of the energy with respect to momentum and can be written as
	\begin{eqnarray}
	\begin{array}{rl}
	\diamondsuit=&\frac{1}{8}\frac{\partial \mathcal{E}}{\partial k_{\rho}} B_{\nu\rho}\Omega^{\mu\nu}B_{\xi\omega}B_{\lambda\iota} \mathbf{Q}^{\xi\omega\lambda\iota}
	+\frac{1}{2}\frac{\partial \mathcal{E}}{\partial k_{\sigma}} B_{\epsilon\eta}\Omega^{\epsilon\eta}B_{\nu\rho}\Omega^{\mu\nu}B_{\lambda\sigma}\Omega^{\rho\lambda}+\frac{\partial \mathcal{E}}{\partial k_{\epsilon}} B_{\omega\epsilon}\Omega^{\delta\omega}B_{\lambda\delta}\Omega^{\rho\lambda}B_{\nu\rho}\Omega^{\mu\nu}\,,\\\\
	\heartsuit=& \frac{1}{48^2}\frac{\partial \mathcal{E}}{\partial k_{\mu}} \left(\epsilon^{\rho\sigma\xi\omega\delta\iota}B_{\rho\sigma}B_{\xi\omega}B_{\delta\iota}\right)\left(\epsilon_{\rho\sigma\xi\omega\delta\iota}\Omega^{\rho\sigma}\Omega^{\xi\omega}\Omega^{\delta\iota}\right)
	=\frac{1}{48}\frac{\partial \mathcal{E}}{\partial k_{\mu}} B_{\rho\sigma}B_{\xi\omega}B_{\delta\iota}\mathbf{Q}^{\rho\sigma\xi\omega\delta\iota}\,,
	\end{array}
	\end{eqnarray}
	where $
	\mathbf{Q}^{\mu\nu\rho\xi\omega\lambda} := \Omega^{\mu\nu}\mathbf{Q}^{\rho\xi\omega\lambda}
	+\Omega^{\mu\rho}\mathbf{Q}^{\xi\nu\omega\lambda}
	+\Omega^{\mu\lambda}\mathbf{Q}^{\xi\omega\nu\rho}
	+\Omega^{\mu\xi}\mathbf{Q}^{\nu\omega\lambda\rho}
	+\Omega^{\mu\omega}\mathbf{Q}^{\xi\nu\lambda\rho}
	$. These terms vanish due to Bianchi's identity, as seen by the following calculation. To start, we combine the first three terms to obtain
	\begin{eqnarray}
	\begin{array}{rl}
	\diamondsuit =&\frac{1}{8}\frac{\partial \mathcal{E}}{\partial k_{\rho}} B_{\nu\rho}\Omega^{\mu\nu}B_{\xi\omega}B_{\lambda\iota} \mathbf{Q}^{\xi\omega\lambda\iota}
	+\frac{1}{2}\frac{\partial \mathcal{E}}{\partial k_{\sigma}} B_{\epsilon\eta}\Omega^{\epsilon\eta}B_{\nu\rho}\Omega^{\mu\nu}B_{\lambda\sigma}\Omega^{\rho\lambda}+\frac{\partial \mathcal{E}}{\partial k_{\epsilon}} B_{\omega\epsilon}\Omega^{\delta\omega}B_{\lambda\delta}\Omega^{\rho\lambda}B_{\nu\rho}\Omega^{\mu\nu}\\\\
	=& B_{\nu\rho}\Omega^{\mu\nu}\left(\frac{1}{8}\frac{\partial \mathcal{E}}{\partial k_{\rho}}B_{\xi\omega}B_{\lambda\iota} \mathbf{Q}^{\xi\omega\lambda\iota}
	+\frac{1}{2}\frac{\partial \mathcal{E}}{\partial k_{\sigma}} B_{\epsilon\eta}\Omega^{\epsilon\eta}B_{\lambda\sigma}\Omega^{\rho\lambda}+\frac{\partial \mathcal{E}}{\partial k_{\epsilon}} B_{\omega\epsilon}\Omega^{\delta\omega}B_{\lambda\delta}\Omega^{\rho\lambda}\right)\,.
	\end{array}
	\label{eq:correction3current}
	\end{eqnarray}
	Let
	\begin{eqnarray}
	\begin{array}{cc}
	\Lambda_1&=\frac{1}{8}\frac{\partial \mathcal{E}}{\partial k_{\rho}}B_{\xi\omega}B_{\lambda\iota} \mathbf{Q}^{\xi\omega\lambda\iota}\,,\\\\
	\Lambda_2&=\frac{1}{2}\frac{\partial \mathcal{E}}{\partial k_{\xi}} B_{\lambda\xi}B_{\omega\iota}\Omega^{\omega\iota}\Omega^{\rho\lambda}\,,\\\\
	\Lambda_3&=\frac{\partial \mathcal{E}}{\partial k_{\omega}} B_{\xi\omega}B_{\lambda\iota}\Omega^{\iota\xi}\Omega^{\rho\lambda}\,.
	\end{array}
	\end{eqnarray}
	We can re-write $\Lambda_2$ and $\Lambda_3$ as
	\begin{align}
	\displaystyle\Lambda_2\displaystyle= &\frac{1}{2}\cdot\frac{1}{4}B_{\lambda\xi}B_{\omega\iota}\displaystyle\left[\left(\partial_\xi \mathcal{E}\right) \Omega^{\omega\iota}\Omega^{\rho\lambda} 
	+\left(\partial_\lambda \mathcal{E}\right) \Omega^{\omega\iota}\Omega^{\xi\rho} 
	+ \left(\partial_\iota \mathcal{E}\right) \Omega^{\lambda\xi}\Omega^{\rho\omega} 
	+ \left(\partial_\omega \mathcal{E}\right) \Omega^{\xi\lambda}\Omega^{\iota\rho}\right]\,,\notag\\
	\Lambda_3=&\displaystyle  \frac{1}{8} B_{\lambda\iota}B_{\xi\omega}\displaystyle\left[
	\left(\partial_\omega \mathcal{E}\right) \Omega^{\iota\xi}\Omega^{\rho\lambda} 
	+\left(\partial_\omega \mathcal{E}\right) \Omega^{\lambda\xi}\Omega^{\iota\rho}
	+\left(\partial_\xi \mathcal{E}\right) \Omega^{\iota\omega}\Omega^{\lambda\rho}
	+\left(\partial_\xi \mathcal{E} \right)\Omega^{\lambda\omega}\Omega^{\rho\iota}
	+\left(\partial_\iota \mathcal{E}\right) \Omega^{\omega\lambda}\Omega^{\rho\xi}\right.\\
	&\displaystyle\left.+\left(\partial_\iota \mathcal{E}\right) \Omega^{\xi\lambda}\Omega^{\omega\rho}
	+\left(\partial_\lambda \mathcal{E}\right) \Omega^{\omega\iota}\Omega^{\xi\rho}
	+\left(\partial_{\lambda} \mathcal{E}\right) \Omega^{\xi\iota}\Omega^{\rho\omega}
	\right]\,.\notag
	\end{align}
	Summing the three and using the definition of $\mathbf{Q}^{\alpha\beta\gamma\delta}$, we obtain
	\begin{eqnarray}
	\begin{array}{c}
	\Lambda_1+\Lambda_2+\Lambda_3 =\displaystyle \frac{1}{8} B_{\lambda\iota} B_{\xi\omega} \left[
	\left(\partial_\rho \mathcal{E} \right)\mathbf{Q}^{\xi\omega\lambda\iota} +
	\left(\partial_\xi \mathcal{E} \right)\mathbf{Q}^{\omega\lambda\iota\rho}+
	\left(\partial_\omega \mathcal{E} \right)\mathbf{Q}^{\lambda\iota\rho\xi}+
	\left(\partial_\lambda \mathcal{E} \right)\mathbf{Q}^{\iota\rho\xi\omega}+
	\left(\partial_\iota \mathcal{E} \right)\mathbf{Q}^{\rho\xi\omega\lambda}
	\right]\,.
	\end{array}
	\end{eqnarray}
	Substituting back to Equation $\diamondsuit$ and using the freedom of re-naming the indices  
	\begin{eqnarray}
	\begin{array}{ll}
	\diamondsuit&=\displaystyle\frac{1}{8} B_{\nu\rho}B_{\lambda\iota} B_{\xi\omega}\Omega^{\mu\nu}\left[
	\left(\partial_\rho \mathcal{E} \right)\mathbf{Q}^{\xi\omega\lambda\iota} +
	\left(\partial_\xi \mathcal{E} \right)\mathbf{Q}^{\omega\lambda\iota\rho}+
	\left(\partial_\omega \mathcal{E} \right)\mathbf{Q}^{\lambda\iota\rho\xi}+
	\left(\partial_\lambda \mathcal{E} \right)\mathbf{Q}^{\iota\rho\xi\omega}+
	\left(\partial_\iota \mathcal{E} \right)\mathbf{Q}^{\rho\xi\omega\lambda}
	\right]\\\\
	&=\displaystyle\frac{1}{8\cdot 3}\left( 
	B_{\nu\rho}B_{\lambda\iota} B_{\xi\omega}\left[		
	\left(\partial_\rho \mathcal{E} \right)\Omega^{\mu\nu}\mathbf{Q}^{\xi\omega\lambda\iota} +
	\left(\partial_\xi \mathcal{E} \right)\Omega^{\mu\nu}\mathbf{Q}^{\omega\lambda\iota\rho}+
	\left(\partial_\omega \mathcal{E} \right)\Omega^{\mu\nu}\mathbf{Q}^{\lambda\iota\rho\xi}\right.\right.\\\\
	&\displaystyle\hspace{250pt}+\left.
	\left(\partial_\lambda \mathcal{E} \right)\Omega^{\mu\nu}\mathbf{Q}^{\iota\rho\xi\omega}+
	\left(\partial_\iota \mathcal{E} \right)\Omega^{\mu\nu}\mathbf{Q}^{\rho\xi\omega\lambda}
	\right]

	\\\\
	&\displaystyle\hspace{18pt}+B_{\lambda\iota}B_{\nu\rho} B_{\xi\omega}\left[
	\left(\partial_\iota \mathcal{E} \right)\Omega^{\mu\lambda}\mathbf{Q}^{\xi\omega\nu\rho} +
	\left(\partial_\xi \mathcal{E} \right)\Omega^{\mu\lambda}\mathbf{Q}^{\omega\nu\rho\iota}+
	\left(\partial_\omega \mathcal{E} \right)\Omega^{\mu\lambda}\mathbf{Q}^{\nu\rho\iota\xi}\right.\\\\
	&\displaystyle\hspace{250pt}+\left.
	\left(\partial_\nu \mathcal{E} \right)\Omega^{\mu\lambda}\mathbf{Q}^{\rho\iota\xi\omega}+
	\left(\partial_\rho \mathcal{E} \right)\Omega^{\mu\lambda}\mathbf{Q}^{\iota\xi\omega\nu}
	\right]
	
	\\\\
	&\displaystyle\hspace{18pt}+B_{\xi\omega}B_{\lambda\iota} B_{\nu\rho}\left[
	\left(\partial_\iota \mathcal{E} \right)\Omega^{\mu\xi}\mathbf{Q}^{\omega\nu\rho\lambda} +
	\left(\partial_\nu \mathcal{E} \right)\Omega^{\mu\xi}\mathbf{Q}^{\rho\lambda\iota\omega}+
	\left(\partial_\omega \mathcal{E} \right)\Omega^{\mu\xi}\mathbf{Q}^{\nu\rho\lambda\iota}\right.\\\\
	&\displaystyle\left.\hspace{250pt}+\left.
	\left(\partial_\lambda \mathcal{E} \right)\Omega^{\mu\xi}\mathbf{Q}^{\iota\omega\nu\rho}+
	\left(\partial_\rho \mathcal{E} \right)\Omega^{\mu\xi}\mathbf{Q}^{\lambda\iota\omega\nu}
	\right]\right)
	
	\\\\
	
	&=\displaystyle\frac{1}{8\cdot 3\cdot 2}\left( 
	B_{\nu\rho}B_{\lambda\iota} B_{\xi\omega}\left[		
	\left(\partial_\rho \mathcal{E} \right)\Omega^{\mu\nu}\mathbf{Q}^{\xi\omega\lambda\iota} +
	\left(\partial_\xi \mathcal{E} \right)\Omega^{\mu\nu}\mathbf{Q}^{\omega\lambda\iota\rho}+
	\left(\partial_\omega \mathcal{E} \right)\Omega^{\mu\nu}\mathbf{Q}^{\lambda\iota\rho\xi}\right.\right.\\\\
	&\displaystyle\hspace{250pt}+\left.
	\left(\partial_\lambda \mathcal{E} \right)\Omega^{\mu\nu}\mathbf{Q}^{\iota\rho\xi\omega}+
	\left(\partial_\iota \mathcal{E} \right)\Omega^{\mu\nu}\mathbf{Q}^{\rho\xi\omega\lambda}
	\right]

	\\\\
	&\displaystyle\hspace{35pt}+B_{\rho\nu}B_{\lambda\iota} B_{\xi\omega}\left[		
	\left(\partial_\nu \mathcal{E} \right)\Omega^{\mu\rho}\mathbf{Q}^{\xi\omega\lambda\iota} +
	\left(\partial_\xi \mathcal{E} \right)\Omega^{\mu\rho}\mathbf{Q}^{\omega\lambda\iota\nu}+
	\left(\partial_\omega \mathcal{E} \right)\Omega^{\mu\rho}\mathbf{Q}^{\lambda\iota\nu\xi}\right.\\\\
	&\displaystyle\hspace{250pt}+\left.
	\left(\partial_\lambda \mathcal{E} \right)\Omega^{\mu\rho}\mathbf{Q}^{\iota\nu\xi\omega}+
	\left(\partial_\iota \mathcal{E} \right)\Omega^{\mu\rho}\mathbf{Q}^{\nu\xi\omega\lambda}
	\right]

	\\\\
	&\displaystyle\hspace{35pt}+B_{\lambda\iota}B_{\nu\rho} B_{\xi\omega}\left[
	\left(\partial_\iota \mathcal{E} \right)\Omega^{\mu\lambda}\mathbf{Q}^{\xi\omega\nu\rho} +
	\left(\partial_\xi \mathcal{E} \right)\Omega^{\mu\lambda}\mathbf{Q}^{\omega\nu\rho\iota}+
	\left(\partial_\omega \mathcal{E} \right)\Omega^{\mu\lambda}\mathbf{Q}^{\nu\rho\iota\xi}\right.\\\\
	&\displaystyle\hspace{250pt}+\left.
	\left(\partial_\nu \mathcal{E} \right)\Omega^{\mu\lambda}\mathbf{Q}^{\rho\iota\xi\omega}+
	\left(\partial_\rho \mathcal{E} \right)\Omega^{\mu\lambda}\mathbf{Q}^{\iota\xi\omega\nu}
	\right]
	
	\\\\

	&\displaystyle\hspace{35pt}+B_{\iota\lambda}B_{\nu\rho} B_{\xi\omega}\left[
	\left(\partial_\lambda \mathcal{E} \right)\Omega^{\mu\iota}\mathbf{Q}^{\xi\omega\nu\rho} +
	\left(\partial_\xi \mathcal{E} \right)\Omega^{\mu\iota}\mathbf{Q}^{\omega\nu\rho\lambda}+
	\left(\partial_\omega \mathcal{E} \right)\Omega^{\mu\iota}\mathbf{Q}^{\nu\rho\lambda\xi}\right.\\\\
	&\displaystyle\hspace{250pt}+\left.
	\left(\partial_\nu \mathcal{E} \right)\Omega^{\mu\iota}\mathbf{Q}^{\rho\lambda\xi\omega}+
	\left(\partial_\rho \mathcal{E} \right)\Omega^{\mu\iota}\mathbf{Q}^{\lambda\xi\omega\nu}
	\right]
	
	\\\\
	&\displaystyle\hspace{35pt}+B_{\xi\omega}B_{\lambda\iota} B_{\nu\rho}\left[
	\left(\partial_\iota \mathcal{E} \right)\Omega^{\mu\xi}\mathbf{Q}^{\omega\nu\rho\lambda} +
	\left(\partial_\nu \mathcal{E} \right)\Omega^{\mu\xi}\mathbf{Q}^{\rho\lambda\iota\omega}+
	\left(\partial_\omega \mathcal{E} \right)\Omega^{\mu\xi}\mathbf{Q}^{\nu\rho\lambda\iota}\right.\\\\
	&\displaystyle\hspace{250pt}+\left.
	\left(\partial_\lambda \mathcal{E} \right)\Omega^{\mu\xi}\mathbf{Q}^{\iota\omega\nu\rho}+
	\left(\partial_\rho \mathcal{E} \right)\Omega^{\mu\xi}\mathbf{Q}^{\lambda\iota\omega\nu}\right]
	\\\\
	&\displaystyle\hspace{35pt}+B_{\omega\xi}B_{\lambda\iota} B_{\nu\rho}\left[
	\left(\partial_\iota \mathcal{E} \right)\Omega^{\mu\omega}\mathbf{Q}^{\xi\nu\rho\lambda} +
	\left(\partial_\nu \mathcal{E} \right)\Omega^{\mu\omega}\mathbf{Q}^{\rho\lambda\iota\xi}+
	\left(\partial_\xi \mathcal{E} \right)\Omega^{\mu\omega}\mathbf{Q}^{\nu\rho\lambda\iota}\right.\\\\
	&\displaystyle\left.\left.\hspace{250pt}+
	\left(\partial_\lambda \mathcal{E} \right)\Omega^{\mu\omega}\mathbf{Q}^{\iota\xi\nu\rho}+
	\left(\partial_\rho \mathcal{E} \right)\Omega^{\mu\omega}\mathbf{Q}^{\lambda\iota\xi\nu}
	\right]\right)\,.
	\end{array}
	\end{eqnarray}
	Taking the B-fields out of the parenthesis as a common factor and grouping the terms with respect to the derivatives of the energy
	\begin{eqnarray}
	\begin{array}{rl}
	\diamondsuit&\displaystyle=\frac{1}{48}B_{\nu\rho}B_{\lambda\iota}B_{\xi\omega}\left(
	\partial_\rho \mathcal{E} \left[
	\Omega^{\mu\nu}\mathbf{Q}^{\xi\omega\lambda\iota}
	+\Omega^{\mu\lambda}\mathbf{Q}^{\iota\xi\omega\nu}
	+\Omega^{\mu\iota}\mathbf{Q}^{\lambda\xi\nu\omega}
	+\Omega^{\mu\xi}\mathbf{Q}^{\lambda\iota\omega\nu}
	+\Omega^{\mu\omega}\mathbf{Q}^{\iota\xi\nu\lambda}
	\right] \right.\\\\
	
	&\displaystyle\hspace{77pt}+\partial_\nu \mathcal{E} \left[
	\Omega^{\mu\rho}\mathbf{Q}^{\omega\xi\lambda\iota}
	+\Omega^{\mu\lambda}\mathbf{Q}^{\rho\iota\xi\omega}
	+\Omega^{\mu\iota}\mathbf{Q}^{\rho\lambda\omega\xi}
	+\Omega^{\mu\xi}\mathbf{Q}^{\rho\lambda\iota\omega}
	+\Omega^{\mu\omega}\mathbf{Q}^{\lambda\rho\iota\xi}
	\right]\\\\
	
	&\displaystyle\hspace{77pt}+\partial_\xi \mathcal{E} \left[
	\Omega^{\mu\nu}\mathbf{Q}^{\omega\lambda\iota\rho}
	+\Omega^{\mu\rho}\mathbf{Q}^{\omega\iota\lambda\nu}
	+\Omega^{\mu\lambda}\mathbf{Q}^{\omega\nu\rho\iota}
	+\Omega^{\mu\iota}\mathbf{Q}^{\omega\nu\lambda\rho}
	+\Omega^{\mu\omega}\mathbf{Q}^{\rho\nu\lambda\iota}
	\right]\\\\
	
	&\displaystyle\hspace{77pt}+\partial_\omega \mathcal{E} \left[
	\Omega^{\mu\nu}\mathbf{Q}^{\lambda\iota\rho\xi}
	+\Omega^{\mu\rho}\mathbf{Q}^{\lambda\iota\xi\nu}
	+\Omega^{\mu\lambda}\mathbf{Q}^{\nu\rho\iota\xi}
	+\Omega^{\mu\iota}\mathbf{Q}^{\nu\lambda\rho\xi}
	+\Omega^{\mu\xi}\mathbf{Q}^{\nu\rho\lambda\iota}
	\right]\\\\
	
	&\displaystyle\hspace{77pt}+\partial_\lambda \mathcal{E} \left[
	\Omega^{\mu\nu}\mathbf{Q}^{\iota\rho\xi\omega}
	+\Omega^{\mu\rho}\mathbf{Q}^{\iota\xi\nu\omega}
	+\Omega^{\mu\iota}\mathbf{Q}^{\omega\xi\nu\rho}
	+\Omega^{\mu\xi}\mathbf{Q}^{\iota\omega\nu\rho}
	+\Omega^{\mu\omega}\mathbf{Q}^{\xi\iota\nu\rho}
	\right]\\\\
	
	&\hspace{77pt}+\displaystyle	\left.\partial_\iota \mathcal{E} \left[
	\Omega^{\mu\nu}\mathbf{Q}^{\rho\xi\omega\lambda}
	+\Omega^{\mu\rho}\mathbf{Q}^{\xi\nu\omega\lambda}
	+\Omega^{\mu\lambda}\mathbf{Q}^{\xi\omega\nu\rho}
	+\Omega^{\mu\xi}\mathbf{Q}^{\nu\omega\rho\lambda}
	+\Omega^{\mu\omega}\mathbf{Q}^{\xi\nu\lambda\rho}
	\right]		\right)\,.
	\end{array}
	\end{eqnarray}
	Using the definition 
	\begin{eqnarray}
	\mathbf{Q}^{\mu\nu\rho\xi\omega\lambda} := \Omega^{\mu\nu}\mathbf{Q}^{\rho\xi\omega\lambda}
	+\Omega^{\mu\rho}\mathbf{Q}^{\xi\nu\omega\lambda}
	+\Omega^{\mu\lambda}\mathbf{Q}^{\xi\omega\nu\rho}
	+\Omega^{\mu\xi}\mathbf{Q}^{\nu\omega\rho\lambda}
	+\Omega^{\mu\omega}\mathbf{Q}^{\xi\nu\lambda\rho}\,,
	\end{eqnarray}
	with $\mathbf{Q}^{\mu\nu\rho\xi\omega\lambda}$ being antisymmetric in all of its indices, we can recast $\diamondsuit$ to	
	\begin{eqnarray}
	\begin{array}{rl}
	\displaystyle\diamondsuit =&\displaystyle \frac{1}{48}B_{\nu\rho}B_{\lambda\iota}B_{\xi\omega}\left[
	\left(\partial_\rho \mathcal{E}\right)\mathbf{Q}^{\mu\nu\xi\omega\lambda\iota} 
	+\left(\partial_\nu \mathcal{E}\right)\mathbf{Q}^{\mu\rho\omega\xi\lambda\iota} 
	+\left(\partial_\xi \mathcal{E}\right)\mathbf{Q}^{\mu\nu\omega\lambda\iota\rho} \right.\\\\
	&\hspace{70pt}\left.+\left(\partial_\omega \mathcal{E}\right)\mathbf{Q}^{\mu\nu\lambda\iota\rho\xi} 
	+\left(\partial_\lambda \mathcal{E}\right)\mathbf{Q}^{\mu\nu\iota\rho\xi\omega}
	+\left(\partial_\iota \mathcal{E}\right)\mathbf{Q}^{\mu\lambda\xi\omega\nu\rho}
	\right]\,.
	\end{array}
	\end{eqnarray}
	Summing $\diamondsuit$ and $\heartsuit$ we arrive at 
	\begin{eqnarray}
	\begin{array}{ll}
	\displaystyle\heartsuit + \diamondsuit&\displaystyle=\frac{1}{48}B_{\nu\rho}B_{\lambda\iota}B_{\xi\omega}\left[
	\left(\partial_\rho \mathcal{E}\right)\mathbf{Q}^{\mu\nu\xi\omega\lambda\iota} 
	+\left(\partial_\nu \mathcal{E}\right)\mathbf{Q}^{\mu\rho\omega\xi\lambda\iota} 
	
	+\left(\partial_\xi \mathcal{E}\right)\mathbf{Q}^{\mu\nu\omega\lambda\iota\rho} 
	+\left(\partial_\omega \mathcal{E}\right)\mathbf{Q}^{\mu\nu\lambda\iota\rho\xi} \right.\\\\
	&\displaystyle\hspace{80pt}+\left(\partial_\lambda \mathcal{E}\right)\mathbf{Q}^{\mu\nu\iota\rho\xi\omega}
	\left.+\left(\partial_\iota \mathcal{E}\right)\mathbf{Q}^{\mu\lambda\xi\omega\nu\rho}
	+\left(\partial_\mu \mathcal{E}\right)\mathbf{Q}^{\nu\rho\lambda\iota\xi\omega}
	\right]\\\\
	&\displaystyle\longrightarrow\frac{1}{48}B_{\nu\rho}B_{\lambda\iota}B_{\xi\omega}\mathcal{E}\left(
	\partial_\rho \mathbf{Q}^{\mu\nu\xi\omega\lambda\iota} + \partial_\nu \mathbf{Q}^{\mu\rho\omega\xi\lambda\iota} 
	+\partial_\xi\mathbf{Q}^{\mu\nu\omega\lambda\iota\rho} 
	+\partial_\omega \mathbf{Q}^{\mu\nu\lambda\iota\rho\xi} \right.\\\\
	&\displaystyle\hspace{94pt}+\partial_\lambda \mathbf{Q}^{\mu\nu\iota\rho\xi\omega}
	\left.+\partial_\iota \mathbf{Q}^{\mu\lambda\xi\omega\nu\rho}
	+\partial_\mu \mathbf{Q}^{\nu\rho\lambda\iota\xi\omega}
	\right)\\\\
	
	&\displaystyle=\frac{1}{48}B_{\nu\rho}B_{\lambda\iota}B_{\xi\omega}\mathcal{E}\left(
	\partial_\rho \mathbf{Q}^{\mu\nu\xi\omega\lambda\iota} + cycl(\rho\mu\nu\xi\omega\lambda\iota) 
	\right)\,,\\\\
	\end{array}
	\end{eqnarray}
	where the arrow is obtained using the (not shown) integral over the Brillouin zone, which is integrated by parts. The last line is obtained using the antisymmetry of $ \mathbf{Q}^{\iota\rho\mu\nu\xi\omega}$. The terms in the parenthesis vanish due to the generalized Bianchi identity for six-forms [see Eq.~(\ref{eq:3rd BI sup})]. 
	

\begin{thebibliography}{86}%
		\makeatletter
		\providecommand \@ifxundefined [1]{%
			\@ifx{#1\undefined}
		}%
		\providecommand \@ifnum [1]{%
			\ifnum #1\expandafter \@firstoftwo
			\else \expandafter \@secondoftwo
			\fi
		}%
		\providecommand \@ifx [1]{%
			\ifx #1\expandafter \@firstoftwo
			\else \expandafter \@secondoftwo
			\fi
		}%
		\providecommand \natexlab [1]{#1}%
		\providecommand \enquote  [1]{``#1''}%
		\providecommand \bibnamefont  [1]{#1}%
		\providecommand \bibfnamefont [1]{#1}%
		\providecommand \citenamefont [1]{#1}%
		\providecommand \href@noop [0]{\@secondoftwo}%
		\providecommand \href [0]{\begingroup \@sanitize@url \@href}%
		\providecommand \@href[1]{\@@startlink{#1}\@@href}%
		\providecommand \@@href[1]{\endgroup#1\@@endlink}%
		\providecommand \@sanitize@url [0]{\catcode `\\12\catcode `\$12\catcode
			`\&12\catcode `\#12\catcode `\^12\catcode `\_12\catcode `\%12\relax}%
		\providecommand \@@startlink[1]{}%
		\providecommand \@@endlink[0]{}%
		\providecommand \url  [0]{\begingroup\@sanitize@url \@url }%
		\providecommand \@url [1]{\endgroup\@href {#1}{\urlprefix }}%
		\providecommand \urlprefix  [0]{URL }%
		\providecommand \Eprint [0]{\href }%
		\providecommand \doibase [0]{http://dx.doi.org/}%
		\providecommand \selectlanguage [0]{\@gobble}%
		\providecommand \bibinfo  [0]{\@secondoftwo}%
		\providecommand \bibfield  [0]{\@secondoftwo}%
		\providecommand \translation [1]{[#1]}%
		\providecommand \BibitemOpen [0]{}%
		\providecommand \bibitemStop [0]{}%
		\providecommand \bibitemNoStop [0]{.\EOS\space}%
		\providecommand \EOS [0]{\spacefactor3000\relax}%
		\providecommand \BibitemShut  [1]{\csname bibitem#1\endcsname}%
		\let\auto@bib@innerbib\@empty
		\bibitem [{\citenamefont {Hasan}\ and\ \citenamefont {Kane}(2010)}]{RMP_TI}%
		\BibitemOpen
		\bibfield  {author} {\bibinfo {author} {\bibfnamefont {M.~Z.}\ \bibnamefont
				{Hasan}}\ and\ \bibinfo {author} {\bibfnamefont {C.~L.}\ \bibnamefont
				{Kane}},\ }\href@noop {} {\bibfield  {journal} {\bibinfo  {journal} {Rev.
					Mod. Phys.}\ }\textbf {\bibinfo {volume} {82}},\ \bibinfo {pages} {3045}
			(\bibinfo {year} {2010})}\BibitemShut {NoStop}%
		\bibitem [{\citenamefont {Qi}\ and\ \citenamefont {Zhang}(2011)}]{RMP_TI2}%
		\BibitemOpen
		\bibfield  {author} {\bibinfo {author} {\bibfnamefont {X.-L.}\ \bibnamefont
				{Qi}}\ and\ \bibinfo {author} {\bibfnamefont {S.-C.}\ \bibnamefont {Zhang}},\
		}\href@noop {} {\bibfield  {journal} {\bibinfo  {journal} {Rev. Mod. Phys.}\
			}\textbf {\bibinfo {volume} {83}},\ \bibinfo {pages} {1057} (\bibinfo {year}
			{2011})}\BibitemShut {NoStop}%
		\bibitem [{\citenamefont {Ozawa}\ \emph {et~al.}(2018)\citenamefont {Ozawa},
			\citenamefont {Price}, \citenamefont {Amo}, \citenamefont {Goldman},
			\citenamefont {Hafezi}, \citenamefont {Lu}, \citenamefont {Rechtsman},
			\citenamefont {Schuster}, \citenamefont {Simon}, \citenamefont {Zilberberg}
			\emph {et~al.}}]{ozawareview}%
		\BibitemOpen
		\bibfield  {author} {\bibinfo {author} {\bibfnamefont {T.}~\bibnamefont
				{Ozawa}}, \bibinfo {author} {\bibfnamefont {H.~M.}\ \bibnamefont {Price}},
			\bibinfo {author} {\bibfnamefont {A.}~\bibnamefont {Amo}}, \bibinfo {author}
			{\bibfnamefont {N.}~\bibnamefont {Goldman}}, \bibinfo {author} {\bibfnamefont
				{M.}~\bibnamefont {Hafezi}}, \bibinfo {author} {\bibfnamefont
				{L.}~\bibnamefont {Lu}}, \bibinfo {author} {\bibfnamefont {M.}~\bibnamefont
				{Rechtsman}}, \bibinfo {author} {\bibfnamefont {D.}~\bibnamefont {Schuster}},
			\bibinfo {author} {\bibfnamefont {J.}~\bibnamefont {Simon}}, \bibinfo
			{author} {\bibfnamefont {O.}~\bibnamefont {Zilberberg}},  \emph {et~al.},\
		}\href@noop {} {\bibfield  {journal} {\bibinfo  {journal} {arXiv preprint
					arXiv:1802.04173}\ } (\bibinfo {year} {2018})}\BibitemShut {NoStop}%
		\bibitem [{\citenamefont {Schulz-Baldes}\ \emph {et~al.}(2000)\citenamefont
			{Schulz-Baldes}, \citenamefont {Kellendonk},\ and\ \citenamefont
			{Richter}}]{schulz2000simultaneous}%
		\BibitemOpen
		\bibfield  {author} {\bibinfo {author} {\bibfnamefont {H.}~\bibnamefont
				{Schulz-Baldes}}, \bibinfo {author} {\bibfnamefont {J.}~\bibnamefont
				{Kellendonk}}, \ and\ \bibinfo {author} {\bibfnamefont {T.}~\bibnamefont
				{Richter}},\ }\href@noop {} {\bibfield  {journal} {\bibinfo  {journal}
				{Journal of Physics A: Mathematical and General}\ }\textbf {\bibinfo {volume}
				{33}},\ \bibinfo {pages} {L27} (\bibinfo {year} {2000})}\BibitemShut
		{NoStop}%
		\bibitem [{\citenamefont {Kellendonk}\ \emph {et~al.}(2002)\citenamefont
			{Kellendonk}, \citenamefont {Richter},\ and\ \citenamefont
			{Schulz-Baldes}}]{kellendonk2002edge}%
		\BibitemOpen
		\bibfield  {author} {\bibinfo {author} {\bibfnamefont {J.}~\bibnamefont
				{Kellendonk}}, \bibinfo {author} {\bibfnamefont {T.}~\bibnamefont {Richter}},
			\ and\ \bibinfo {author} {\bibfnamefont {H.}~\bibnamefont {Schulz-Baldes}},\
		}\href@noop {} {\bibfield  {journal} {\bibinfo  {journal} {Reviews in
					Mathematical Physics}\ }\textbf {\bibinfo {volume} {14}},\ \bibinfo {pages}
			{87} (\bibinfo {year} {2002})}\BibitemShut {NoStop}%
		\bibitem [{\citenamefont {Klitzing}\ \emph {et~al.}(1980)\citenamefont
			{Klitzing}, \citenamefont {Dorda},\ and\ \citenamefont
			{Pepper}}]{Klitzing:1980PRL}%
		\BibitemOpen
		\bibfield  {author} {\bibinfo {author} {\bibfnamefont {K.~v.}\ \bibnamefont
				{Klitzing}}, \bibinfo {author} {\bibfnamefont {G.}~\bibnamefont {Dorda}}, \
			and\ \bibinfo {author} {\bibfnamefont {M.}~\bibnamefont {Pepper}},\
		}\href@noop {} {\bibfield  {journal} {\bibinfo  {journal} {Phys. Rev. Lett.}\
			}\textbf {\bibinfo {volume} {45}},\ \bibinfo {pages} {494} (\bibinfo {year}
			{1980})}\BibitemShut {NoStop}%
		\bibitem [{\citenamefont {Thouless}\ \emph {et~al.}(1982)\citenamefont
			{Thouless}, \citenamefont {Kohmoto}, \citenamefont {Nightingale},\ and\
			\citenamefont {den Nijs}}]{TKNN}%
		\BibitemOpen
		\bibfield  {author} {\bibinfo {author} {\bibfnamefont {D.~J.}\ \bibnamefont
				{Thouless}}, \bibinfo {author} {\bibfnamefont {M.}~\bibnamefont {Kohmoto}},
			\bibinfo {author} {\bibfnamefont {M.~P.}\ \bibnamefont {Nightingale}}, \ and\
			\bibinfo {author} {\bibfnamefont {M.}~\bibnamefont {den Nijs}},\ }\href@noop
		{} {\bibfield  {journal} {\bibinfo  {journal} {Phys. Rev. Lett.}\ }\textbf
			{\bibinfo {volume} {49}},\ \bibinfo {pages} {405} (\bibinfo {year}
			{1982})}\BibitemShut {NoStop}%
		\bibitem [{\citenamefont {Goldman}\ \emph {et~al.}(2016)\citenamefont
			{Goldman}, \citenamefont {Budich},\ and\ \citenamefont
			{Zoller}}]{Goldman:2016NatPhys}%
		\BibitemOpen
		\bibfield  {author} {\bibinfo {author} {\bibfnamefont {N.}~\bibnamefont
				{Goldman}}, \bibinfo {author} {\bibfnamefont {J.}~\bibnamefont {Budich}}, \
			and\ \bibinfo {author} {\bibfnamefont {P.}~\bibnamefont {Zoller}},\
		}\href@noop {} {\bibfield  {journal} {\bibinfo  {journal} {Nature Physics}\
			}\textbf {\bibinfo {volume} {12}},\ \bibinfo {pages} {639} (\bibinfo {year}
			{2016})}\BibitemShut {NoStop}%
		\bibitem [{\citenamefont {{Cooper}}\ \emph {et~al.}(2018)\citenamefont
			{{Cooper}}, \citenamefont {{Dalibard}},\ and\ \citenamefont
			{{Spielman}}}]{cooper2018}%
		\BibitemOpen
		\bibfield  {author} {\bibinfo {author} {\bibfnamefont {N.~R.}\ \bibnamefont
				{{Cooper}}}, \bibinfo {author} {\bibfnamefont {J.}~\bibnamefont
				{{Dalibard}}}, \ and\ \bibinfo {author} {\bibfnamefont {I.~B.}\ \bibnamefont
				{{Spielman}}},\ }\href@noop {} {\bibfield  {journal} {\bibinfo  {journal}
				{ArXiv e-prints}\ } (\bibinfo {year} {2018})},\ \Eprint
		{http://arxiv.org/abs/1803.00249} {arXiv:1803.00249 [cond-mat.quant-gas]}
		\BibitemShut {NoStop}%
		\bibitem [{\citenamefont {Lu}\ \emph {et~al.}(2016)\citenamefont {Lu},
			\citenamefont {Joannopoulos},\ and\ \citenamefont
			{Solja{\v{c}}i{\'c}}}]{lu2016topological}%
		\BibitemOpen
		\bibfield  {author} {\bibinfo {author} {\bibfnamefont {L.}~\bibnamefont
				{Lu}}, \bibinfo {author} {\bibfnamefont {J.~D.}\ \bibnamefont
				{Joannopoulos}}, \ and\ \bibinfo {author} {\bibfnamefont {M.}~\bibnamefont
				{Solja{\v{c}}i{\'c}}},\ }\href@noop {} {\bibfield  {journal} {\bibinfo
				{journal} {Nature Physics}\ }\textbf {\bibinfo {volume} {12}},\ \bibinfo
			{pages} {626} (\bibinfo {year} {2016})}\BibitemShut {NoStop}%
		\bibitem [{\citenamefont {Khanikaev}\ and\ \citenamefont
			{Shvets}(2017)}]{khanikaev2017two}%
		\BibitemOpen
		\bibfield  {author} {\bibinfo {author} {\bibfnamefont {A.~B.}\ \bibnamefont
				{Khanikaev}}\ and\ \bibinfo {author} {\bibfnamefont {G.}~\bibnamefont
				{Shvets}},\ }\href@noop {} {\bibfield  {journal} {\bibinfo  {journal} {Nature
					Photonics}\ }\textbf {\bibinfo {volume} {11}},\ \bibinfo {pages} {763}
			(\bibinfo {year} {2017})}\BibitemShut {NoStop}%
		\bibitem [{\citenamefont {Chiu}\ \emph {et~al.}(2016)\citenamefont {Chiu},
			\citenamefont {Teo}, \citenamefont {Schnyder},\ and\ \citenamefont
			{Ryu}}]{Chiu:2016RMP}%
		\BibitemOpen
		\bibfield  {author} {\bibinfo {author} {\bibfnamefont {C.-K.}\ \bibnamefont
				{Chiu}}, \bibinfo {author} {\bibfnamefont {J.~C.~Y.}\ \bibnamefont {Teo}},
			\bibinfo {author} {\bibfnamefont {A.~P.}\ \bibnamefont {Schnyder}}, \ and\
			\bibinfo {author} {\bibfnamefont {S.}~\bibnamefont {Ryu}},\ }\href@noop {}
		{\bibfield  {journal} {\bibinfo  {journal} {Rev. Mod. Phys.}\ }\textbf
			{\bibinfo {volume} {88}},\ \bibinfo {pages} {035005} (\bibinfo {year}
			{2016})}\BibitemShut {NoStop}%
		\bibitem [{\citenamefont {Avron}\ \emph {et~al.}(1988)\citenamefont {Avron},
			\citenamefont {Sadun}, \citenamefont {Segert},\ and\ \citenamefont
			{Simon}}]{Avron1988}%
		\BibitemOpen
		\bibfield  {author} {\bibinfo {author} {\bibfnamefont {J.~E.}\ \bibnamefont
				{Avron}}, \bibinfo {author} {\bibfnamefont {L.}~\bibnamefont {Sadun}},
			\bibinfo {author} {\bibfnamefont {J.}~\bibnamefont {Segert}}, \ and\ \bibinfo
			{author} {\bibfnamefont {B.}~\bibnamefont {Simon}},\ }\href@noop {}
		{\bibfield  {journal} {\bibinfo  {journal} {Phys. Rev. Lett.}\ }\textbf
			{\bibinfo {volume} {61}},\ \bibinfo {pages} {1329} (\bibinfo {year}
			{1988})}\BibitemShut {NoStop}%
		\bibitem [{\citenamefont {Fr{\"o}hlich}\ and\ \citenamefont
			{Perdini}()}]{frohlich2000}%
		\BibitemOpen
		\bibfield  {author} {\bibinfo {author} {\bibfnamefont {J.}~\bibnamefont
				{Fr{\"o}hlich}}\ and\ \bibinfo {author} {\bibfnamefont {B.}~\bibnamefont
				{Perdini}},\ }\enquote {\bibinfo {title} {New applications of the chiral
				anomaly},}\ in\ \href@noop {} {\emph {\bibinfo {booktitle} {Mathematical
					Physics 2000}}}\ (\bibinfo  {publisher} {Imperial College Press, London,
			United Kingdom})\ pp.\ \bibinfo {pages} {9--47}\BibitemShut {NoStop}%
		\bibitem [{\citenamefont {Zhang}\ and\ \citenamefont {Hu}(2001)}]{IQHE4D}%
		\BibitemOpen
		\bibfield  {author} {\bibinfo {author} {\bibfnamefont {S.-C.}\ \bibnamefont
				{Zhang}}\ and\ \bibinfo {author} {\bibfnamefont {J.}~\bibnamefont {Hu}},\
		}\href@noop {} {\bibfield  {journal} {\bibinfo  {journal} {Science}\ }\textbf
			{\bibinfo {volume} {294}},\ \bibinfo {pages} {823} (\bibinfo {year}
			{2001})}\BibitemShut {NoStop}%
		\bibitem [{\citenamefont {Qi}\ \emph {et~al.}(2008)\citenamefont {Qi},
			\citenamefont {Hughes},\ and\ \citenamefont {Zhang}}]{QiZhang}%
		\BibitemOpen
		\bibfield  {author} {\bibinfo {author} {\bibfnamefont {X.-L.}\ \bibnamefont
				{Qi}}, \bibinfo {author} {\bibfnamefont {T.~L.}\ \bibnamefont {Hughes}}, \
			and\ \bibinfo {author} {\bibfnamefont {S.-C.}\ \bibnamefont {Zhang}},\
		}\href@noop {} {\bibfield  {journal} {\bibinfo  {journal} {Phys. Rev. B}\
			}\textbf {\bibinfo {volume} {78}},\ \bibinfo {pages} {195424} (\bibinfo
			{year} {2008})}\BibitemShut {NoStop}%
		\bibitem [{\citenamefont {Price}\ \emph {et~al.}(2015)\citenamefont {Price},
			\citenamefont {Zilberberg}, \citenamefont {Ozawa}, \citenamefont
			{Carusotto},\ and\ \citenamefont {Goldman}}]{Price2015}%
		\BibitemOpen
		\bibfield  {author} {\bibinfo {author} {\bibfnamefont {H.~M.}\ \bibnamefont
				{Price}}, \bibinfo {author} {\bibfnamefont {O.}~\bibnamefont {Zilberberg}},
			\bibinfo {author} {\bibfnamefont {T.}~\bibnamefont {Ozawa}}, \bibinfo
			{author} {\bibfnamefont {I.}~\bibnamefont {Carusotto}}, \ and\ \bibinfo
			{author} {\bibfnamefont {N.}~\bibnamefont {Goldman}},\ }\href@noop {}
		{\bibfield  {journal} {\bibinfo  {journal} {Physical Review Letters}\
			}\textbf {\bibinfo {volume} {115}},\ \bibinfo {pages} {195303} (\bibinfo
			{year} {2015})}\BibitemShut {NoStop}%
		
		\bibitem [{\citenamefont {Prodan}\ and\ \citenamefont
			{Schulz-Baldes}(2016)}]{prodan2016bulk}%
		\bibfield  {author} {\bibinfo {author} {\bibfnamefont {E.}~\bibnamefont
				{Prodan}}\ and\ \bibinfo {author} {\bibfnamefont {H.}~\bibnamefont
				{Schulz-Baldes}},\ }\href@noop {} {\bibfield  {journal}{\bibinfo  {title}
				{\textit{Bulk and Boundary Invariants for Complex Topological Insulators: From K-Theory to Physics,}}\ } {\bibinfo  {journal}
				{Springer, Cham}\ } (\bibinfo {year} {2016})}\BibitemShut {NoStop}%
		\bibitem [{\citenamefont {Kraus}\ \emph {et~al.}(2013)\citenamefont {Kraus},
			\citenamefont {Ringel},\ and\ \citenamefont {Zilberberg}}]{Kraus2013}%
		\BibitemOpen
		\bibfield  {author} {\bibinfo {author} {\bibfnamefont {Y.~E.}\ \bibnamefont
				{Kraus}}, \bibinfo {author} {\bibfnamefont {Z.}~\bibnamefont {Ringel}}, \
			and\ \bibinfo {author} {\bibfnamefont {O.}~\bibnamefont {Zilberberg}},\
		}\href@noop {} {\bibfield  {journal} {\bibinfo  {journal} {Phys. Rev. Lett.}\
			}\textbf {\bibinfo {volume} {111}},\ \bibinfo {pages} {226401} (\bibinfo
			{year} {2013})}\BibitemShut {NoStop}%
		\bibitem [{\citenamefont {Kraus}\ and\ \citenamefont
			{Zilberberg}(2016)}]{kraus2016quasiperiodicity}%
		\BibitemOpen
		\bibfield  {author} {\bibinfo {author} {\bibfnamefont {Y.~E.}\ \bibnamefont
				{Kraus}}\ and\ \bibinfo {author} {\bibfnamefont {O.}~\bibnamefont
				{Zilberberg}},\ }\href@noop {} {\bibfield  {journal} {\bibinfo  {journal}
				{Nature Physics}\ }\textbf {\bibinfo {volume} {12}},\ \bibinfo {pages} {624}
			(\bibinfo {year} {2016})}\BibitemShut {NoStop}%
		\bibitem [{\citenamefont {Li}\ and\ \citenamefont {Wu}(2013)}]{li2013high}%
		\BibitemOpen
		\bibfield  {author} {\bibinfo {author} {\bibfnamefont {Y.}~\bibnamefont
				{Li}}\ and\ \bibinfo {author} {\bibfnamefont {C.}~\bibnamefont {Wu}},\
		}\href@noop {} {\bibfield  {journal} {\bibinfo  {journal} {Physical Review
					Letters}\ }\textbf {\bibinfo {volume} {110}},\ \bibinfo {pages} {216802}
			(\bibinfo {year} {2013})}\BibitemShut {NoStop}%
		\bibitem [{\citenamefont {Li}\ \emph {et~al.}(2013)\citenamefont {Li},
			\citenamefont {Zhang},\ and\ \citenamefont {Wu}}]{li2013topological}%
		\BibitemOpen
		\bibfield  {author} {\bibinfo {author} {\bibfnamefont {Y.}~\bibnamefont
				{Li}}, \bibinfo {author} {\bibfnamefont {S.-C.}\ \bibnamefont {Zhang}}, \
			and\ \bibinfo {author} {\bibfnamefont {C.}~\bibnamefont {Wu}},\ }\href@noop
		{} {\bibfield  {journal} {\bibinfo  {journal} {Physical review letters}\
			}\textbf {\bibinfo {volume} {111}},\ \bibinfo {pages} {186803} (\bibinfo
			{year} {2013})}\BibitemShut {NoStop}%
		\bibitem [{\citenamefont {Prodan}(2015)}]{prodan2015virtual}%
		\BibitemOpen
		\bibfield  {author} {\bibinfo {author} {\bibfnamefont {E.}~\bibnamefont
				{Prodan}},\ }\href@noop {} {\bibfield  {journal} {\bibinfo  {journal}
				{Physical Review B}\ }\textbf {\bibinfo {volume} {91}},\ \bibinfo {pages}
			{245104} (\bibinfo {year} {2015})}\BibitemShut {NoStop}%
		\bibitem [{\citenamefont {Ozawa}\ \emph {et~al.}(2016)\citenamefont {Ozawa},
			\citenamefont {Price}, \citenamefont {Goldman}, \citenamefont {Zilberberg},\
			and\ \citenamefont {Carusotto}}]{Ozawa2016}%
		\BibitemOpen
		\bibfield  {author} {\bibinfo {author} {\bibfnamefont {T.}~\bibnamefont
				{Ozawa}}, \bibinfo {author} {\bibfnamefont {H.~M.}\ \bibnamefont {Price}},
			\bibinfo {author} {\bibfnamefont {N.}~\bibnamefont {Goldman}}, \bibinfo
			{author} {\bibfnamefont {O.}~\bibnamefont {Zilberberg}}, \ and\ \bibinfo
			{author} {\bibfnamefont {I.}~\bibnamefont {Carusotto}},\ }\href@noop {}
		{\bibfield  {journal} {\bibinfo  {journal} {Physical Review A}\ }\textbf
			{\bibinfo {volume} {93}},\ \bibinfo {pages} {043827} (\bibinfo {year}
			{2016})}\BibitemShut {NoStop}%
		\bibitem [{\citenamefont {Price}\ \emph {et~al.}(2016)\citenamefont {Price},
			\citenamefont {Zilberberg}, \citenamefont {Ozawa}, \citenamefont
			{Carusotto},\ and\ \citenamefont {Goldman}}]{Price2016}%
		\BibitemOpen
		\bibfield  {author} {\bibinfo {author} {\bibfnamefont {H.~M.}\ \bibnamefont
				{Price}}, \bibinfo {author} {\bibfnamefont {O.}~\bibnamefont {Zilberberg}},
			\bibinfo {author} {\bibfnamefont {T.}~\bibnamefont {Ozawa}}, \bibinfo
			{author} {\bibfnamefont {I.}~\bibnamefont {Carusotto}}, \ and\ \bibinfo
			{author} {\bibfnamefont {N.}~\bibnamefont {Goldman}},\ }\href@noop {}
		{\bibfield  {journal} {\bibinfo  {journal} {Physical Review B}\ }\textbf
			{\bibinfo {volume} {93}},\ \bibinfo {pages} {245113} (\bibinfo {year}
			{2016})}\BibitemShut {NoStop}%
		\bibitem [{\citenamefont {Boada}\ \emph {et~al.}(2012)\citenamefont {Boada},
			\citenamefont {Celi}, \citenamefont {Latorre},\ and\ \citenamefont
			{Lewenstein}}]{Boada2012}%
		\BibitemOpen
		\bibfield  {author} {\bibinfo {author} {\bibfnamefont {O.}~\bibnamefont
				{Boada}}, \bibinfo {author} {\bibfnamefont {A.}~\bibnamefont {Celi}},
			\bibinfo {author} {\bibfnamefont {J.~I.}\ \bibnamefont {Latorre}}, \ and\
			\bibinfo {author} {\bibfnamefont {M.}~\bibnamefont {Lewenstein}},\
		}\href@noop {} {\bibfield  {journal} {\bibinfo  {journal} {Phys. Rev. Lett.}\
			}\textbf {\bibinfo {volume} {108}},\ \bibinfo {pages} {133001} (\bibinfo
			{year} {2012})}\BibitemShut {NoStop}%
		\bibitem [{\citenamefont {Celi}\ \emph {et~al.}(2014)\citenamefont {Celi},
			\citenamefont {Massignan}, \citenamefont {Ruseckas}, \citenamefont {Goldman},
			\citenamefont {Spielman}, \citenamefont {Juzeli\ifmmode~\bar{u}\else
				\={u}\fi{}nas},\ and\ \citenamefont {Lewenstein}}]{Celi:2012PRL}%
		\BibitemOpen
		\bibfield  {author} {\bibinfo {author} {\bibfnamefont {A.}~\bibnamefont
				{Celi}}, \bibinfo {author} {\bibfnamefont {P.}~\bibnamefont {Massignan}},
			\bibinfo {author} {\bibfnamefont {J.}~\bibnamefont {Ruseckas}}, \bibinfo
			{author} {\bibfnamefont {N.}~\bibnamefont {Goldman}}, \bibinfo {author}
			{\bibfnamefont {I.~B.}\ \bibnamefont {Spielman}}, \bibinfo {author}
			{\bibfnamefont {G.}~\bibnamefont {Juzeliunas}}, \ and\ \bibinfo {author} {\bibfnamefont {M.}~\bibnamefont
				{Lewenstein}},\ }\href@noop {} {\bibfield  {journal} {\bibinfo  {journal}
				{Phys. Rev. Lett.}\ }\textbf {\bibinfo {volume} {112}},\ \bibinfo {pages}
			{043001} (\bibinfo {year} {2014})}\BibitemShut {NoStop}%
		\bibitem [{\citenamefont {Boada}\ \emph {et~al.}(2015)\citenamefont {Boada},
			\citenamefont {Celi}, \citenamefont {Rodr{\'\i}guez-Laguna}, \citenamefont
			{Latorre},\ and\ \citenamefont {Lewenstein}}]{boada2015quantum}%
		\BibitemOpen
		\bibfield  {author} {\bibinfo {author} {\bibfnamefont {O.}~\bibnamefont
				{Boada}}, \bibinfo {author} {\bibfnamefont {A.}~\bibnamefont {Celi}},
			\bibinfo {author} {\bibfnamefont {J.}~\bibnamefont {Rodr{\'\i}guez-Laguna}},
			\bibinfo {author} {\bibfnamefont {J.~I.}\ \bibnamefont {Latorre}}, \ and\
			\bibinfo {author} {\bibfnamefont {M.}~\bibnamefont {Lewenstein}},\
		}\href@noop {} {\bibfield  {journal} {\bibinfo  {journal} {New Journal of
					Physics}\ }\textbf {\bibinfo {volume} {17}},\ \bibinfo {pages} {045007}
			(\bibinfo {year} {2015})}\BibitemShut {NoStop}%
		\bibitem [{\citenamefont {Mancini}\ \emph {et~al.}(2015)\citenamefont
			{Mancini}, \citenamefont {Pagano}, \citenamefont {Cappellini}, \citenamefont
			{Livi}, \citenamefont {Rider}, \citenamefont {Catani}, \citenamefont {Sias},
			\citenamefont {Zoller}, \citenamefont {Inguscio}, \citenamefont {Dalmonte},\
			and\ \citenamefont {Fallani}}]{Mancini:2015Science}%
		\BibitemOpen
		\bibfield  {author} {\bibinfo {author} {\bibfnamefont {M.}~\bibnamefont
				{Mancini}}, \bibinfo {author} {\bibfnamefont {G.}~\bibnamefont {Pagano}},
			\bibinfo {author} {\bibfnamefont {G.}~\bibnamefont {Cappellini}}, \bibinfo
			{author} {\bibfnamefont {L.}~\bibnamefont {Livi}}, \bibinfo {author}
			{\bibfnamefont {M.}~\bibnamefont {Rider}}, \bibinfo {author} {\bibfnamefont
				{J.}~\bibnamefont {Catani}}, \bibinfo {author} {\bibfnamefont
				{C.}~\bibnamefont {Sias}}, \bibinfo {author} {\bibfnamefont {P.}~\bibnamefont
				{Zoller}}, \bibinfo {author} {\bibfnamefont {M.}~\bibnamefont {Inguscio}},
			\bibinfo {author} {\bibfnamefont {M.}~\bibnamefont {Dalmonte}}, \ and\
			\bibinfo {author} {\bibfnamefont {L.}~\bibnamefont {Fallani}},\ }\href@noop
		{} {\bibfield  {journal} {\bibinfo  {journal} {Science}\ }\textbf {\bibinfo
				{volume} {349}},\ \bibinfo {pages} {1510} (\bibinfo {year}
			{2015})}\BibitemShut {NoStop}%
		\bibitem [{\citenamefont {Stuhl}\ \emph {et~al.}(2015)\citenamefont {Stuhl},
			\citenamefont {Lu}, \citenamefont {Aycock}, \citenamefont {Genkina},\ and\
			\citenamefont {Spielman}}]{Stuhl:2015Science}%
		\BibitemOpen
		\bibfield  {author} {\bibinfo {author} {\bibfnamefont {B.~K.}\ \bibnamefont
				{Stuhl}}, \bibinfo {author} {\bibfnamefont {H.-I.}\ \bibnamefont {Lu}},
			\bibinfo {author} {\bibfnamefont {L.~M.}\ \bibnamefont {Aycock}}, \bibinfo
			{author} {\bibfnamefont {D.}~\bibnamefont {Genkina}}, \ and\ \bibinfo
			{author} {\bibfnamefont {I.~B.}\ \bibnamefont {Spielman}},\ }\href@noop {}
		{\bibfield  {journal} {\bibinfo  {journal} {Science}\ }\textbf {\bibinfo
				{volume} {349}},\ \bibinfo {pages} {1514} (\bibinfo {year}
			{2015})}\BibitemShut {NoStop}%
		\bibitem [{\citenamefont {Luo}\ \emph {et~al.}(2015)\citenamefont {Luo},
			\citenamefont {Zhou}, \citenamefont {Li}, \citenamefont {Xu}, \citenamefont
			{Guo},\ and\ \citenamefont {Zhou}}]{luo2015quantum}%
		\BibitemOpen
		\bibfield  {author} {\bibinfo {author} {\bibfnamefont {X.-W.}\ \bibnamefont
				{Luo}}, \bibinfo {author} {\bibfnamefont {X.}~\bibnamefont {Zhou}}, \bibinfo
			{author} {\bibfnamefont {C.-F.}\ \bibnamefont {Li}}, \bibinfo {author}
			{\bibfnamefont {J.-S.}\ \bibnamefont {Xu}}, \bibinfo {author} {\bibfnamefont
				{G.-C.}\ \bibnamefont {Guo}}, \ and\ \bibinfo {author} {\bibfnamefont
				{Z.-W.}\ \bibnamefont {Zhou}},\ }\href@noop {} {\bibfield  {journal}
			{\bibinfo  {journal} {Nature communications}\ }\textbf {\bibinfo {volume}
				{6}},\ \bibinfo {pages} {7704} (\bibinfo {year} {2015})}\BibitemShut
		{NoStop}%
		\bibitem [{\citenamefont {Livi}\ \emph {et~al.}(2016)\citenamefont {Livi},
			\citenamefont {Cappellini}, \citenamefont {Diem}, \citenamefont {Franchi},
			\citenamefont {Clivati}, \citenamefont {Frittelli}, \citenamefont {Levi},
			\citenamefont {Calonico}, \citenamefont {Catani}, \citenamefont {Inguscio},\
			and\ \citenamefont {Fallani}}]{Livi:2016PRL}%
		\BibitemOpen
		\bibfield  {author} {\bibinfo {author} {\bibfnamefont {L.~F.}\ \bibnamefont
				{Livi}}, \bibinfo {author} {\bibfnamefont {G.}~\bibnamefont {Cappellini}},
			\bibinfo {author} {\bibfnamefont {M.}~\bibnamefont {Diem}}, \bibinfo {author}
			{\bibfnamefont {L.}~\bibnamefont {Franchi}}, \bibinfo {author} {\bibfnamefont
				{C.}~\bibnamefont {Clivati}}, \bibinfo {author} {\bibfnamefont
				{M.}~\bibnamefont {Frittelli}}, \bibinfo {author} {\bibfnamefont
				{F.}~\bibnamefont {Levi}}, \bibinfo {author} {\bibfnamefont {D.}~\bibnamefont
				{Calonico}}, \bibinfo {author} {\bibfnamefont {J.}~\bibnamefont {Catani}},
			\bibinfo {author} {\bibfnamefont {M.}~\bibnamefont {Inguscio}}, \ and\
			\bibinfo {author} {\bibfnamefont {L.}~\bibnamefont {Fallani}},\ }\href@noop
		{} {\bibfield  {journal} {\bibinfo  {journal} {Phys. Rev. Lett.}\ }\textbf
			{\bibinfo {volume} {117}},\ \bibinfo {pages} {220401} (\bibinfo {year}
			{2016})}\BibitemShut {NoStop}%
		\bibitem [{\citenamefont {Yuan}\ \emph {et~al.}(2016)\citenamefont {Yuan},
			\citenamefont {Shi},\ and\ \citenamefont {Fan}}]{yuan2016photonic}%
		\BibitemOpen
		\bibfield  {author} {\bibinfo {author} {\bibfnamefont {L.}~\bibnamefont
				{Yuan}}, \bibinfo {author} {\bibfnamefont {Y.}~\bibnamefont {Shi}}, \ and\
			\bibinfo {author} {\bibfnamefont {S.}~\bibnamefont {Fan}},\ }\href@noop {}
		{\bibfield  {journal} {\bibinfo  {journal} {Optics letters}\ }\textbf
			{\bibinfo {volume} {41}},\ \bibinfo {pages} {741} (\bibinfo {year}
			{2016})}\BibitemShut {NoStop}%
		\bibitem [{\citenamefont {Ozawa}\ and\ \citenamefont
			{Carusotto}(2017)}]{ozawa2017synthetic}%
		\BibitemOpen
		\bibfield  {author} {\bibinfo {author} {\bibfnamefont {T.}~\bibnamefont
				{Ozawa}}\ and\ \bibinfo {author} {\bibfnamefont {I.}~\bibnamefont
				{Carusotto}},\ }\href@noop {} {\bibfield  {journal} {\bibinfo  {journal}
				{Physical review letters}\ }\textbf {\bibinfo {volume} {118}},\ \bibinfo
			{pages} {013601} (\bibinfo {year} {2017})}\BibitemShut {NoStop}%
		\bibitem [{\citenamefont {Price}\ \emph {et~al.}(2017)\citenamefont {Price},
			\citenamefont {Ozawa},\ and\ \citenamefont {Goldman}}]{Price:2017PRA}%
		\BibitemOpen
		\bibfield  {author} {\bibinfo {author} {\bibfnamefont {H.~M.}\ \bibnamefont
				{Price}}, \bibinfo {author} {\bibfnamefont {T.}~\bibnamefont {Ozawa}}, \ and\
			\bibinfo {author} {\bibfnamefont {N.}~\bibnamefont {Goldman}},\ }\href@noop
		{} {\bibfield  {journal} {\bibinfo  {journal} {Phys. Rev. A}\ }\textbf
			{\bibinfo {volume} {95}},\ \bibinfo {pages} {023607} (\bibinfo {year}
			{2017})}\BibitemShut {NoStop}%
		\bibitem [{\citenamefont {An}\ \emph {et~al.}(2017)\citenamefont {An},
			\citenamefont {Meier},\ and\ \citenamefont {Gadway}}]{An:2017SciAdv}%
		\BibitemOpen
		\bibfield  {author} {\bibinfo {author} {\bibfnamefont {F.~A.}\ \bibnamefont
				{An}}, \bibinfo {author} {\bibfnamefont {E.~J.}\ \bibnamefont {Meier}}, \
			and\ \bibinfo {author} {\bibfnamefont {B.}~\bibnamefont {Gadway}},\
		}\href@noop {} {\bibfield  {journal} {\bibinfo  {journal} {Science Advances}\
			}\textbf {\bibinfo {volume} {3}} (\bibinfo {year} {2017})}\BibitemShut
		{NoStop}%
		\bibitem [{\citenamefont {Thouless}(1983)}]{Thouless83}%
		\BibitemOpen
		\bibfield  {author} {\bibinfo {author} {\bibfnamefont {D.~J.}\ \bibnamefont
				{Thouless}},\ }\href@noop {} {\bibfield  {journal} {\bibinfo  {journal}
				{Phys. Rev. B}\ }\textbf {\bibinfo {volume} {27}},\ \bibinfo {pages} {6083}
			(\bibinfo {year} {1983})}\BibitemShut {NoStop}%
		\bibitem [{\citenamefont {Kunz}(1986)}]{Kunz1986}%
		\BibitemOpen
		\bibfield  {author} {\bibinfo {author} {\bibfnamefont {H.}~\bibnamefont
				{Kunz}},\ }\href@noop {} {\bibfield  {journal} {\bibinfo  {journal} {Phys.
					Rev. Lett.}\ }\textbf {\bibinfo {volume} {57}},\ \bibinfo {pages} {1095}
			(\bibinfo {year} {1986})}\BibitemShut {NoStop}%
		\bibitem [{\citenamefont {Kraus}\ \emph {et~al.}(2012)\citenamefont {Kraus},
			\citenamefont {Lahini}, \citenamefont {Ringel}, \citenamefont {Verbin},\ and\
			\citenamefont {Zilberberg}}]{Kraus:2012a}%
		\BibitemOpen
		\bibfield  {author} {\bibinfo {author} {\bibfnamefont {Y.~E.}\ \bibnamefont
				{Kraus}}, \bibinfo {author} {\bibfnamefont {Y.}~\bibnamefont {Lahini}},
			\bibinfo {author} {\bibfnamefont {Z.}~\bibnamefont {Ringel}}, \bibinfo
			{author} {\bibfnamefont {M.}~\bibnamefont {Verbin}}, \ and\ \bibinfo {author}
			{\bibfnamefont {O.}~\bibnamefont {Zilberberg}},\ }\href@noop {} {\bibfield
			{journal} {\bibinfo  {journal} {Phys. Rev. Lett.}\ }\textbf {\bibinfo
				{volume} {109}},\ \bibinfo {pages} {106402} (\bibinfo {year}
			{2012})}\BibitemShut {NoStop}%
		\bibitem [{\citenamefont {Kraus}\ and\ \citenamefont
			{Zilberberg}(2012)}]{Kraus:2012b}%
		\BibitemOpen
		\bibfield  {author} {\bibinfo {author} {\bibfnamefont {Y.~E.}\ \bibnamefont
				{Kraus}}\ and\ \bibinfo {author} {\bibfnamefont {O.}~\bibnamefont
				{Zilberberg}},\ }\href@noop {} {\bibfield  {journal} {\bibinfo  {journal}
				{Phys. Rev. Lett.}\ }\textbf {\bibinfo {volume} {109}},\ \bibinfo {pages}
			{116404} (\bibinfo {year} {2012})}\BibitemShut {NoStop}%
		\bibitem [{\citenamefont {Verbin}\ \emph {et~al.}(2015)\citenamefont {Verbin},
			\citenamefont {Zilberberg}, \citenamefont {Lahini}, \citenamefont {Kraus},\
			and\ \citenamefont {Silberberg}}]{Verbin:2015}%
		\BibitemOpen
		\bibfield  {author} {\bibinfo {author} {\bibfnamefont {M.}~\bibnamefont
				{Verbin}}, \bibinfo {author} {\bibfnamefont {O.}~\bibnamefont {Zilberberg}},
			\bibinfo {author} {\bibfnamefont {Y.}~\bibnamefont {Lahini}}, \bibinfo
			{author} {\bibfnamefont {Y.~E.}\ \bibnamefont {Kraus}}, \ and\ \bibinfo
			{author} {\bibfnamefont {Y.}~\bibnamefont {Silberberg}},\ }\href@noop {}
		{\bibfield  {journal} {\bibinfo  {journal} {Phys. Rev. B}\ }\textbf {\bibinfo
				{volume} {91}},\ \bibinfo {pages} {064201} (\bibinfo {year}
			{2015})}\BibitemShut {NoStop}%
		\bibitem [{\citenamefont {Lohse}\ \emph {et~al.}(2016)\citenamefont {Lohse},
			\citenamefont {Schweizer}, \citenamefont {Zilberberg}, \citenamefont
			{Aidelsburger},\ and\ \citenamefont {Bloch}}]{Lohse2016}%
		\BibitemOpen
		\bibfield  {author} {\bibinfo {author} {\bibfnamefont {M.}~\bibnamefont
				{Lohse}}, \bibinfo {author} {\bibfnamefont {C.}~\bibnamefont {Schweizer}},
			\bibinfo {author} {\bibfnamefont {O.}~\bibnamefont {Zilberberg}}, \bibinfo
			{author} {\bibfnamefont {M.}~\bibnamefont {Aidelsburger}}, \ and\ \bibinfo
			{author} {\bibfnamefont {I.}~\bibnamefont {Bloch}},\ }\href@noop {}
		{\bibfield  {journal} {\bibinfo  {journal} {Nature Physics}\ }\textbf
			{\bibinfo {volume} {12}},\ \bibinfo {pages} {350} (\bibinfo {year}
			{2016})}\BibitemShut {NoStop}%
		\bibitem [{\citenamefont {Nakajima}\ \emph {et~al.}(2016)\citenamefont
			{Nakajima}, \citenamefont {Tomita}, \citenamefont {Taie}, \citenamefont
			{Ichinose}, \citenamefont {Ozawa}, \citenamefont {Wang}, \citenamefont
			{Troyer},\ and\ \citenamefont {Takahashi}}]{Nakajima2016}%
		\BibitemOpen
		\bibfield  {author} {\bibinfo {author} {\bibfnamefont {S.}~\bibnamefont
				{Nakajima}}, \bibinfo {author} {\bibfnamefont {T.}~\bibnamefont {Tomita}},
			\bibinfo {author} {\bibfnamefont {S.}~\bibnamefont {Taie}}, \bibinfo {author}
			{\bibfnamefont {T.}~\bibnamefont {Ichinose}}, \bibinfo {author}
			{\bibfnamefont {H.}~\bibnamefont {Ozawa}}, \bibinfo {author} {\bibfnamefont
				{L.}~\bibnamefont {Wang}}, \bibinfo {author} {\bibfnamefont {M.}~\bibnamefont
				{Troyer}}, \ and\ \bibinfo {author} {\bibfnamefont {Y.}~\bibnamefont
				{Takahashi}},\ }\href@noop {} {\bibfield  {journal} {\bibinfo  {journal}
				{Nature Physics}\ }\textbf {\bibinfo {volume} {12}},\ \bibinfo {pages} {296}
			(\bibinfo {year} {2016})}\BibitemShut {NoStop}%
		\bibitem [{\citenamefont {Lohse}\ \emph {et~al.}(2018)\citenamefont {Lohse},
			\citenamefont {Schweizer}, \citenamefont {Price}, \citenamefont
			{Zilberberg},\ and\ \citenamefont {Bloch}}]{Lohse2018}%
		\BibitemOpen
		\bibfield  {author} {\bibinfo {author} {\bibfnamefont {M.}~\bibnamefont
				{Lohse}}, \bibinfo {author} {\bibfnamefont {C.}~\bibnamefont {Schweizer}},
			\bibinfo {author} {\bibfnamefont {H.~M.}\ \bibnamefont {Price}}, \bibinfo
			{author} {\bibfnamefont {O.}~\bibnamefont {Zilberberg}}, \ and\ \bibinfo
			{author} {\bibfnamefont {I.}~\bibnamefont {Bloch}},\ }\href@noop {}
		{\bibfield  {journal} {\bibinfo  {journal} {Nature}\ }\textbf {\bibinfo
				{volume} {553}},\ \bibinfo {pages} {55} (\bibinfo {year} {2018})}\BibitemShut
		{NoStop}%
		\bibitem [{\citenamefont {Zilberberg}\ \emph {et~al.}(2018)\citenamefont
			{Zilberberg}, \citenamefont {Huang}, \citenamefont {Guglielmon},
			\citenamefont {Wang}, \citenamefont {Chen}, \citenamefont {Kraus},\ and\
			\citenamefont {Rechtsman}}]{Zilberberg2018}%
		\BibitemOpen
		\bibfield  {author} {\bibinfo {author} {\bibfnamefont {O.}~\bibnamefont
				{Zilberberg}}, \bibinfo {author} {\bibfnamefont {S.}~\bibnamefont {Huang}},
			\bibinfo {author} {\bibfnamefont {J.}~\bibnamefont {Guglielmon}}, \bibinfo
			{author} {\bibfnamefont {M.}~\bibnamefont {Wang}}, \bibinfo {author}
			{\bibfnamefont {K.~P.}\ \bibnamefont {Chen}}, \bibinfo {author}
			{\bibfnamefont {Y.~E.}\ \bibnamefont {Kraus}}, \ and\ \bibinfo {author}
			{\bibfnamefont {M.~C.}\ \bibnamefont {Rechtsman}},\ }\href@noop {} {\bibfield
			{journal} {\bibinfo  {journal} {Nature}\ }\textbf {\bibinfo {volume}
				{553}},\ \bibinfo {pages} {59} (\bibinfo {year} {2018})}\BibitemShut
		{NoStop}%
		\bibitem [{\citenamefont {Nakahara}(2003)}]{Nakahara}%
		\BibitemOpen
		\bibfield  {author} {\bibinfo {author} {\bibfnamefont {M.}~\bibnamefont
				{Nakahara}},\ }\href@noop {} {\emph {\bibinfo {title} {Geometry, Topology and
					Physics}}}\ (\bibinfo  {publisher} {IOP Publishing Ltd.},\ \bibinfo {address}
		{Bristol and Philadelphia},\ \bibinfo {year} {2003})\BibitemShut {NoStop}%
		\bibitem [{\citenamefont {Harper}(1955)}]{Harper:1955PPSA}%
		\BibitemOpen
		\bibfield  {author} {\bibinfo {author} {\bibfnamefont {P.~G.}\ \bibnamefont
				{Harper}},\ }\href@noop {} {\bibfield  {journal} {\bibinfo  {journal} {Proc.
					Phys. Soc. London A}\ }\textbf {\bibinfo {volume} {68}},\ \bibinfo {pages}
			{874} (\bibinfo {year} {1955})}\BibitemShut {NoStop}%
		\bibitem [{\citenamefont {Azbel}(1964)}]{Azbel:1964JETP}%
		\BibitemOpen
		\bibfield  {author} {\bibinfo {author} {\bibfnamefont {M.~Y.}\ \bibnamefont
				{Azbel}},\ }\href@noop {} {\bibfield  {journal} {\bibinfo  {journal} {JETP
					Lett.}\ }\textbf {\bibinfo {volume} {19}},\ \bibinfo {pages} {634} (\bibinfo
			{year} {1964})}\BibitemShut {NoStop}%
		\bibitem [{\citenamefont {Hofstadter}(1976)}]{Hofstadter76}%
		\BibitemOpen
		\bibfield  {author} {\bibinfo {author} {\bibfnamefont {D.~R.}\ \bibnamefont
				{Hofstadter}},\ }\href@noop {} {\bibfield  {journal} {\bibinfo  {journal}
				{Phys. Rev. B}\ }\textbf {\bibinfo {volume} {14}},\ \bibinfo {pages} {2239}
			(\bibinfo {year} {1976})}\BibitemShut {NoStop}%
		\bibitem [{\citenamefont {Niu}\ and\ \citenamefont
			{Thouless}(1984)}]{Niu:1984}%
		\BibitemOpen
		\bibfield  {author} {\bibinfo {author} {\bibfnamefont {Q.}~\bibnamefont
				{Niu}}\ and\ \bibinfo {author} {\bibfnamefont {D.}~\bibnamefont {Thouless}},\
		}\href@noop {} {\bibfield  {journal} {\bibinfo  {journal} {Journal of Physics
					A: Mathematical and General}\ }\textbf {\bibinfo {volume} {17}},\ \bibinfo
			{pages} {2453} (\bibinfo {year} {1984})}\BibitemShut {NoStop}%
		\bibitem [{\citenamefont {Xiao}\ \emph {et~al.}(2010)\citenamefont {Xiao},
			\citenamefont {Chang},\ and\ \citenamefont {Niu}}]{Xiao2010}%
		\BibitemOpen
		\bibfield  {author} {\bibinfo {author} {\bibfnamefont {D.}~\bibnamefont
				{Xiao}}, \bibinfo {author} {\bibfnamefont {M.~C.}\ \bibnamefont {Chang}}, \
			and\ \bibinfo {author} {\bibfnamefont {Q.}~\bibnamefont {Niu}},\ }\href@noop
		{} {\bibfield  {journal} {\bibinfo  {journal} {Reviews of Modern Physics}\
			}\textbf {\bibinfo {volume} {82}},\ \bibinfo {pages} {1959} (\bibinfo {year}
			{2010})}\BibitemShut {NoStop}%
		\bibitem [{\citenamefont {Berry}(1985)}]{Berry1985}%
		\BibitemOpen
		\bibfield  {author} {\bibinfo {author} {\bibfnamefont {M.~V.}\ \bibnamefont
				{Berry}},\ }\href@noop {} {\bibfield  {journal} {\bibinfo  {journal} {Journal
					of Physics A: Mathematical and General}\ }\textbf {\bibinfo {volume} {18}},\
			\bibinfo {pages} {15} (\bibinfo {year} {1985})}\BibitemShut {NoStop}%
		\bibitem [{\citenamefont {Nagaosa}\ \emph {et~al.}(2010)\citenamefont
			{Nagaosa}, \citenamefont {Sinova}, \citenamefont {Onoda}, \citenamefont
			{MacDonald},\ and\ \citenamefont {Ong}}]{Nagaosa}%
		\BibitemOpen
		\bibfield  {author} {\bibinfo {author} {\bibfnamefont {N.}~\bibnamefont
				{Nagaosa}}, \bibinfo {author} {\bibfnamefont {J.}~\bibnamefont {Sinova}},
			\bibinfo {author} {\bibfnamefont {S.}~\bibnamefont {Onoda}}, \bibinfo
			{author} {\bibfnamefont {A.~H.}\ \bibnamefont {MacDonald}}, \ and\ \bibinfo
			{author} {\bibfnamefont {N.~P.}\ \bibnamefont {Ong}},\ }\href@noop {}
		{\bibfield  {journal} {\bibinfo  {journal} {Rev. Mod. Phys.}\ }\textbf
			{\bibinfo {volume} {82}},\ \bibinfo {pages} {1539} (\bibinfo {year}
			{2010})}\BibitemShut {NoStop}%
		\bibitem [{\citenamefont {Bliokh}\ and\ \citenamefont
			{Bliokh}(2005)}]{bliokh2005spin}%
		\BibitemOpen
		\bibfield  {author} {\bibinfo {author} {\bibfnamefont {K.~Y.}\ \bibnamefont
				{Bliokh}}\ and\ \bibinfo {author} {\bibfnamefont {Y.~P.}\ \bibnamefont
				{Bliokh}},\ }\href@noop {} {\bibfield  {journal} {\bibinfo  {journal} {Annals
					of Physics}\ }\textbf {\bibinfo {volume} {319}},\ \bibinfo {pages} {13}
			(\bibinfo {year} {2005})}\BibitemShut {NoStop}%
		\bibitem [{\citenamefont {Price}\ \emph {et~al.}(2014)\citenamefont {Price},
			\citenamefont {Ozawa},\ and\ \citenamefont {Carusotto}}]{Price:2014PRL}%
		\BibitemOpen
		\bibfield  {author} {\bibinfo {author} {\bibfnamefont {H.~M.}\ \bibnamefont
				{Price}}, \bibinfo {author} {\bibfnamefont {T.}~\bibnamefont {Ozawa}}, \ and\
			\bibinfo {author} {\bibfnamefont {I.}~\bibnamefont {Carusotto}},\ }\href@noop
		{} {\bibfield  {journal} {\bibinfo  {journal} {Phys. Rev. Lett.}\ }\textbf
			{\bibinfo {volume} {113}},\ \bibinfo {pages} {190403} (\bibinfo {year}
			{2014})}\BibitemShut {NoStop}%
		\bibitem [{\citenamefont {Price}\ and\ \citenamefont
			{Cooper}(2012)}]{PricePRA2012}%
		\BibitemOpen
		\bibfield  {author} {\bibinfo {author} {\bibfnamefont {H.~M.}\ \bibnamefont
				{Price}}\ and\ \bibinfo {author} {\bibfnamefont {N.~R.}\ \bibnamefont
				{Cooper}},\ }\href@noop {} {\bibfield  {journal} {\bibinfo  {journal} {Phys.
					Rev. A}\ }\textbf {\bibinfo {volume} {85}},\ \bibinfo {pages} {033620}
			(\bibinfo {year} {2012})}\BibitemShut {NoStop}%
		\bibitem [{\citenamefont {Wimmer}\ \emph {et~al.}(2017)\citenamefont {Wimmer},
			\citenamefont {Price}, \citenamefont {Carusotto},\ and\ \citenamefont
			{Peschel}}]{Wimmer:2017NatPhys}%
		\BibitemOpen
		\bibfield  {author} {\bibinfo {author} {\bibfnamefont {M.}~\bibnamefont
				{Wimmer}}, \bibinfo {author} {\bibfnamefont {H.~M.}\ \bibnamefont {Price}},
			\bibinfo {author} {\bibfnamefont {I.}~\bibnamefont {Carusotto}}, \ and\
			\bibinfo {author} {\bibfnamefont {U.}~\bibnamefont {Peschel}},\ }\href@noop
		{} {\bibfield  {journal} {\bibinfo  {journal} {Nature Physics}\ }\textbf
			{\bibinfo {volume} {13}},\ \bibinfo {pages} {545} (\bibinfo {year}
			{2017})}\BibitemShut {NoStop}%
		\bibitem [{\citenamefont {Fl{\"a}schner}\ \emph {et~al.}(2016)\citenamefont
			{Fl{\"a}schner}, \citenamefont {Rem}, \citenamefont {Tarnowski},
			\citenamefont {Vogel}, \citenamefont {L{\"u}hmann}, \citenamefont
			{Sengstock},\ and\ \citenamefont {Weitenberg}}]{flaschner2016experimental}%
		\BibitemOpen
		\bibfield  {author} {\bibinfo {author} {\bibfnamefont {N.}~\bibnamefont
				{Fl{\"a}schner}}, \bibinfo {author} {\bibfnamefont {B.}~\bibnamefont {Rem}},
			\bibinfo {author} {\bibfnamefont {M.}~\bibnamefont {Tarnowski}}, \bibinfo
			{author} {\bibfnamefont {D.}~\bibnamefont {Vogel}}, \bibinfo {author}
			{\bibfnamefont {D.-S.}\ \bibnamefont {L{\"u}hmann}}, \bibinfo {author}
			{\bibfnamefont {K.}~\bibnamefont {Sengstock}}, \ and\ \bibinfo {author}
			{\bibfnamefont {C.}~\bibnamefont {Weitenberg}},\ }\href@noop {} {\bibfield
			{journal} {\bibinfo  {journal} {Science}\ }\textbf {\bibinfo {volume}
				{352}},\ \bibinfo {pages} {1091} (\bibinfo {year} {2016})}\BibitemShut
		{NoStop}%
		\bibitem [{\citenamefont {Li}\ \emph {et~al.}(2016)\citenamefont {Li},
			\citenamefont {Duca}, \citenamefont {Reitter}, \citenamefont {Grusdt},
			\citenamefont {Demler}, \citenamefont {Endres}, \citenamefont
			{Schleier-Smith}, \citenamefont {Bloch},\ and\ \citenamefont
			{Schneider}}]{li2016bloch}%
		\BibitemOpen
		\bibfield  {author} {\bibinfo {author} {\bibfnamefont {T.}~\bibnamefont
				{Li}}, \bibinfo {author} {\bibfnamefont {L.}~\bibnamefont {Duca}}, \bibinfo
			{author} {\bibfnamefont {M.}~\bibnamefont {Reitter}}, \bibinfo {author}
			{\bibfnamefont {F.}~\bibnamefont {Grusdt}}, \bibinfo {author} {\bibfnamefont
				{E.}~\bibnamefont {Demler}}, \bibinfo {author} {\bibfnamefont
				{M.}~\bibnamefont {Endres}}, \bibinfo {author} {\bibfnamefont
				{M.}~\bibnamefont {Schleier-Smith}}, \bibinfo {author} {\bibfnamefont
				{I.}~\bibnamefont {Bloch}}, \ and\ \bibinfo {author} {\bibfnamefont
				{U.}~\bibnamefont {Schneider}},\ }\href@noop {} {\bibfield  {journal}
			{\bibinfo  {journal} {Science}\ }\textbf {\bibinfo {volume} {352}},\ \bibinfo
			{pages} {1094} (\bibinfo {year} {2016})}\BibitemShut {NoStop}%
		\bibitem [{\citenamefont {Price}\ and\ \citenamefont
			{Cooper}(2013)}]{PhysRevLett.111.220407}%
		\BibitemOpen
		\bibfield  {author} {\bibinfo {author} {\bibfnamefont {H.~M.}\ \bibnamefont
				{Price}}\ and\ \bibinfo {author} {\bibfnamefont {N.~R.}\ \bibnamefont
				{Cooper}},\ }\href@noop {} {\bibfield  {journal} {\bibinfo  {journal} {Phys.
					Rev. Lett.}\ }\textbf {\bibinfo {volume} {111}},\ \bibinfo {pages} {220407}
			(\bibinfo {year} {2013})}\BibitemShut {NoStop}%
		\bibitem [{\citenamefont {Ozawa}\ and\ \citenamefont
			{Carusotto}(2014)}]{ozawa2014anomalous}%
		\BibitemOpen
		\bibfield  {author} {\bibinfo {author} {\bibfnamefont {T.}~\bibnamefont
				{Ozawa}}\ and\ \bibinfo {author} {\bibfnamefont {I.}~\bibnamefont
				{Carusotto}},\ }\href@noop {} {\bibfield  {journal} {\bibinfo  {journal}
				{Physical review letters}\ }\textbf {\bibinfo {volume} {112}},\ \bibinfo
			{pages} {133902} (\bibinfo {year} {2014})}\BibitemShut {NoStop}%
		\bibitem [{\citenamefont {Aidelsburger}\ \emph {et~al.}(2015)\citenamefont
			{Aidelsburger}, \citenamefont {Lohse}, \citenamefont {Schweizer},
			\citenamefont {Atala}, \citenamefont {Barreiro}, \citenamefont {Nascimbene},
			\citenamefont {Cooper}, \citenamefont {Bloch},\ and\ \citenamefont
			{Goldman}}]{Aidelsburger:2015NatPhys}%
		\BibitemOpen
		\bibfield  {author} {\bibinfo {author} {\bibfnamefont {M.}~\bibnamefont
				{Aidelsburger}}, \bibinfo {author} {\bibfnamefont {M.}~\bibnamefont {Lohse}},
			\bibinfo {author} {\bibfnamefont {C.}~\bibnamefont {Schweizer}}, \bibinfo
			{author} {\bibfnamefont {M.}~\bibnamefont {Atala}}, \bibinfo {author}
			{\bibfnamefont {J.~T.}\ \bibnamefont {Barreiro}}, \bibinfo {author}
			{\bibfnamefont {S.}~\bibnamefont {Nascimbene}}, \bibinfo {author}
			{\bibfnamefont {N.}~\bibnamefont {Cooper}}, \bibinfo {author} {\bibfnamefont
				{I.}~\bibnamefont {Bloch}}, \ and\ \bibinfo {author} {\bibfnamefont
				{N.}~\bibnamefont {Goldman}},\ }\href@noop {} {\bibfield  {journal} {\bibinfo
				{journal} {Nature Physics}\ }\textbf {\bibinfo {volume} {11}},\ \bibinfo
			{pages} {162} (\bibinfo {year} {2015})}\BibitemShut {NoStop}%
		\bibitem [{\citenamefont {Tarnowski}\ \emph {et~al.}(2017)\citenamefont
			{Tarnowski}, \citenamefont {{\"U}nal}, \citenamefont {Fl{\"a}schner},
			\citenamefont {Rem}, \citenamefont {Eckardt}, \citenamefont {Sengstock},\
			and\ \citenamefont {Weitenberg}}]{tarnowski2017characterizing}%
		\BibitemOpen
		\bibfield  {author} {\bibinfo {author} {\bibfnamefont {M.}~\bibnamefont
				{Tarnowski}}, \bibinfo {author} {\bibfnamefont {F.~N.}\ \bibnamefont
				{{\"U}nal}}, \bibinfo {author} {\bibfnamefont {N.}~\bibnamefont
				{Fl{\"a}schner}}, \bibinfo {author} {\bibfnamefont {B.~S.}\ \bibnamefont
				{Rem}}, \bibinfo {author} {\bibfnamefont {A.}~\bibnamefont {Eckardt}},
			\bibinfo {author} {\bibfnamefont {K.}~\bibnamefont {Sengstock}}, \ and\
			\bibinfo {author} {\bibfnamefont {C.}~\bibnamefont {Weitenberg}},\
		}\href@noop {} {\bibfield  {journal} {\bibinfo  {journal} {arXiv preprint
					arXiv:1709.01046}\ } (\bibinfo {year} {2017})}\BibitemShut {NoStop}%
		\bibitem [{\citenamefont {{Tran}}\ \emph {et~al.}(2017)\citenamefont {{Tran}},
			\citenamefont {{Dauphin}}, \citenamefont {{Grushin}}, \citenamefont
			{{Zoller}},\ and\ \citenamefont {{Goldman}}}]{Tran2017}%
		\BibitemOpen
		\bibfield  {author} {\bibinfo {author} {\bibfnamefont {D.~T.}\ \bibnamefont
				{{Tran}}}, \bibinfo {author} {\bibfnamefont {A.}~\bibnamefont {{Dauphin}}},
			\bibinfo {author} {\bibfnamefont {A.~G.}\ \bibnamefont {{Grushin}}}, \bibinfo
			{author} {\bibfnamefont {P.}~\bibnamefont {{Zoller}}}, \ and\ \bibinfo
			{author} {\bibfnamefont {N.}~\bibnamefont {{Goldman}}},\ }\href@noop {}
		{\bibfield  {journal} {\bibinfo  {journal} {Science Advances}\ }\textbf
			{\bibinfo {volume} {3}},\ \bibinfo {pages} {e1701207} (\bibinfo {year}
			{2017})} \BibitemShut {NoStop}%
		\bibitem [{\citenamefont {Tran}\ \emph {et~al.}(2018)\citenamefont {Tran},
			\citenamefont {Cooper},\ and\ \citenamefont {Goldman}}]{tran2018quantized}%
		\BibitemOpen
		\bibfield  {author} {\bibinfo {author} {\bibfnamefont {D.~T.}\ \bibnamefont
				{Tran}}, \bibinfo {author} {\bibfnamefont {N.~R.}\ \bibnamefont {Cooper}}, \
			and\ \bibinfo {author} {\bibfnamefont {N.}~\bibnamefont {Goldman}},\
		}\href@noop {} {\bibfield  {journal} {\bibinfo
				{journal} {Phys. Rev. A}\ }\textbf {\bibinfo {volume} {97}},\ \bibinfo
			{pages} {061602} (\bibinfo {year} {2018})}\BibitemShut {NoStop}%
		\bibitem [{\citenamefont {Asteria}\ \emph {et~al.}(2018)\citenamefont
			{Asteria}, \citenamefont {Tran}, \citenamefont {Ozawa}, \citenamefont
			{Tarnowski}, \citenamefont {Rem}, \citenamefont {Fl{\"a}schner},
			\citenamefont {Sengstock}, \citenamefont {Goldman},\ and\ \citenamefont
			{Weitenberg}}]{asteria2018measuring}%
		\BibitemOpen
		\bibfield  {author} {\bibinfo {author} {\bibfnamefont {L.}~\bibnamefont
				{Asteria}}, \bibinfo {author} {\bibfnamefont {D.~T.}\ \bibnamefont {Tran}},
			\bibinfo {author} {\bibfnamefont {T.}~\bibnamefont {Ozawa}}, \bibinfo
			{author} {\bibfnamefont {M.}~\bibnamefont {Tarnowski}}, \bibinfo {author}
			{\bibfnamefont {B.~S.}\ \bibnamefont {Rem}}, \bibinfo {author} {\bibfnamefont
				{N.}~\bibnamefont {Fl{\"a}schner}}, \bibinfo {author} {\bibfnamefont
				{K.}~\bibnamefont {Sengstock}}, \bibinfo {author} {\bibfnamefont
				{N.}~\bibnamefont {Goldman}}, \ and\ \bibinfo {author} {\bibfnamefont
				{C.}~\bibnamefont {Weitenberg}},\ }\href@noop {} {\bibfield  {journal}
			{\bibinfo  {journal} {arXiv preprint arXiv:1805.11077}\ } (\bibinfo {year}
			{2018})}\BibitemShut {NoStop}%
		\bibitem [{\citenamefont {Salerno}\ \emph {et~al.}(2016)\citenamefont
			{Salerno}, \citenamefont {Ozawa}, \citenamefont {Price},\ and\ \citenamefont
			{Carusotto}}]{Salerno:2016PRB}%
		\BibitemOpen
		\bibfield  {author} {\bibinfo {author} {\bibfnamefont {G.}~\bibnamefont
				{Salerno}}, \bibinfo {author} {\bibfnamefont {T.}~\bibnamefont {Ozawa}},
			\bibinfo {author} {\bibfnamefont {H.~M.}\ \bibnamefont {Price}}, \ and\
			\bibinfo {author} {\bibfnamefont {I.}~\bibnamefont {Carusotto}},\ }\href@noop
		{} {\bibfield  {journal} {\bibinfo  {journal} {Phys. Rev. B}\ }\textbf
			{\bibinfo {volume} {93}},\ \bibinfo {pages} {085105} (\bibinfo {year}
			{2016})}\BibitemShut {NoStop}%
		\bibitem [{\citenamefont {Sugawa}\ \emph {et~al.}(2016)\citenamefont {Sugawa},
			\citenamefont {Salces-Carcoba}, \citenamefont {Perry}, \citenamefont {Yue},\
			and\ \citenamefont {Spielman}}]{Sugawa:2016arXiv}%
		\BibitemOpen
		\bibfield  {author} {\bibinfo {author} {\bibfnamefont {S.}~\bibnamefont
				{Sugawa}}, \bibinfo {author} {\bibfnamefont {F.}~\bibnamefont
				{Salces-Carcoba}}, \bibinfo {author} {\bibfnamefont {A.~R.}\ \bibnamefont
				{Perry}}, \bibinfo {author} {\bibfnamefont {Y.}~\bibnamefont {Yue}}, \ and\
			\bibinfo {author} {\bibfnamefont {I.~B.}\ \bibnamefont {Spielman}},\
		}\href@noop {} {\bibfield  {journal} {\bibinfo  {journal} {arXiv preprint
					arXiv:1610.06228}\ } (\bibinfo {year} {2016})}\BibitemShut {NoStop}%
		\bibitem [{\citenamefont {Bardyn}\ \emph {et~al.}(2014)\citenamefont {Bardyn},
			\citenamefont {Huber},\ and\ \citenamefont {Zilberberg}}]{Bardyn:2014}%
		\BibitemOpen
		\bibfield  {author} {\bibinfo {author} {\bibfnamefont {C.-E.}\ \bibnamefont
				{Bardyn}}, \bibinfo {author} {\bibfnamefont {S.~D.}\ \bibnamefont {Huber}}, \
			and\ \bibinfo {author} {\bibfnamefont {O.}~\bibnamefont {Zilberberg}},\
		}\href@noop {} {\bibfield  {journal} {\bibinfo  {journal} {New Journal of
					Physics}\ }\textbf {\bibinfo {volume} {16}},\ \bibinfo {pages} {123013}
			(\bibinfo {year} {2014})}\BibitemShut {NoStop}%
		\bibitem [{\citenamefont {{Mochol-Grzelak}}\ \emph {et~al.}(2018)\citenamefont
			{{Mochol-Grzelak}}, \citenamefont {{Dauphin}}, \citenamefont {{Celi}},\ and\
			\citenamefont {{Lewenstein}}}]{Mochol2018}%
		\BibitemOpen
		\bibfield  {author} {\bibinfo {author} {\bibfnamefont {M.}~\bibnamefont
				{{Mochol-Grzelak}}}, \bibinfo {author} {\bibfnamefont {A.}~\bibnamefont
				{{Dauphin}}}, \bibinfo {author} {\bibfnamefont {A.}~\bibnamefont {{Celi}}}, \
			and\ \bibinfo {author} {\bibfnamefont {M.}~\bibnamefont {{Lewenstein}}},\
		}\href@noop {} {\bibfield  {journal} {\bibinfo  {journal} {ArXiv e-prints}\ }
			(\bibinfo {year} {2018})},\ \Eprint {http://arxiv.org/abs/1803.07003}
		{arXiv:1803.07003} \BibitemShut {NoStop}%
		\bibitem [{\citenamefont {{M.C. Chang and Qian Niu}}(1995)}]{Niu1995}%
		\BibitemOpen
		\bibfield  {author} {\bibinfo {author} {\bibnamefont {{M.C. Chang and Qian
						Niu}}},\ }\href@noop {} {\bibfield  {journal} {\bibinfo  {journal} {Physical
					Review Letters}\ }\textbf {\bibinfo {volume} {75}},\ \bibinfo {pages} {1348}
			(\bibinfo {year} {1995})}\BibitemShut {NoStop}%
		\bibitem [{\citenamefont {Sundaram}\ and\ \citenamefont
			{Niu}(1999)}]{Sundaram1999}%
		\BibitemOpen
		\bibfield  {author} {\bibinfo {author} {\bibfnamefont {G.}~\bibnamefont
				{Sundaram}}\ and\ \bibinfo {author} {\bibfnamefont {Q.}~\bibnamefont {Niu}},\
		}\href@noop {} {\bibfield  {journal} {\bibinfo  {journal} {Physical Review B
					- Condensed Matter and Materials Physics}\ }\textbf {\bibinfo {volume}
				{59}},\ \bibinfo {pages} {14915} (\bibinfo {year} {1999})}\BibitemShut
		{NoStop}%
		\bibitem [{\citenamefont {Xiao}\ \emph {et~al.}(2009)\citenamefont {Xiao},
			\citenamefont {Shi}, \citenamefont {Clougherty},\ and\ \citenamefont
			{Niu}}]{Xiao2009}%
		\BibitemOpen
		\bibfield  {author} {\bibinfo {author} {\bibfnamefont {D.}~\bibnamefont
				{Xiao}}, \bibinfo {author} {\bibfnamefont {J.}~\bibnamefont {Shi}}, \bibinfo
			{author} {\bibfnamefont {D.~P.}\ \bibnamefont {Clougherty}}, \ and\ \bibinfo
			{author} {\bibfnamefont {Q.}~\bibnamefont {Niu}},\ }\href@noop {} {\bibfield
			{journal} {\bibinfo  {journal} {Physical Review Letters}\ }\textbf {\bibinfo
				{volume} {102}},\ \bibinfo {pages} {087602} (\bibinfo {year} {2009})}\BibitemShut
		{NoStop}%
		\bibitem [{\citenamefont {Gao}\ \emph {et~al.}(2014)\citenamefont {Gao},
			\citenamefont {Yang},\ and\ \citenamefont {Niu}}]{Gao2014}%
		\BibitemOpen
		\bibfield  {author} {\bibinfo {author} {\bibfnamefont {Y.}~\bibnamefont
				{Gao}}, \bibinfo {author} {\bibfnamefont {S.~A.}\ \bibnamefont {Yang}}, \
			and\ \bibinfo {author} {\bibfnamefont {Q.}~\bibnamefont {Niu}},\ }\href@noop
		{} {\bibfield  {journal} {\bibinfo  {journal} {Physical Review Letters}\
			}\textbf {\bibinfo {volume} {112}},\ \bibinfo {pages} {166601} (\bibinfo {year}
			{2014})}\BibitemShut {NoStop}%
		\bibitem [{\citenamefont {Chang}\ and\ \citenamefont {Niu}(1996)}]{Chang1995}%
		\BibitemOpen
		\bibfield  {author} {\bibinfo {author} {\bibfnamefont {M.~C.}\ \bibnamefont
				{Chang}}\ and\ \bibinfo {author} {\bibfnamefont {Q.}~\bibnamefont {Niu}},\
		}\href@noop {} {\bibfield  {journal} {\bibinfo  {journal} {Physical Review
					B}\ }\textbf {\bibinfo {volume} {53}},\ \bibinfo {pages} {7010} (\bibinfo
			{year} {1996})}\BibitemShut {NoStop}%
		\bibitem [{\citenamefont {Karplus}\ and\ \citenamefont
			{Luttinger}(1954)}]{Karplus:1954PR}%
		\BibitemOpen
		\bibfield  {author} {\bibinfo {author} {\bibfnamefont {R.}~\bibnamefont
				{Karplus}}\ and\ \bibinfo {author} {\bibfnamefont {J.}~\bibnamefont
				{Luttinger}},\ }\href@noop {} {\bibfield  {journal} {\bibinfo  {journal}
				{Physical Review}\ }\textbf {\bibinfo {volume} {95}},\ \bibinfo {pages}
			{1154} (\bibinfo {year} {1954})}\BibitemShut {NoStop}%
		\bibitem [{\citenamefont {Cominotti}\ and\ \citenamefont
			{Carusotto}(2013)}]{Cominotti2013}%
		\BibitemOpen
		\bibfield  {author} {\bibinfo {author} {\bibfnamefont {M.}~\bibnamefont
				{Cominotti}}\ and\ \bibinfo {author} {\bibfnamefont {I.}~\bibnamefont
				{Carusotto}},\ }\href@noop {} {\bibfield  {journal} {\bibinfo  {journal}
				{Epl}\ }\textbf {\bibinfo {volume} {103}} (\bibinfo {year}
			{2013})}\BibitemShut {NoStop}%
		\bibitem [{\citenamefont {Gao}\ \emph {et~al.}(2015)\citenamefont {Gao},
			\citenamefont {Yang},\ and\ \citenamefont {Niu}}]{Gao2015}%
		\BibitemOpen
		\bibfield  {author} {\bibinfo {author} {\bibfnamefont {Y.}~\bibnamefont
				{Gao}}, \bibinfo {author} {\bibfnamefont {S.~A.}\ \bibnamefont {Yang}}, \
			and\ \bibinfo {author} {\bibfnamefont {Q.}~\bibnamefont {Niu}},\ }\href@noop
		{} {\bibfield  {journal} {\bibinfo  {journal} {Physical Review B}\ }\textbf {\bibinfo {volume} {91}},\ \bibinfo
			{pages} {214405} (\bibinfo {year} {2015})}\BibitemShut {NoStop}%
		\bibitem [{sup()}]{supmat}%
		\BibitemOpen
		\href@noop {} {}\bibinfo {note} {See Supplemental Material for additional
			details.}\BibitemShut {Stop}%
		\bibitem [{\citenamefont {Xiao}\ \emph {et~al.}(2005)\citenamefont {Xiao},
			\citenamefont {Shi},\ and\ \citenamefont {Niu}}]{Xiao2005}%
		\BibitemOpen
		\bibfield  {author} {\bibinfo {author} {\bibfnamefont {D.}~\bibnamefont
				{Xiao}}, \bibinfo {author} {\bibfnamefont {J.}~\bibnamefont {Shi}}, \ and\
			\bibinfo {author} {\bibfnamefont {Q.}~\bibnamefont {Niu}},\ }\href@noop {}
		{\bibfield  {journal} {\bibinfo  {journal} {Physical Review Letters}\
			}\textbf {\bibinfo {volume} {95}},\ \bibinfo {pages} {137204} (\bibinfo {year}
			{2005})}\BibitemShut {NoStop}%
		\bibitem [{\citenamefont {Duval}\ \emph {et~al.}(2005)\citenamefont {Duval},
			\citenamefont {Horvath}, \citenamefont {Horvathy}, \citenamefont {Martina},\
			and\ \citenamefont {Stichel}}]{Duval2005}%
		\BibitemOpen
		\bibfield  {author} {\bibinfo {author} {\bibfnamefont {C.}~\bibnamefont
				{Duval}}, \bibinfo {author} {\bibfnamefont {Z.}~\bibnamefont {Horvath}},
			\bibinfo {author} {\bibfnamefont {P.~A.}\ \bibnamefont {Horvathy}}, \bibinfo
			{author} {\bibfnamefont {L.}~\bibnamefont {Martina}}, \ and\ \bibinfo
			{author} {\bibfnamefont {P.}~\bibnamefont {Stichel}},\ }\href@noop {} {\bibfield  {journal} {\bibinfo  {journal}
				{Mod. Phys. Lett. B }\
			}\textbf {\bibinfo {volume} {20}},\ \bibinfo {pages} {373} (\bibinfo {year}
			{2006})}\BibitemShut {NoStop}%
		\bibitem [{\citenamefont {Bliokh}(2006)}]{Bliokh2006}%
		\BibitemOpen
		\bibfield  {author} {\bibinfo {author} {\bibfnamefont {K.~Y.}\ \bibnamefont
				{Bliokh}},\ }\href@noop {} {\bibfield  {journal} {\bibinfo  {journal}
				{Physics Letters, Section A: General, Atomic and Solid State Physics}\
			}\textbf {\bibinfo {volume} {351}},\ \bibinfo {pages} {123} (\bibinfo {year}
			{2006})}\BibitemShut {NoStop}%
		\bibitem [{\citenamefont {Gosselin}\ \emph {et~al.}(2006)\citenamefont
			{Gosselin}, \citenamefont {M{\'{e}}nas}, \citenamefont {B{\'{e}}rard},\ and\
			\citenamefont {Mohrbach}}]{Gosselin2006}%
		\BibitemOpen
		\bibfield  {author} {\bibinfo {author} {\bibfnamefont {P.}~\bibnamefont
				{Gosselin}}, \bibinfo {author} {\bibfnamefont {F.}~\bibnamefont
				{M{\'{e}}nas}}, \bibinfo {author} {\bibfnamefont {A.}~\bibnamefont
				{B{\'{e}}rard}}, \ and\ \bibinfo {author} {\bibfnamefont {H.}~\bibnamefont
				{Mohrbach}},\ }\href@noop {} {\bibfield  {journal} {\bibinfo  {journal}
				{Europhysics Letters}\ }\textbf {\bibinfo {volume} {76}},\ \bibinfo {pages}
			{651} (\bibinfo {year} {2006})}\BibitemShut {NoStop}%
		\bibitem [{\citenamefont {Peierls}(1933)}]{Peierls:1933ZPhys}%
		\BibitemOpen
		\bibfield  {author} {\bibinfo {author} {\bibfnamefont {R.}~\bibnamefont
				{Peierls}},\ }\href@noop {} {\bibfield  {journal} {\bibinfo  {journal}
				{Zeitschrift f{\"u}r Physik}\ }\textbf {\bibinfo {volume} {80}},\ \bibinfo
			{pages} {763} (\bibinfo {year} {1933})}\BibitemShut {NoStop}%
		\bibitem [{\citenamefont {Kraus}\ \emph {et~al.}(2014)\citenamefont {Kraus},
			\citenamefont {Zilberberg},\ and\ \citenamefont
			{Berkovits}}]{Zilberberg2014}%
		\BibitemOpen
		\bibfield  {author} {\bibinfo {author} {\bibfnamefont {Y.~E.}\ \bibnamefont
				{Kraus}}, \bibinfo {author} {\bibfnamefont {O.}~\bibnamefont {Zilberberg}}, \
			and\ \bibinfo {author} {\bibfnamefont {R.}~\bibnamefont {Berkovits}},\
		}\href@noop {} {\bibfield  {journal} {\bibinfo  {journal} {Phys. Rev. B}\
			}\textbf {\bibinfo {volume} {89}},\ \bibinfo {pages} {161106} (\bibinfo
			{year} {2014})}\BibitemShut {NoStop}%
		\bibitem [{\citenamefont {{Lee}}\ \emph {et~al.}(2018)\citenamefont {{Lee}},
			\citenamefont {{Wang}}, \citenamefont {{Chen}},\ and\ \citenamefont
			{{Zhang}}}]{Lee2018}%
		\BibitemOpen
		\bibfield  {author} {\bibinfo {author} {\bibfnamefont {C.~H.}\ \bibnamefont
				{{Lee}}}, \bibinfo {author} {\bibfnamefont {Y.}~\bibnamefont {{Wang}}},
			\bibinfo {author} {\bibfnamefont {Y.}~\bibnamefont {{Chen}}}, \ and\ \bibinfo
			{author} {\bibfnamefont {X.}~\bibnamefont {{Zhang}}},\ }\href@noop {}
		{\bibfield  {journal} {\bibinfo  {journal} {Phys. Rev. B}\
			}\textbf {\bibinfo {volume} {98}},\ \bibinfo {pages} {094434} (\bibinfo
			{year} {2018})}\BibitemShut {NoStop}%
	\end{thebibliography}
	
\end{document}